\documentclass[onecolumn,showpacs,amsmath,amssymb,prd,nofootinbib]{revtex4}
\usepackage{epsfig}
\usepackage{enumerate}
\def\beq{\begin{equation}}
\def\eeqno#1{\label{#1}\end{equation}}

\def\rarrow{\rightarrow }

\def\dleft{\rlap{{\it D}}\raise 8pt
\hbox{$\scriptscriptstyle\Leftarrow$}}
\def\dright{\rlap{{\it
D}}\raise 8pt\hbox{$\scriptscriptstyle\Rightarrow$}}

\def\kms{~{\rm km~s^{-1}}}
\def\cmss{~{\rm cm~s^{-2}}}
\def\kpc{~{\rm Kpc}}
\def\mpc{~{\rm Mpc}}
\def\grcm{~{\rm gr~cm^{-3}}}

\def\az{a_{0}}

\def\l0{\ell_{0}}

\def\rar{\rightarrow}
\def\s{\sigma}

\def\l{\lambda}

\def\f{\phi}

\def\r{\rho}

\def\z{\zeta}

\def\A{\mathcal{A}}

\def\o{\omega}

\def\xlimin{{x\rarrow\infty \atop{\raise 1pt\hbox to 30pt
{\rightarrowfill}}}}
\def\limlim#1#2{{#1\rarrow #2 \atop{\raise 1pt\hbox to 30pt
{\rightarrowfill}}}}

\def\vr{{\bf r}}

\def\vv{{\bf v}}

\def\vg{{\bf g}}

\def\va{{\bf a}}

\def\M{\mathcal{M}}

\def\_#1{_{\scriptscriptstyle #1}}
\def\^#1{^{\scriptscriptstyle #1}}
\def\baz{\bar a_0}

\def\azg{\A_0}
\def\azgs{\A^*_0}

\def\RM{r\_M}

\begin{document}
\title{Cosmological variation of the MOND constant: Secular effects on galactic systems}
\author{Mordehai Milgrom}
\affiliation{Department of Particle Physics and Astrophysics, Weizmann Institute}

\begin{abstract}
The proximity of the MOND acceleration constant with cosmological accelerations--for example, $\az\approx cH_0/2\pi$--points to its possibly decreasing with cosmic time. I begin to consider the secular changes induced in galactic systems by such presumed variations, assumed adiabatic. It is important to understand these effects, in isolation from other evolutionary influences, in order to identify or constrain $\az$ variations by detection of induced effects, or lack thereof. I find that as long as the system is fully in the deep-MOND regime--as applies to many galactic systems--the adiabatic response of the system obeys simple scaling laws. For example, in a system that would be stationary for fixed $\az$,
the system expands homologously as $\az^{-1/4}$, while internal velocities decrease uniformly as $\az^{1/4}$.
If $\az\propto cH$ at all relevant times, this change amounts to a factor of $\sim 2.5$ since redshift 10. For rotating systems, the angular frequency $\Omega\propto \az^{1/2}$. The accelerations increase relative to $\az$ as $\az^{-1/4}$, pushing the system towards the Newtonian regime. All this follows from the appearance of $\az$ in MOND and the scale invariance of the deep-MOND limit--two basic tenets of MOND.
More complicated evolution ensues when parts of the system become Newtonian, or are so from inception. For example, these parts may become unstable since they are not protected by MOND's stabilizing effects. The existence of such regions also modifies the MONDian regime, since they affect the potential everywhere, and since constituents might migrate between the Newtonian and MONDian regimes. Studying these last effects would require detailed numerical calculations.

\end{abstract}
\maketitle
\section{Introduction}
MOND \cite{milgrom83,fm12,milgrom14,milgrom14a} is a paradigm that contends to supersede Newtonian dynamics (ND) and general relativity, departing from them greatly for very small accelerations: at or below the MOND acceleration constant $\az$. The basic tenets of MOND are: approach to standard dynamics for accelerations much above $\az$, and space-time scale invariance in the opposite limit: the deep-MOND limit (DML) \cite{milgrom09}.
MOND's underlying motivation has been to account for the mass discrepancies in the Universe without ``dark'' components. $\az$, beside its role as an approximate validity boundary of standard dynamics, also appears saliently in many ``MOND laws'' that are predicted to govern galactic dynamics \cite{milgrom14}. Examples of such MOND laws are that the asymptotic circular velocity, $V\_{\infty}$, around any isolated, bounded system of total (baryonic) mass, $M$, is constant, and is given by
\beq V\_{\infty}^4=MG\az. \eeqno{jump}

These predicted ``MOND laws'' are indeed well obeyed by galactic systems, from dwarf-spheroidal and dwarf-spiral galaxies through giant elliptical and spiral galaxies, to galaxy groups, as probed by either constituent motions or gravitational lensing. In galaxy clusters, MOND explains away most, but not all, of the observed mass discrepancy.

It is now quite evident that $\az$ appears ubiquitously in observed galactic phenomenology (e.g., Refs. \cite{fm12,milgrom14,wl14}). Its value, determined consistently from these observations is
$\az\approx 1.2\times 10^{-8}\cmss$. It has been noted from MOND's very advent \cite{milgrom83,milgrom89} that $\az$ is near in value to some cosmologically significant accelerations; For example,
\beq \baz\equiv 2\pi \az\approx a\_H(0)\equiv cH_0,~~~~~~~~~~~\baz\approx a\_{\Lambda}\equiv c^2(\Lambda/3)^{1/2}, \eeqno{coinc}
where $a\_H\equiv cH$ is the acceleration associated with the cosmological expansion rate, $H$ (the Hubble constant), and $a\_H(0)$ is its present value, and $\Lambda$ the observed equivalent of a cosmological constant. This numerical ``coincidence'', if fundamental, may have far reaching ramifications for MOND and for gravity in general (e.g., \cite{milgrom09,milgrom14a}).
\par
One of these potential ramifications is that $\az$ varies with cosmic time \cite{milgrom83,milgrom89,milgrom09}.
If indeed $H_0$ enters galaxy dynamics through $\az$ today, then $H$ should have done so also in earlier times when it was larger. The specific dependence of $\az$ on cosmic time cannot be determined from first principles without a better understanding of MOND's origin than we now have. In addition, the fact that today $a\_{\Lambda}\approx a\_H$ makes it difficult to even know which of these cosmic accelerations $\az$ is linked to.\footnote{Here I speak of $\az$ as it enters dynamics in systems small on cosmological scales. It may well lose its meaning altogether in the more fundamental theory that must underlie present MOND, or in the applications to cosmology. This would be similar to the Earth's free-fall acceleration constant, $g$, which is so useful in describing near-Earth-surface phenomena, but loses its significance in the general context of gravity. In other words, $\az$ may turn out to be only an effective constant in an effective theory that applies in galactic systems.}
It might be linked only with $a\_{\Lambda}$, in which case it could be time independent. Otherwise, it might depend on both $\Lambda$ and $H$, or more generally on the expansion history. For example, we could have $\az\sim a\_H$ at all relevant cosmic times,
and, during the flat-space, matter-dominated era, mostly relevant to galaxy evolution, $H\propto t^{-1}\propto (1+z)^{3/2}$, with $z$ the cosmological redshift. (This breaks down for $z\lesssim 0.5$ when the cosmological constant is important.) For the sake of concreteness, I shall use examples with this time dependence, but the exact dependence of $\az$ on time is not material to our discussion, as long as we can assume that its variation time scale is cosmological, i.e. of order $H^{-1}$.
\par
Using $\az$, it is useful to define a MOND length $\ell\_M\equiv c^2/\az$, a MOND density $\r\_M\equiv \az^2/c^2G$, and a MOND mass
$\M\_M\equiv (4\pi/3)\r\_M\ell\_M^3$ (see, e.g., Ref.  \cite{milgrom14c}). Interestingly, the MOND length and density corresponding to today's value of $\az$ are $\ell\_M(z=0)\approx 2400\mpc$, and
$\r\_M(z=0)\approx 2.4\times 10^{-30}\grcm$, very nearly today's Hubble distance, and the density of cosmological ``dark matter'' that standard dynamics require. The latter is also of the order of the ``dark energy'' density that observations require--an expression of one of the known ``cosmological coincidences''. If $\az$ is  found not to vary with cosmic time, it would be natural to associate it with a cosmological constant, and there are several suggestions on how they may, in fact, be physically connected. If, however, $\az$ is found to vary with cosmic time, especially if indeed $\az\propto H$, then the value of the MOND density
 is always of the order of the the density of cosmological ``dark matter'' during the relevant era.
This fact might underlie (in a way not yet understood) the way MOND explains away the need for  cosmological ``dark matter'' in a similar vein to its obviating the need for galactic dark matter.
\par
Reference \cite{bs08} studied how $\az$ might vary in the relativistic MOND formulation TeVeS \cite{bek04}, and Ref. \cite{deffayet14} studied consequences of variable $\az$ on the expansion history of the Universe in their nonlocal, relativistic formulation of MOND \cite{deffayet11}.

$\az$ variations may, in principle, be tested directly by measuring its value at various $z$ from galaxy dynamics, employing the MOND laws that involve $\az$. For example, Refs. \cite{milgrom08,limbach08} tried to do this looking at the $z$ dependence of relation (\ref{jump}). But it has to be said that the data were then, and are still now, not good enough, and the interpretation rife with such systematics, that this check is quite ineffective at present. Such tests require good knowledge of the baryonic mass of high-$z$ galaxies, which requires understanding of their mass-to-light ratios and how they depend on $z$, and a good determination of the contribution of the gas, both atomic and molecular. They also require a good determination of the asymptotic rotational speed, which requires measurement of extended rotation curves of relatively unperturbed galaxies, and their  inclinations. All these are not readily measurable at high $z$, at present.

There are also indirect approaches (alluded to in Ref. \cite{milgrom89})--less decisive, but requiring less information. They require understanding of the secular effects that $\az$ variations have on the structural attributes of galaxies, such as their size or internal velocities, to see if we can identify  or exclude trends with $z$ that are predicted by variable $\az$.

Clearly, variation of $\az$, which enters galactic dynamics so decidedly,
would induce major variations in the structure of galactic systems, over and beyond other influences, such as mergers, matter accretion and ejection, star formation, instabilities, etc. (reviewed, e.g., in Ref. \cite{kc04}).\footnote{In MOND there are also possible influences of a variable external field, through the external-field effect.} There may be times in the life of a system, when one of these effects dominate, and in any event each may produce its own distinct behavior.
We need to understand each of these effects separately before we put them together in modeling galaxy evolution.
\par
Galaxy evolution is still a little-understood subject. For example, we do not understand well how elliptical galaxies and bulges of spiral galaxies form or what role is played by mergers and secular processes (see Ref. \cite{kc04} for a review). In MOND, even with a constant $\az$, these processes may be rather different than in the $\Lambda$CDM picture. Such differences are discussed, for example, in Refs. \cite{morishima94,tiret07,tiret08,combes10}, which considered the MOND evolution in disc galaxies, in Ref. \cite{combes14}, which looked at bulge formation, and in Ref. \cite{zhao13}, which considered interesting aspects of the history of the Milky Way-Andromeda binary system in MOND, in contrast with what is expected in the $\Lambda$CDM paradigm.
Variations of $\az$ would even accentuate the differences.

Thus, one indirect way to pinpoint variation of $\az$ with cosmic time is to study the redshift evolution of galactic structure and dynamics, to see if some of it can be attributed to such variation.
For example, there are various studies of z evolution of galaxy sizes, e.g., the recent Ref. \cite{curtis14} (and other references therein), where galaxy sizes, $D$, is fitted with $D\propto (1+z)^{n}$ for the redshift range $4\le z\le 8$. They find for brighter galaxies $n=-0.34\pm 0.29$, and for dimmer ones $n=-0.57\pm0.76$. This is consistent with no evolution, but also with the typical power of $n=-3/8$, which I find below for low-acceleration galaxies.\footnote{The exact relevance of these observations to MOND has to be checked since at the radii considered, of $0.2-2 \kpc$, we may still be in the Newtonian, not the low-acceleration, regime.}
\par
Another indirect method to identify variation of $\az$ is to try to identify in the structure of present-day (low-$z$) galaxies telltale traits resulting from past $\az$ variations.
For example, there are MOND arguments about why the mean acceleration of the baryons in isolated systems--such globular clusters, galaxies of all types, galaxy groups and clusters--does not much exceed $\az$ \cite{milgrom89a,milgrom84,milgrom14}.
But, in fact, $\az$ seems to be not only an approximate upper bound on, but an accumulation value of, the mean acceleration in the baryon body (e.g., \cite{wl14}). Why is this? I find, for example, that due to the cosmological decrease of $\az$, the ratio of the mean acceleration to $\az$ increases and might halt when it reaches $\sim 1$.

None of the above indirect tactics are easy to employ, given that there are competing effects that need to be disentangled, and that galaxy evolution is a little-understood subject, even without the added effects of $\az$ variations. But clearly, we need to explore these latter effects in isolation to understand what part they may have in shaping galaxies.

Here I lay the theoretical grounds for such a study and consider these important effects in some detail, for the first time to my knowledge.

In Sec. \ref{present}, I present the general considerations that underlie the effects of $\az$ variation on galactic structure.
Section \ref{DML} deals specifically with systems in the scale-invariant DML, where, it is found, the induced adiabatic evolution obeys some clear-cut scaling.
Section \ref{newt} discusses superficially further evolution under partial Newtonian sway.

\section{\label{present}Secular effects of adiabatic $\az$ variation}

Consider a galactic system, such as a galaxy, that was well formed at redshift $z$ of $\sim 1-10$. During its life since then it may have been subject to various episodic influences. But on top of these it might have undergone secular structure changes due to $\az$ variations, which I concentrate on here and consider in isolation. Such changes must occur in MOND if $\az$ varies, because $\az$ enters strongly various predicted MOND relations between masses, velocities, and sizes. We saw one such predicted relation in Eq. (\ref{jump}). Another is the MOND prediction that transition from standard dynamics to MOND dynamics should always occur around accelerations $\sim\az$. For example, in a disc galaxy the transition is predicted to occur around the radius where $V^2(r)/r=\az$. In a well-concentrated system of mass $M$--a system that is well contained within its MOND radius $\RM\equiv(MG/\az)^{1/2}$-- transition effects will occur at $\RM$.\footnote{Such a transition is absent in systems that are fully in the DML, for which $V^2(r)/r<\az$ everywhere.} Yet another predicted MOND relation is a rather general ``virial relation'' between the mean-squared velocity and the masses, in DML systems of point masses \cite{milgrom14b},\footnote{This is not a contradiction in terms. Even if gravitational accelerations near a point mass are arbitrarily high, what matters in defining the DML are the center-of-mass accelerations to which these masses are subject.}
\beq \langle V^2\rangle=\frac{2}{3}(MG\az)^{1/2}[1-\sum_p (m_p/M)^{3/2}], \eeqno{virial}
where $M=\sum_p m_p$, which applies in modified-gravity, MOND theories.

It is clear from these relations that if $\az$ varies, the asymptotic circular speed for any system, and the root-mean-square speed in DML systems must scale as $\az^{1/4}$, while the MOND and transition radii scale as $\az^{-1/2}$. This should hold whether the $\az$ variations are adiabatic or not.

But rather more can be said if we know that the $\az$ variations are adiabatic, namely that they occur on time scales much larger than the dynamical times in the galaxy.

For a system of characteristic size $R$ and internal velocity $V$, the quantity $\theta_a=V\az/R\dot\az$
may be taken as a measure of adiabaticity.
For example, if $\az\propto H$, and $H$ varies as a power of cosmic time, we have $\dot\az/\az\approx H$, so $\theta_a=V/RH$, which I shall, henceforth, use as the default.
Today, galaxies have $\theta_a\gg 1$. For example, for low-acceleration galaxies, such as
a dwarf spheroidal with $R=0.5\kpc$ and three-dimensional velocity dispersion of $\s_3=8\kms$,
$\theta_a\approx 230$, or a low-surface-brightness disc galaxy with rotational speed of $50\kms$ and $R=5\kpc$, $\theta_a=140$. In a galaxy like the Milky Way, with an asymptotic rotational speed of $\approx 220 \kms$, MOND effects become appreciable beyond a radius of $R\sim 15\kpc$, where $\theta_a\sim 210$.

What was the situation in the cosmological past? We shall see that in low-acceleration systems, the dynamical times
are predicted to scale as $\az^{-1/2}$ under adiabatic variations of $\az$. By the above default choice of $\az$ variations, $\az/\dot\az\propto H^{-1}\propto \az^{-1}$. Thus, $\theta_a\propto \az^{-1/2}\propto (1+z)^{-3/4}$ was smaller in the past. But we see that with the high values of $\theta_a$ today, even as far back as $z$ of a few tens, $\theta_a$ was already much larger than 1; thus, adiabaticity holds to a good approximation in the relevant period.

If we consider the period since $z\sim 10$ as relevant to our discussion, we expect $\az$ to have decreased by a factor of $\sim 35$ since then. We shall see that for low-acceleration systems, this would have induced an increase in size $\propto \az^{-1/4}$, i.e., by a factor of $\sim 2.5$, and the same decrease in internal velocities.

\section{\label{DML}Systems in the Deep-MOND regime}

Systems whose dynamics are characterized by accelerations that are all much smaller than $\az$ are described by the DML of the applicable MOND theory. It is a basic tenet of MOND (e.g., Refs. \cite{milgrom09,milgrom14}) that this limit is space-time scale invariant; i.e., invariant to $(t,\vr)\rar\l(t,\vr)$ for any constant $\l>0$. This is required of the nonrelativistic version of a MOND theory, and more generally of the weak-field limit of its relativistic version.
\par
One consequence of this scale invariance is that $\az$ and $G$ cannot appear separately in a DML theory, only the product $\azg=G\az$ can appear \cite{milgrom09} (besides masses and possibly $c$ if we are dealing with gravitational lensing, for example).\footnote{This is because a theory being scale invariant is tantamount to all its constants retaining their value under a simultaneous change of the units of length and time by the same factor. This holds for masses, $c$, and $\azg$, but not $G$ or $\az$.} Since cosmological variations of $G$ are stringently constrained, I assume that  $\azg\propto\az$.

There are many systems that are today fully in the DML. Examples are dwarf spheroidal satellites of large galaxies such as the Milky Way and the Andromeda, many low-surface-brightness disc galaxies, and loose galaxy groups. I will show that when $\az$ decreases with cosmic time, the accelerations in a DML system decrease, but not as fast as $\az$. Thus, such systems were even deeper in the DML in the past if our discussion here is valid.
And, even some systems that are today partly in the high-acceleration regime might have been fully in the DML in their past. For such systems, the predictions are rather clear-cut.
\par
We restrict ourselves then to the strict DML field equations of some MOND theory, for some fixed value, $\azgs$, of $\azg$, and consider a solution that describes an isolated, nonrelativistic, self-gravitating galactic system made of many point bodies of masses $m_i$, whose trajectories are $\vr_i\^*(t\^*)$, and possibly of a continuous fluid component with a density field $\r\^*(\vr\^*,t\^*)$, velocity field $\vv\^*(\vr\^*,t\^*)$, temperature $T\^*(\vr\^*,t\^*)$, pressure $P\^*(\vr\^*,t\^*)$, and an equation of state of a monatomic ideal gas. The fluid is assumed an ideal gas as interactions may introduce additional constants (through the equation of state) that break the scaling laws we employ below. It is assumed effectively ``monatomic'' so as not to involve internal degrees of freedom in the considerations of adiabatic invariance.
\par
To grasp the scaling involved, consider, as a path to our derivation, a  change of the units of length and time under which the values of lengths change as $\ell\rar \z\ell$ and those of times
$t\rar (\z/\eta)t$. The value of $\azg$ changes from $\azgs$ in the old units to $\azg=\eta^4\azgs$ in the new ones.
All equations are, of course, invariant under any change of units.
Thus, the system with the same masses, but trajectories $\vr_i(t)=\z\vr_i\^*(\eta t/\z)$ (and fluid attributes transforming according to their dimensions), is a DML solution of the theory with the constant $\azg=\eta^4\azgs$, for any choice of $\z,\eta>0$.\footnote{Note the appearance of $\eta t/\z$, not $\z t/\eta$, in the dependent, time variable. This is the correct expression, but it is a possible source of confusion.} Velocities become $\vv_i(t)=\eta\vv_i\^*(\eta t/\z)$.
In the present context we ask whether all the adiabatic invariants of the original system take the same values in the new units. Inasmuch as these invariants have dimensions of actions (i.e., $[\ell]^2[t]^{-1}$), we see that they do remain the same if $\z=\eta^{-1}$.
\par
We conclude from this argument that a system described by
$$\vr_i(t)=\eta^{-1}\vr_i\^*(\eta^2 t),~~~~~~~~\vv_i(t)=\eta\vv_i\^*(\eta^2 t),$$
\beq\va_i(t)=\eta^3\va_i\^*(\eta^2 t),~~~~~~~~\f(\vr,t)=\eta^2\f\^*(\eta\vr,\eta^2 t), \eeqno{uiop} satisfies the DML equations with constant $\azg=\eta^4\azgs$, and has the same values of all the action variables as the original system. Here, $\va_i$ are the accelerations,\footnote{One may wonder, since our arguments are based on dimensional analysis, how it is that $\va_i\propto \eta^3\propto \az^{3/4}$ scale differently from $\az$, when they have the same dimensions. The reason is that our dimensional arguments would indeed apply to the a problem where $\az\propto\eta^3$ as well, but then it requires at the same time that $G\propto\eta$. This is not the situation we want to describe. But in a theory where only $\azg=G\az$ appears, such effects are the same as those with $\az\propto \eta^4$ and $G$ fixed, which is the situation we want to describe.\label{jama}}  and $\f$ the gravitational potential fields. (For concreteness, I consider ``modified gravity'' MOND theories, where the nonrelativistic gravitational potential is modified. ``Modified-inertia'' theories require a separate description.) For the fluid component, the density and acceleration fields transform as
 \beq \r(\vr,t)=\eta^3\r\^*[\eta\vr,\eta^2 t],~~~ \vg(\vr,t)=\eta^3\vg\^*(\eta\vr,\eta^2 t),\eeqno{liput}
and $T$ scales as $\eta^2$; so $T$ scales as $\r^{2/3}$, as expected of adiabatic changes in a monatomic ideal gas.
If the system is well described by a phase-space distribution function, then
\beq f(\vr,\vv,t)=f\^*(\eta\vr,\eta^{-1}\vv,\eta^2t), \eeqno{dista}
the phase-space volume element being invariant.\footnote{We could use atomic physics, electromagnetism, or Newtonian-gravity systems such as stars, to define the units of acceleration in which $\az$ is said to vary. DML gravity alone does not provide us with a yardstick and a clock (e.g., Ref. \cite{milgrom14a}). In other words, had nonrelativistic DML gravity, which concerns us here, been the whole of physics, there would be no sense to speaking of the variations we discuss here. But, ``the rest of physics,'' while not directly involved in DML gravity, does provide us with yardsticks, such as the Bohr, or the Planck, radius, velocity standards, such as the speed of light, etc., relative to which we can measure the variations discussed here. The constants entering these standards, such as the speed of light, the electron charge, etc., are then not to be subject to the unit changes we employ here. Changing units in ``the whole of physics'' can teach us nothing, of course.}
\par
All that is said above concerns the relation between two systems that are governed by different, but fixed, values of $\azg$.
Now consider the putative adiabatic variation of $\azg$ from some initial value $\azgs$, according to $\azg(t)=\eta^4(t)\azgs$.
It is, of course, not possible to write exactly the solutions of the resulting theory in terms of solutions for the constant-$\azg$ problem. However, inspired by the above arguments, I make an educated conjecture regarding the limit of adiabatic changes of $\eta$.
It is based on viewing $\eta(t)$ as the oxymoron ``time-dependent constant,'' which takes into account the fact that $\eta$ is not the same at all times, but neglects its time derivatives when they appear. This treats the variation of $\eta$ as if it is kept constant over many dynamical times but is changed from one constant value to the next. This is the spirit of the adiabatic approximation.
At time $t$, consider a change of time and length units (which it is convenient to take as the inverse of the one above) applied to the time differentials: $dt=\eta^{-2}(t)dt\^*$. (Here, as above, nonstarred  quantities describe the variable-$\azg$ situation, and starred, or ``reduced,'' quantities correspond to the reference, constant-$\azg$ problem.) Thus,
\beq t\^*(t)=\int\_0\^{t}\eta^2(t)dt.   \eeqno{taitai}
Also, change the length units so that $d\vr=\eta^{-1}d\vr\^*$. Then, \beq\vv=\frac{d\vr}{dt}=\eta(t)\frac{d\vr\^*}{dt\^*}=\eta[t(t\^*)]\vv\^*,
\eeqno{kopyt} and quantities of the form $\vv d\vr$, which enter the action variables, are unaffected. Also, the value of $\azg$ in the starred units is $\azgs=\eta^{-4}\azg$, and so is time independent. However, we cannot proceed in the same vein, in an exact manner, to the equation relating $\dot\vv$ to the gravitational acceleration field, since $\dot\vv$, with $\vv$ scaling as in Eq.(\ref{kopyt}), does not have simple scaling with $\eta$. Likewise, we cannot simply write $\vr_i(t)$ in terms of $\vr_i\^*(t\^*)$. We can however do these if we neglect derivatives of $\eta$. Doing this, I end up with the conjecture that under adiabatic changes of $\azg$ the solutions of the problem can be written as in Eqs.(\ref{uiop}-\ref{dista}), but with the independent time variable $\eta^2t$ replaced with $t\^*$ given in Eq.(\ref{taitai}). For example,
\beq \vr_i(t)\approx\eta^{-1}(t)\vr_i\^*[t\^*(t)].\eeqno{mushasha}
It may be seen that, indeed, neglecting $\dot\eta$, this conjectured behavior follows consistently from applying the above changes of units at any given time.
For example, we get $\dot\vr(t)\approx\vv(t)$, and $\dot\vv\approx\vg$, if the equalities hold for the starred quantities for a fixed $\azgs$.
\par
Before discussing tests of the conjecture, note the following generalities.
\par
If the above conjecture is valid, systems retain all their dimensionless attributes, and only increase homologously in size, with the velocities everywhere decreasing uniformly.
Thus, for example, a DML exponential disc of scale length $h$ and scale height $z_0$ remains an exponential disc with $h$ and $z_0$ increasing as $\az^{-1/4}$, and all speeds, rotational as well as dispersions, decreasing as $\az^{-1/4}$. And, a DML isothermal sphere of a given anisotropy ratio, $\beta$ \cite{milgrom84} remains so with $\beta$ fixed, and the velocity dispersions decreasing like $\az^{1/4}$.
\par
The above adiabatic scaling behavior is a result of the fact that $\azg$ is the only relevant constant in the theory that is affected by the scaling. This, in turn, rests heavily on the two basic tenets of MOND: the appearance of $\az$ as the only new constant in MOND and the space-time scale invariance of the DML.
\par
In systems described by a theory that does involve $G$ separately--for example, systems with accelerations of order $\az$ or above--our adiabatic scaling behavior breaks down, because it would produce a change in $G$, as well as $\az$, and so would not account correctly for variation in $\az$ alone.\footnote{The rescaling of length and time units that underlies our derivation takes $G\rar\eta G$, $\az\rar \eta^3\az$. So the adiabatic scaling behavior we find holds in theories in a world where $G$ does not vary, $\az\propto \eta^4$, but only $\azg$ appears in the theory, and $\azg\propto\eta^4$; this is the situation that concerns us here for the DML (see also footnote \ref{jama}). But this behavior applies as well to theories where both constants appear, each varying according to its power, or to a theory such as Newtonian dynamics, where only $G$ appears, and varies in proportion to $\eta$. In MOND, this last situation is useful in describing DML systems embedded in a DML external field (the MOND external-field effect); see below.\label{gvar}}
In the same vein, when we deal with gravitational lensing in the DML, weak-field limit of a relativistic MOND theory, $c$ too appears, and would change to $\eta c$ had we applied the scaling indiscriminately ($\hbar$, if it appeared, would not be affected, having dimensions of action). Since we do not wish $c$ to vary with $\az$, we have to hold it fixed; otherwise, we would have concluded, e.g., that light-bending angles, which are dimensionless, do not vary with the $\azg$ variations. In fact, these angles, which scale as $V^2/c^2$, do vary as $\eta^2$.
\par
As with almost any result on the adiabatic behavior of systems, the ultimate test of our conjecture and its validity limits would be to conduct $N$-body simulations of various DML galactic systems under various
forms of adiabatically varying $\azg$. Here I perform an easier, but still strong, numerical test of the conjecture: Start with a DML potential
$\f\^*(\vr\^*)$, representing the smooth field of a system--a solution in the given MOND theory with a constant $\azg=\azgs$. By and large, I confine myself to time-independent potentials representing stationary systems.\footnote{I also ran successful tests for systems for which $\f\^*$ is time independent in a (slowly) rotating system of frequency $\Omega^*$;
so $\f\^*=\f\^*\_{\Omega^*}(\vr\^*)$. In this case the assumed potential in Eq. (\ref{hutrew}) is replaced by $\f(\vr,t)\approx \eta^2(t)\f\^*\_{\Omega(t)}[\eta(t)\vr]$, where $\Omega(t)=\eta^2(t)\Omega^*$, which is quasistatic in a rotating frame of decreasing frequency $\Omega(t)$.} The system is envisaged as a collection of many particles on various orbits in  $\f\^*(\vr\^*)$ that, in turn, give rise to $\f\^*(\vr\^*)$ self-consistently.
 \par
According to our conjecture, the potential of the real system is \beq\f(\vr,t)\approx \eta^2(t)\f\^*[\eta(t)\vr]. \eeqno{hutrew}
I computed many particle trajectories, $\vr(t)$, in various such spherical, biaxial, and triaxial potentials, letting $\azg$ decrease from beginning to end by large factors.
The aim is to check to what extent these orbits are of the form given in Eq.(\ref{mushasha}), as dictated by our conjecture. If they are, then {\it automatically, and without recourse to adiabaticity}, the potential produced by the collection of particles moving on such orbits, is the one we assumed to begin with, given by Eq. (\ref{hutrew}), sustaining the conjecture.
\par
The reason for this is that in a nonrelativistic situation, the potential responds instantly to changes in the positions of the sources (there are no time derivatives in the equations that determine the potential from the positions of the particles, so we do not even have to neglect $\dot\eta$ in the derivation). All that needs to be used is the invariance to change of the units by the appropriate powers of $\eta(t)$. To recapitulate, adiabaticity enters only in how the variation of the potential affects the orbits, not in how the orbits variation affects the potential.
\par
Full $N$-body calculations would check all this simultaneously for a large number of particles on orbits that self-consistently give the potential. But still, our present test is a strong consistency check of our conjecture.
\par
More specifically, this test is done as follows: For each computed orbit, $\vr(t)$, I calculate the
``unscaled'' solution \beq\bar\vr\^*(t\^*)\equiv\eta[t(t\^*)]\vr[t(t\^*)],
\eeqno{redar}
which inverts Eq.(\ref{mushasha}), between $t\_i$ and $t\_f$.  According to our conjecture this segment of the orbit should be given, within the approximation, by $\vr\^*(t\^*)$, calculated for the same initial conditions, for a fixed $\azg=\azgs$, in the time interval between $t\_i$ and $t\_i+\Delta t\^*$, with the ``reduced'' duration \beq\Delta t\^*=\int_{t\_i}^{t\_f}\eta^2(t)dt.\eeqno{redtime}
\par
\begin{figure}
\begin{tabular}{lll}
\includegraphics[width=0.35\columnwidth]{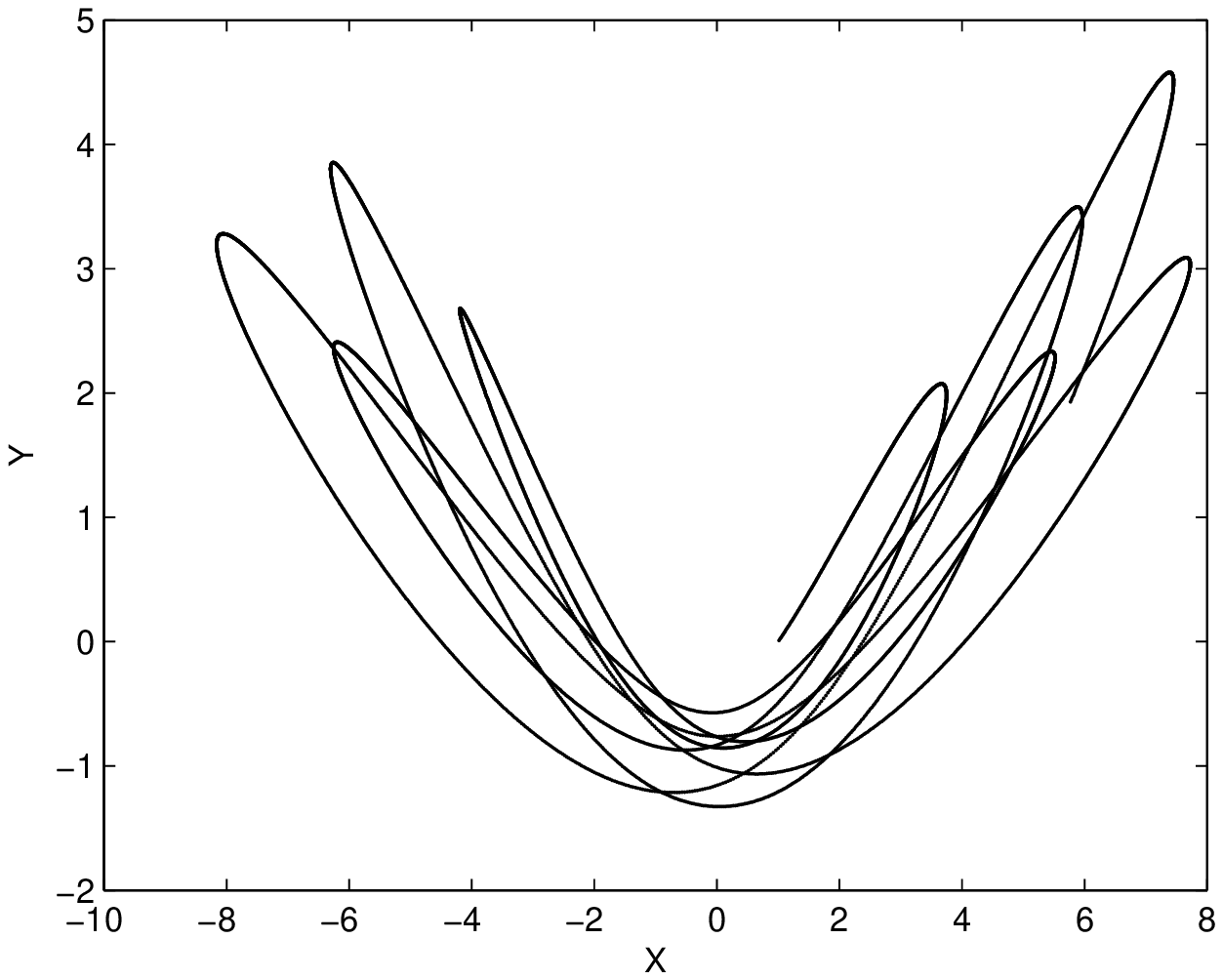}
\includegraphics[width=0.35\columnwidth]{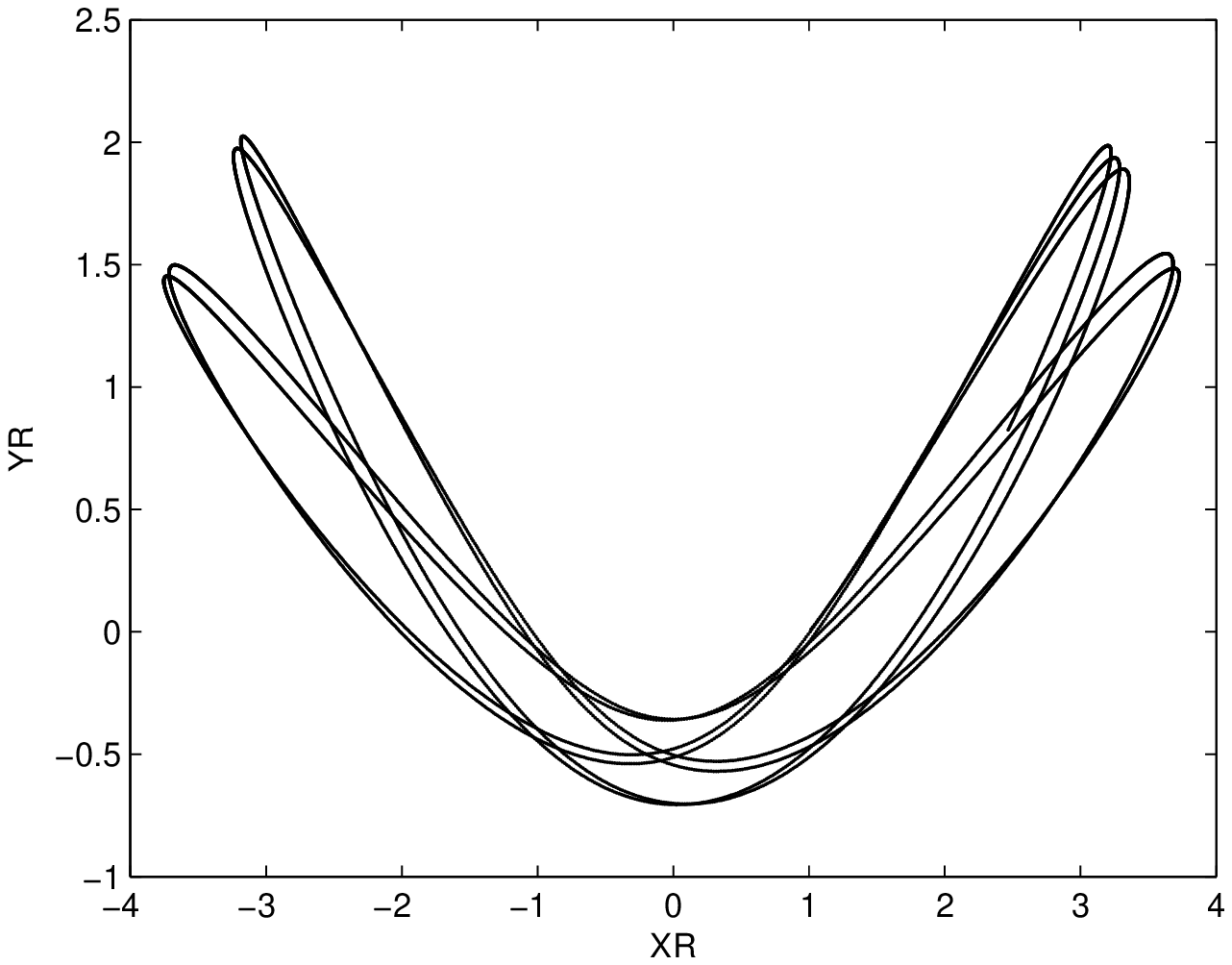}
\includegraphics[width=0.35\columnwidth]{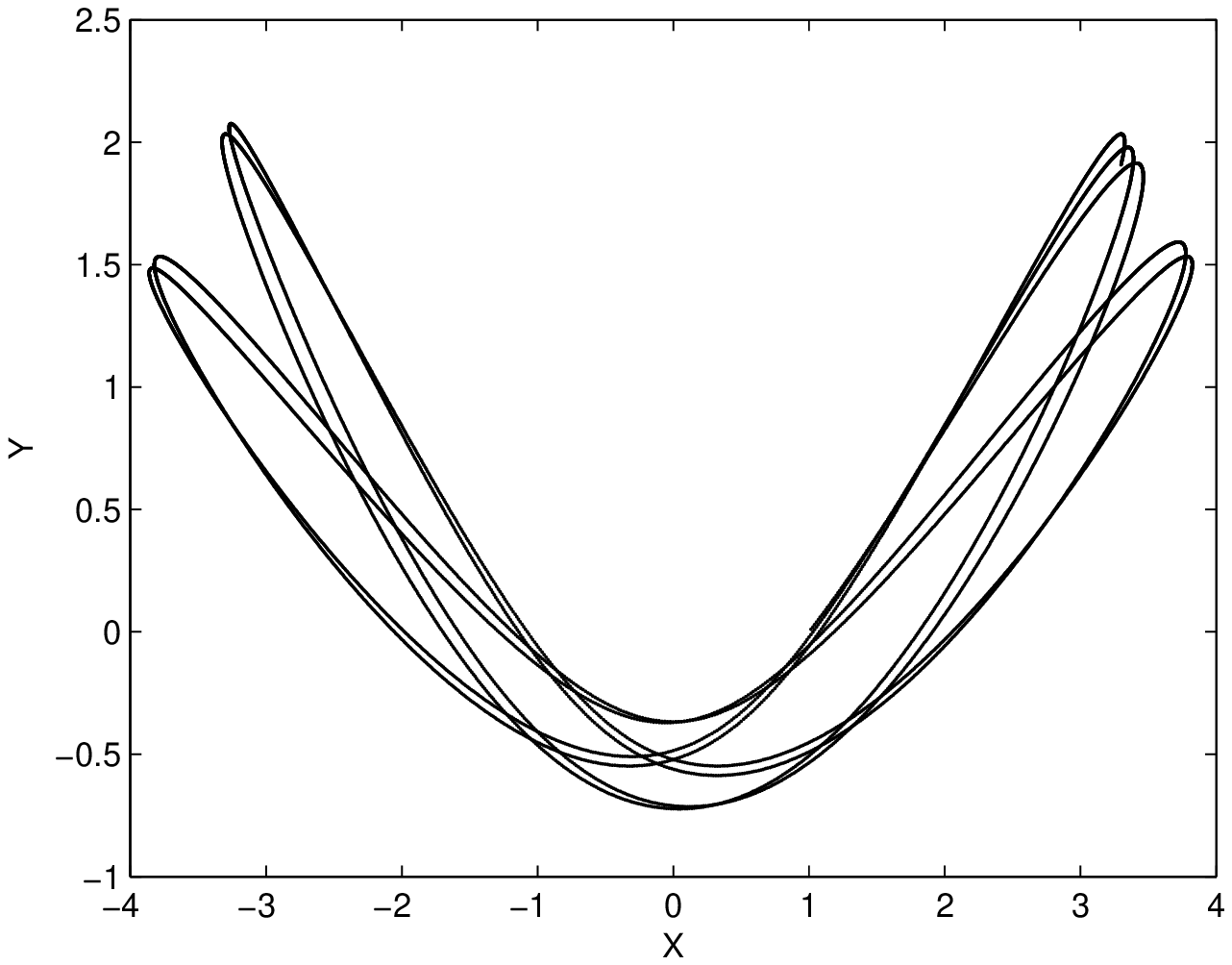}
\end{tabular}
\begin{tabular}{lll}
\includegraphics[width=0.35\columnwidth]{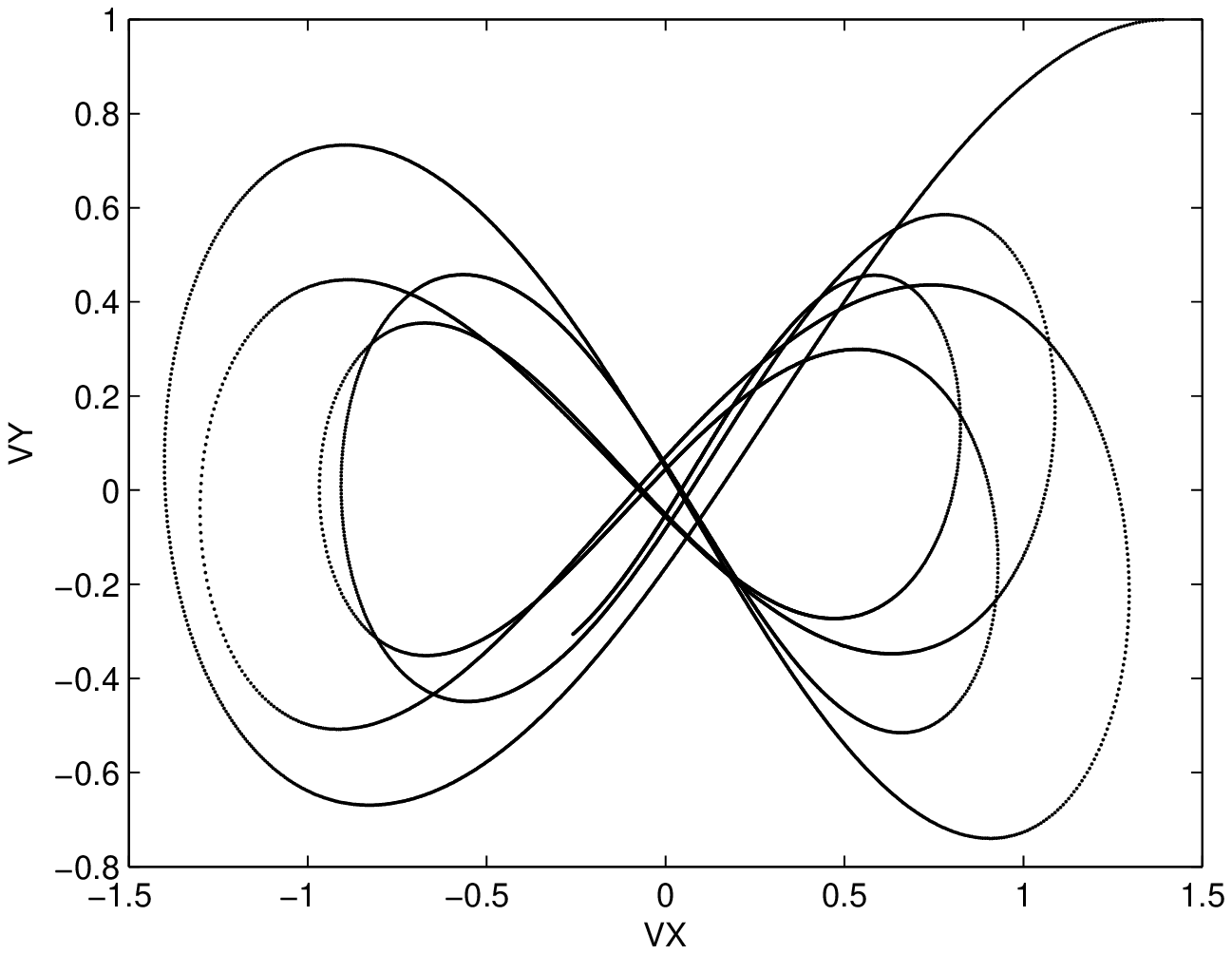}
\includegraphics[width=0.35\columnwidth]{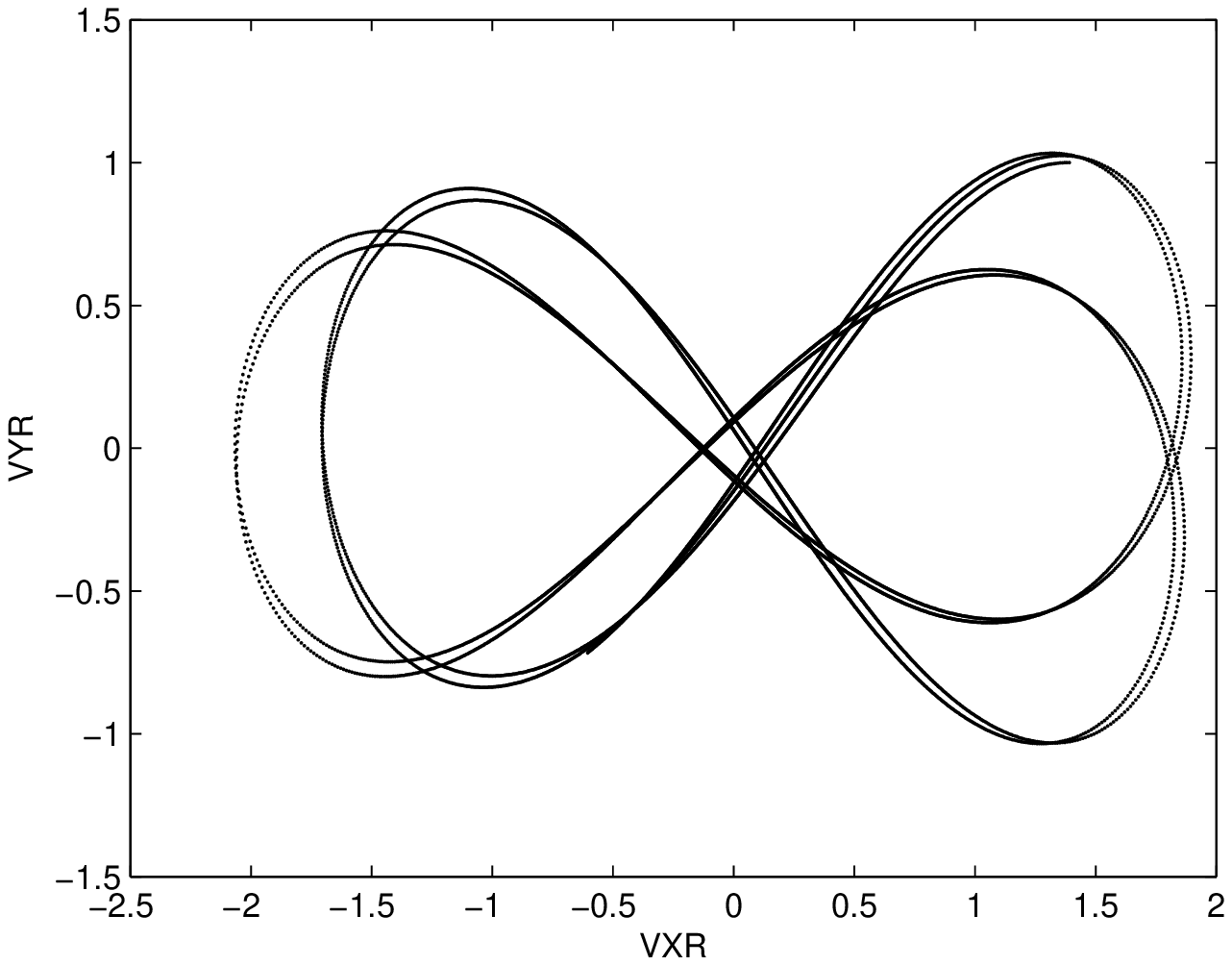}
\includegraphics[width=0.35\columnwidth]{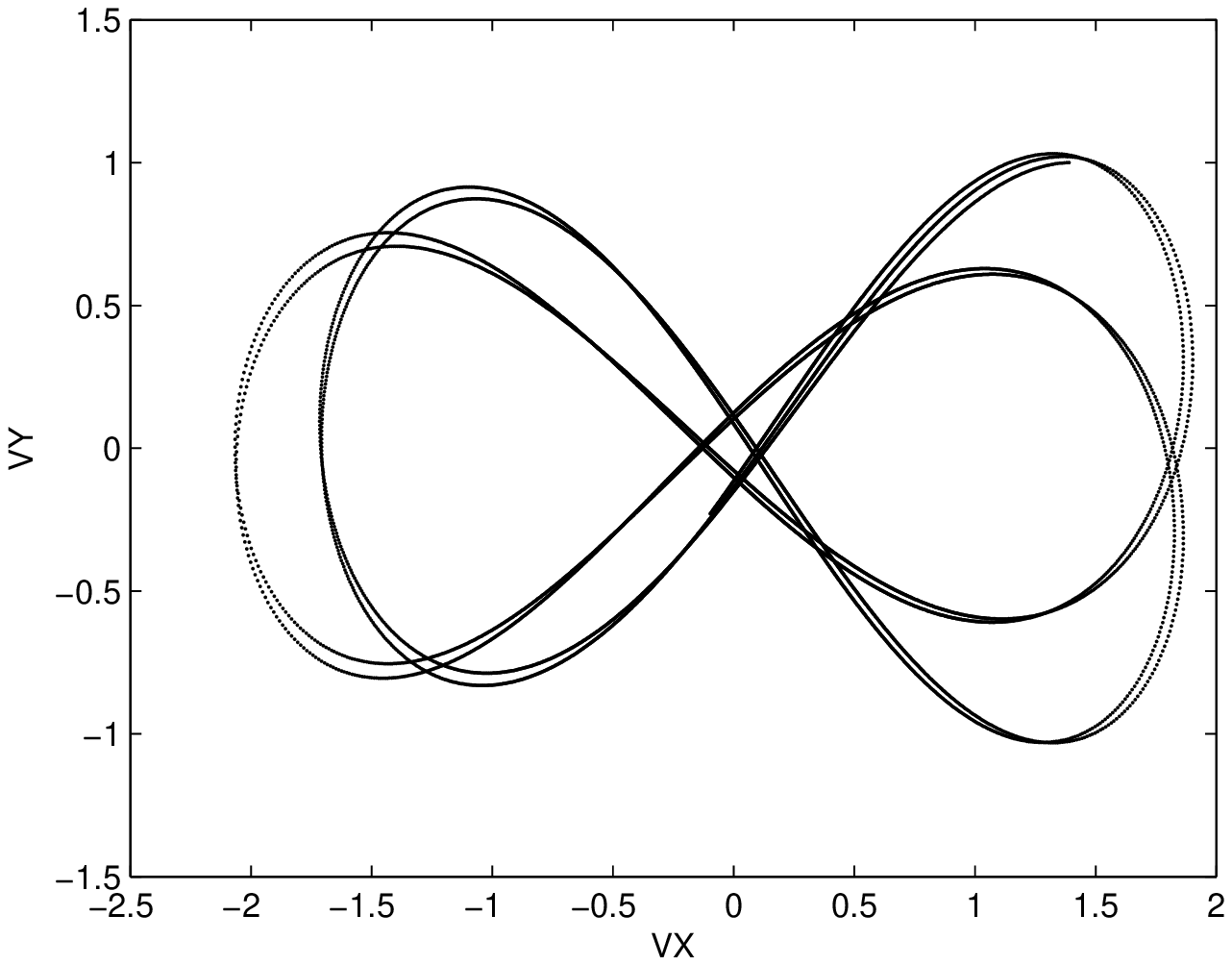}
\end{tabular}
\begin{tabular}{lll}
\includegraphics[width=0.35\columnwidth]{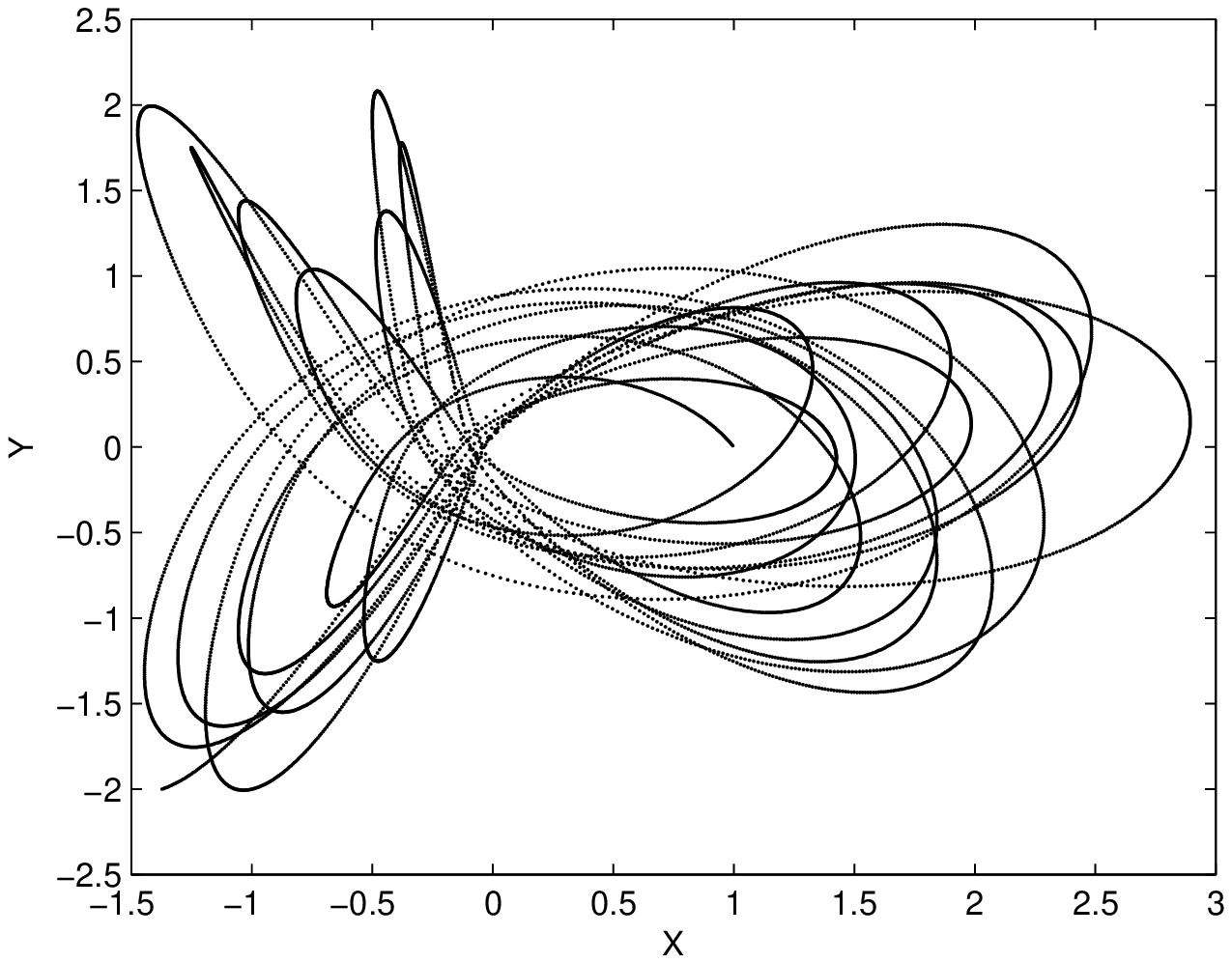}
\includegraphics[width=0.35\columnwidth]{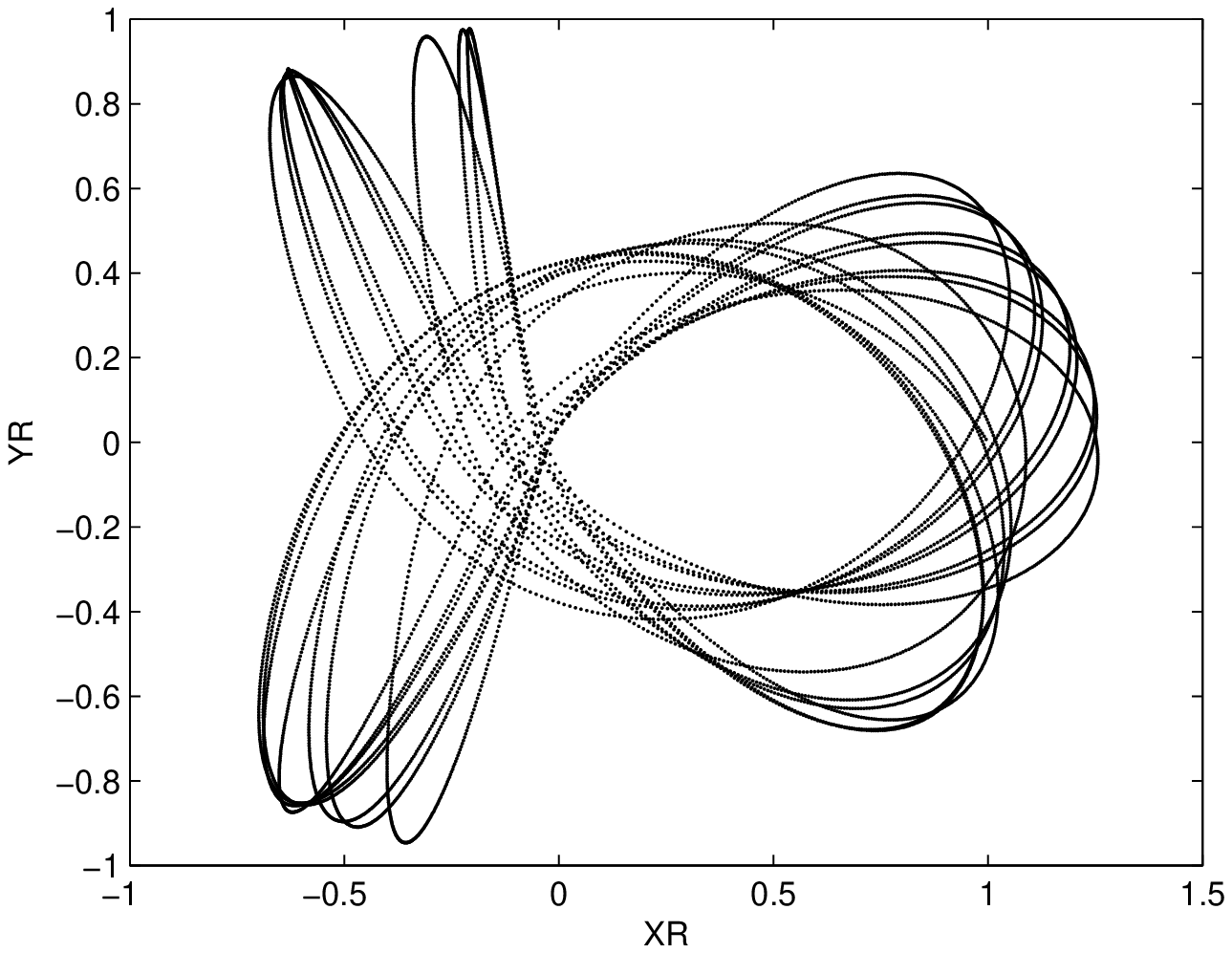}
\includegraphics[width=0.35\columnwidth]{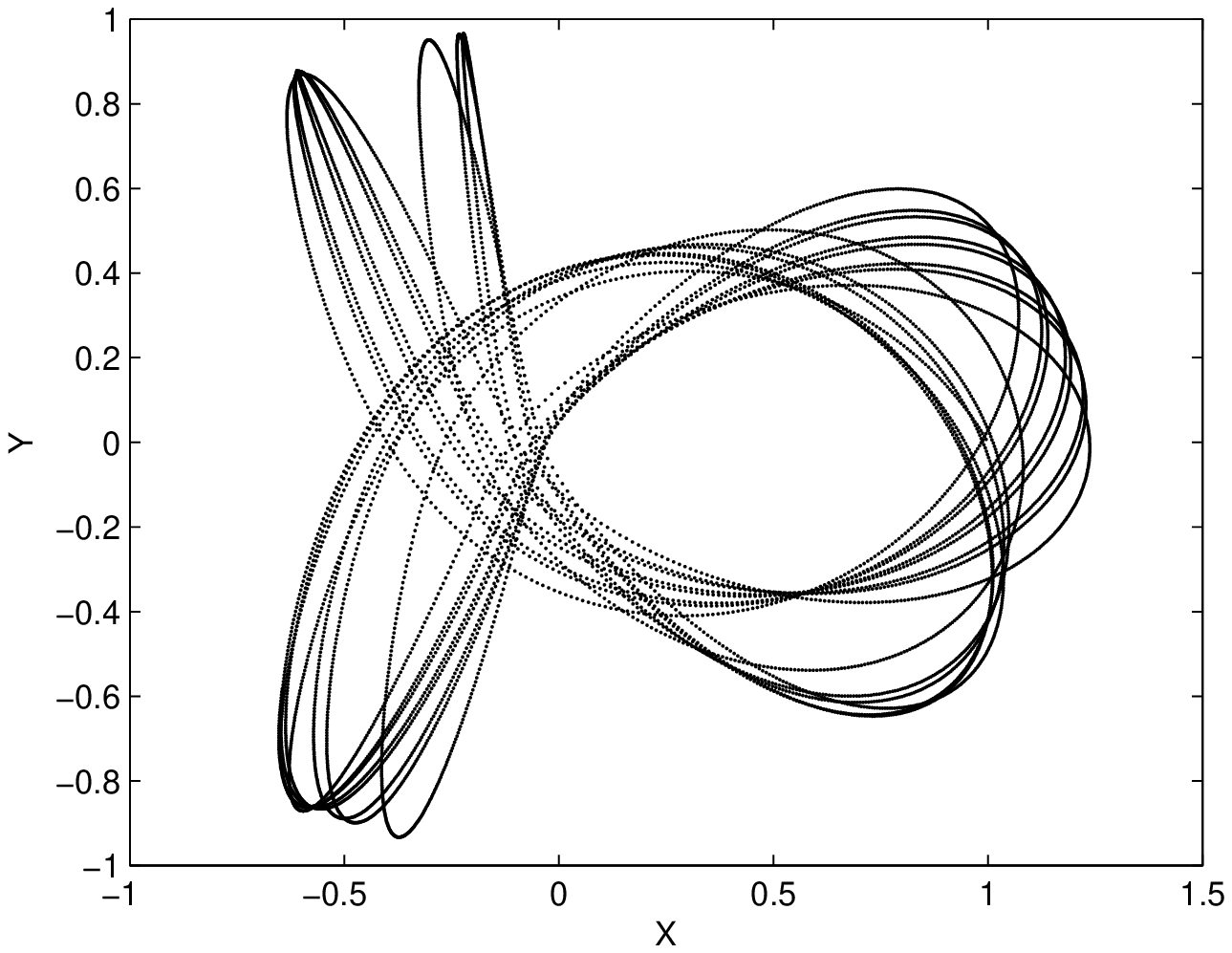}
\end{tabular}
\begin{tabular}{lll}
\includegraphics[width=0.35\columnwidth]{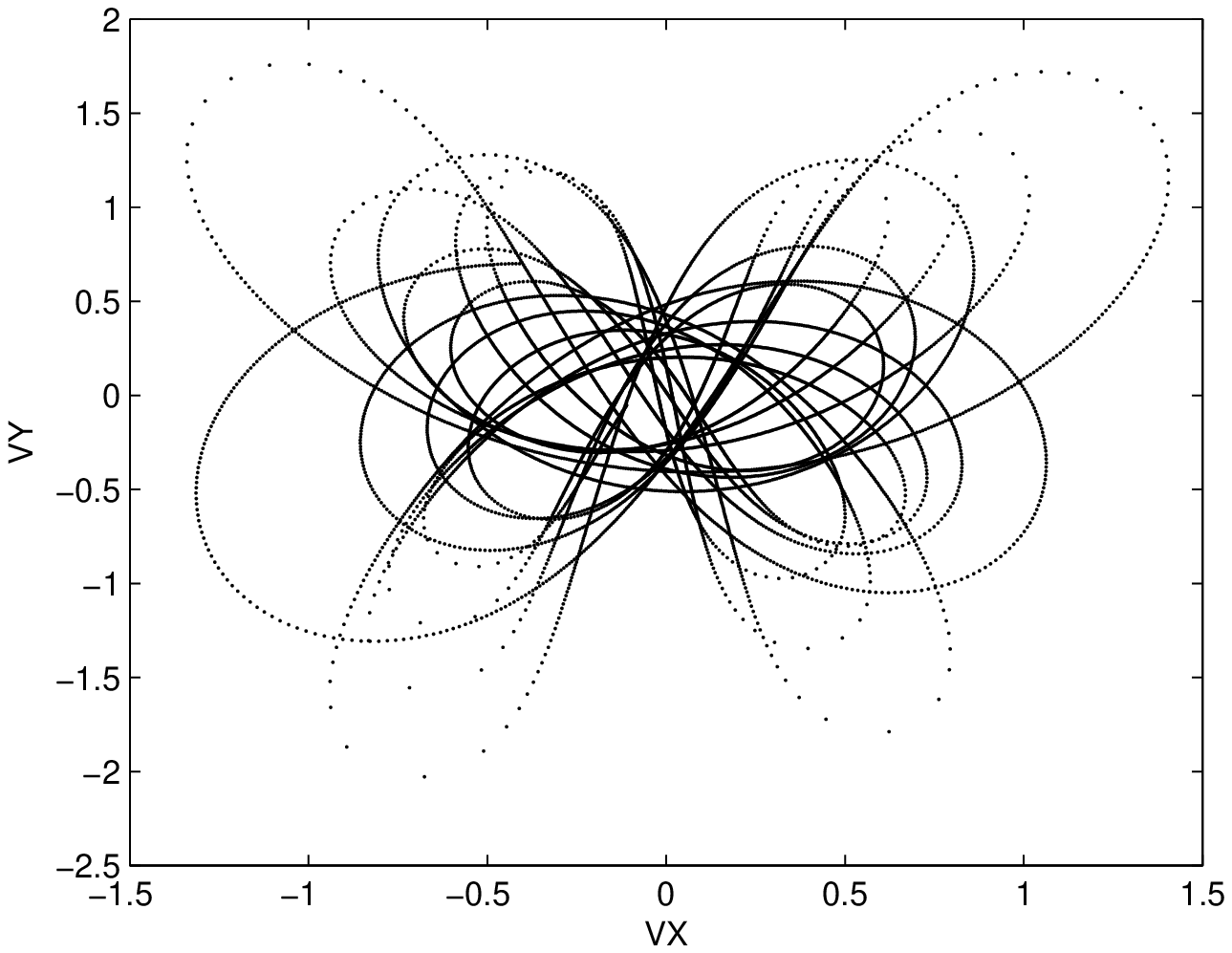}
\includegraphics[width=0.35\columnwidth]{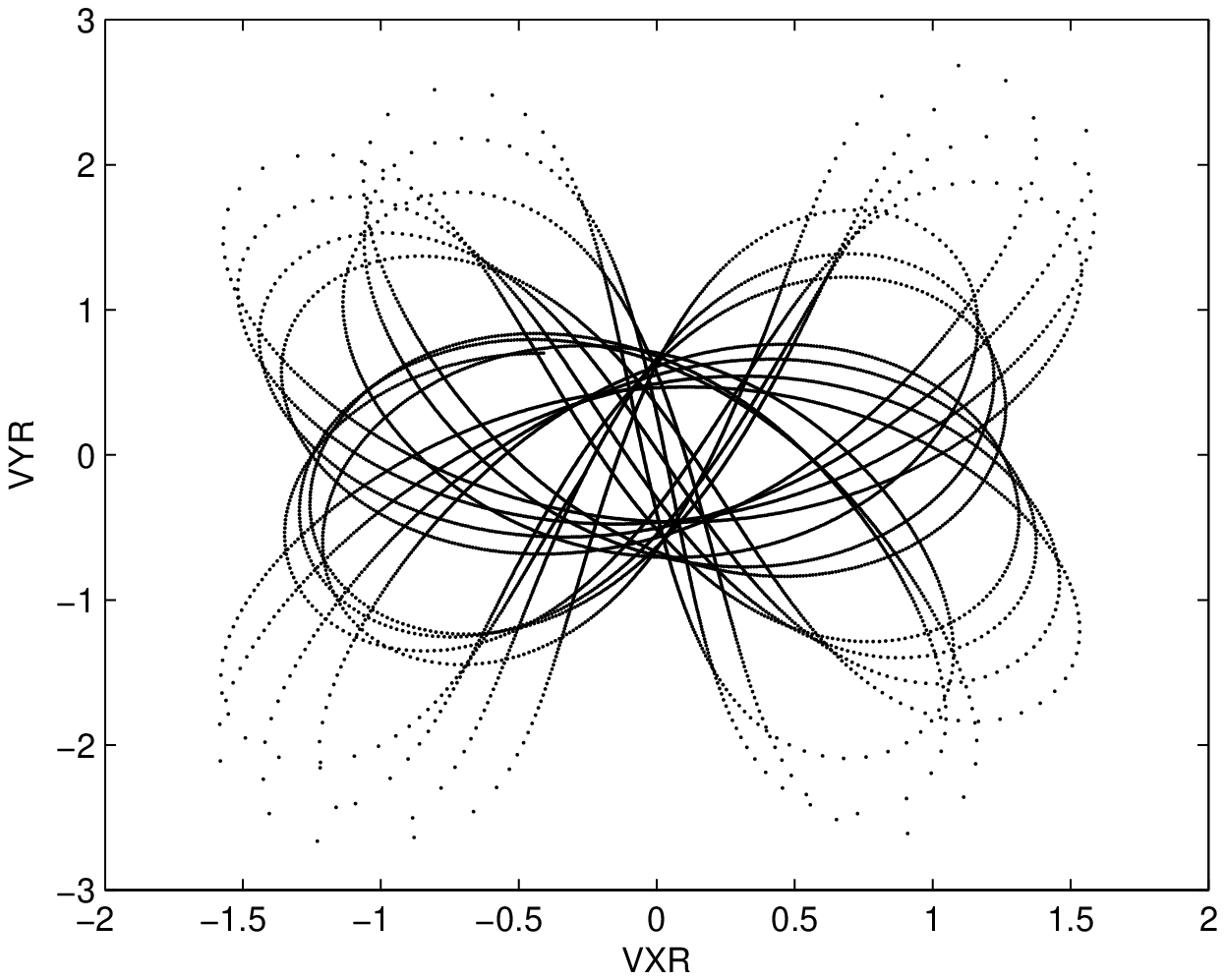}
\includegraphics[width=0.35\columnwidth]{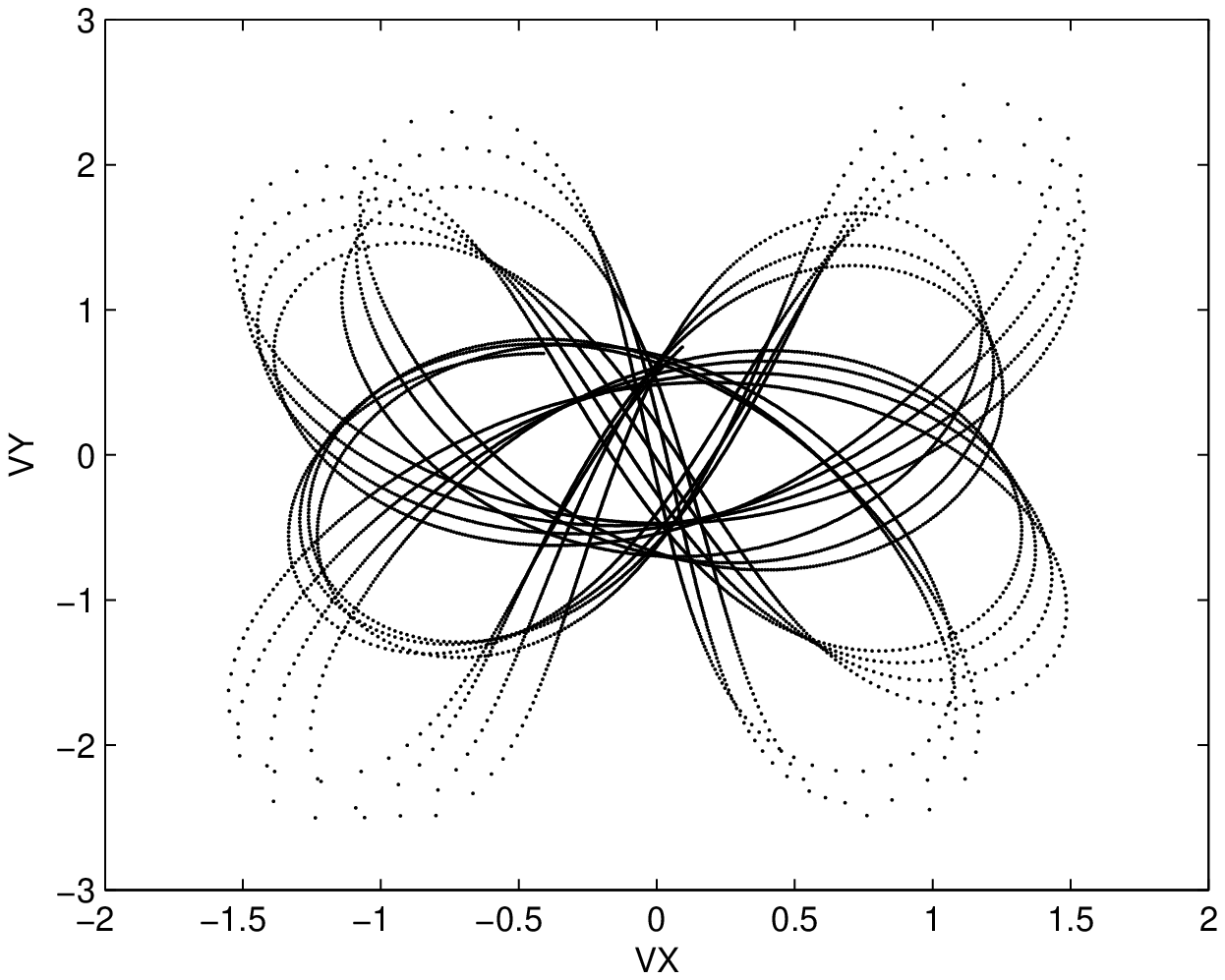}
\end{tabular}
\caption{Planar orbits. The $x-y$, and $V_x-V_y$
projections of two phase-space orbits in a biaxial $\f={\rm ln}(r)$ potential with axes ratio $q_1^{1/2}=\sqrt{2}$ (length and time units are arbitrary.) $\azg= t\_i/t$; $t\_i=10$ and  $t\_f=300$.
Rows 1-2: initial conditions are $x=1,~y=0,~z=0,~V_x=1.4,~V_y=1,~V_z=0$.
Rows 3-4: initial conditions are $x=1,~y=0,~z=0,~V_x=-0.4,~V_y=0.7,~V_z=0$.
Left column: the actual orbit, $\vr$; middle column: $\bar\vr\^*$, the same orbit ``unscaled'' [see Eq.(\ref{redar})], to be compared with the 3rd column: $\vr\^*$, with no $\azg$ variation, and the same initial conditions, run for an ``equivalent'' time $\Delta t\^*$ [see Eq.(\ref{redtime})]. Note the different scales for the first one and last two columns. Points are plotted at equal intervals of $\eta^2(t)\Delta t$.}\label{fig1}
\end{figure}
\begin{figure}

\begin{tabular}{lll}
\includegraphics[width=0.35\columnwidth]{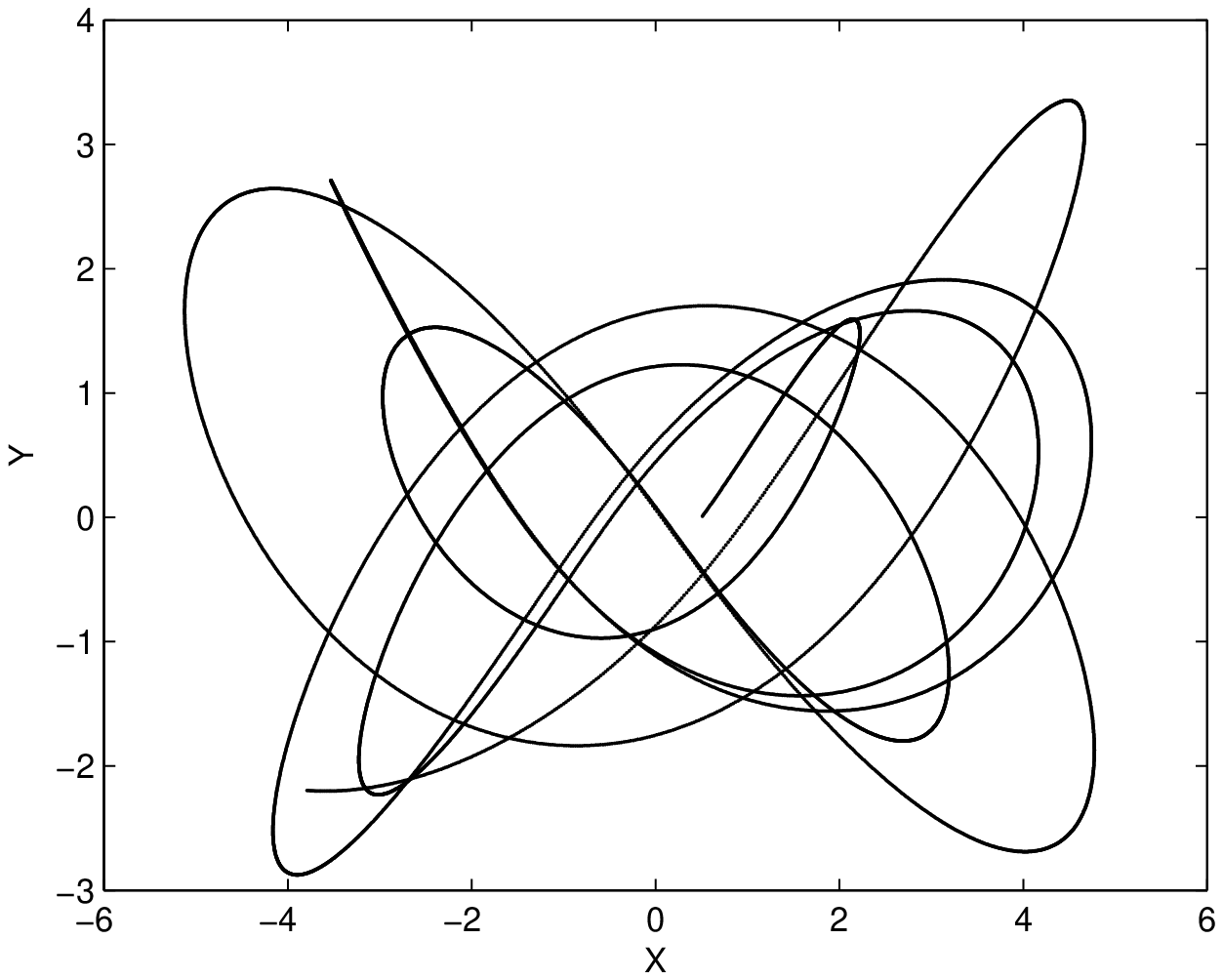}
\includegraphics[width=0.35\columnwidth]{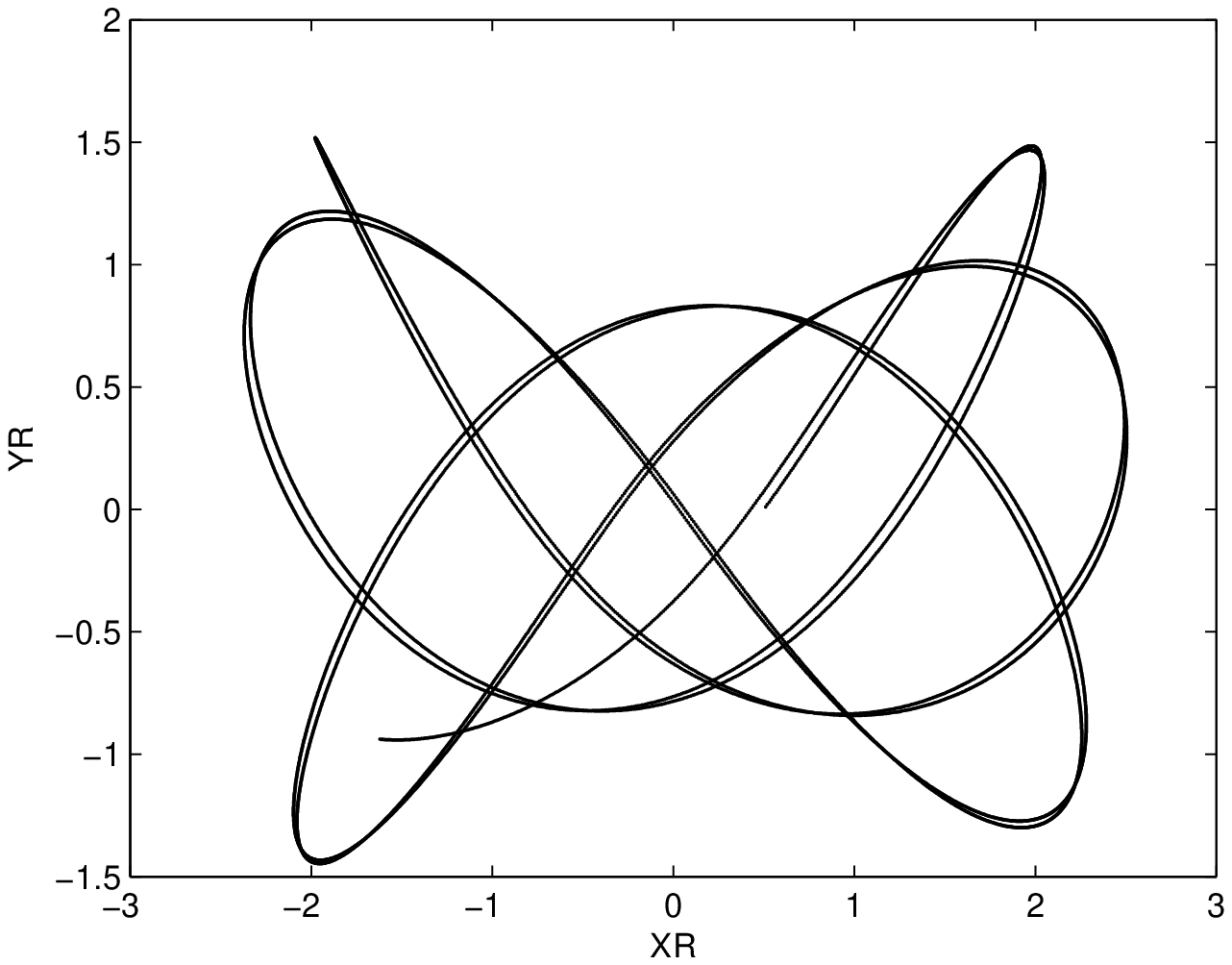}
\includegraphics[width=0.35\columnwidth]{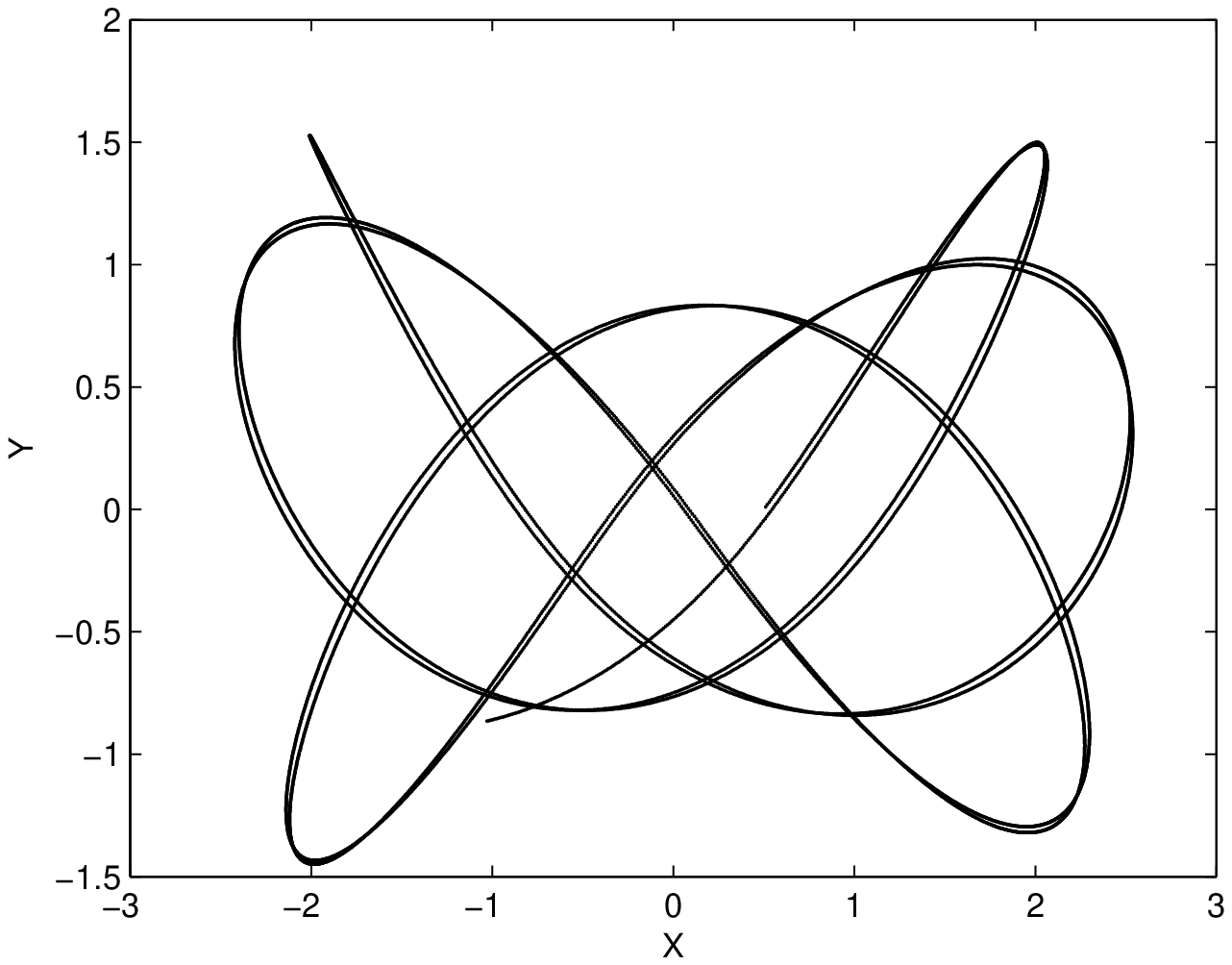}
\end{tabular}
\begin{tabular}{lll}
\includegraphics[width=0.35\columnwidth]{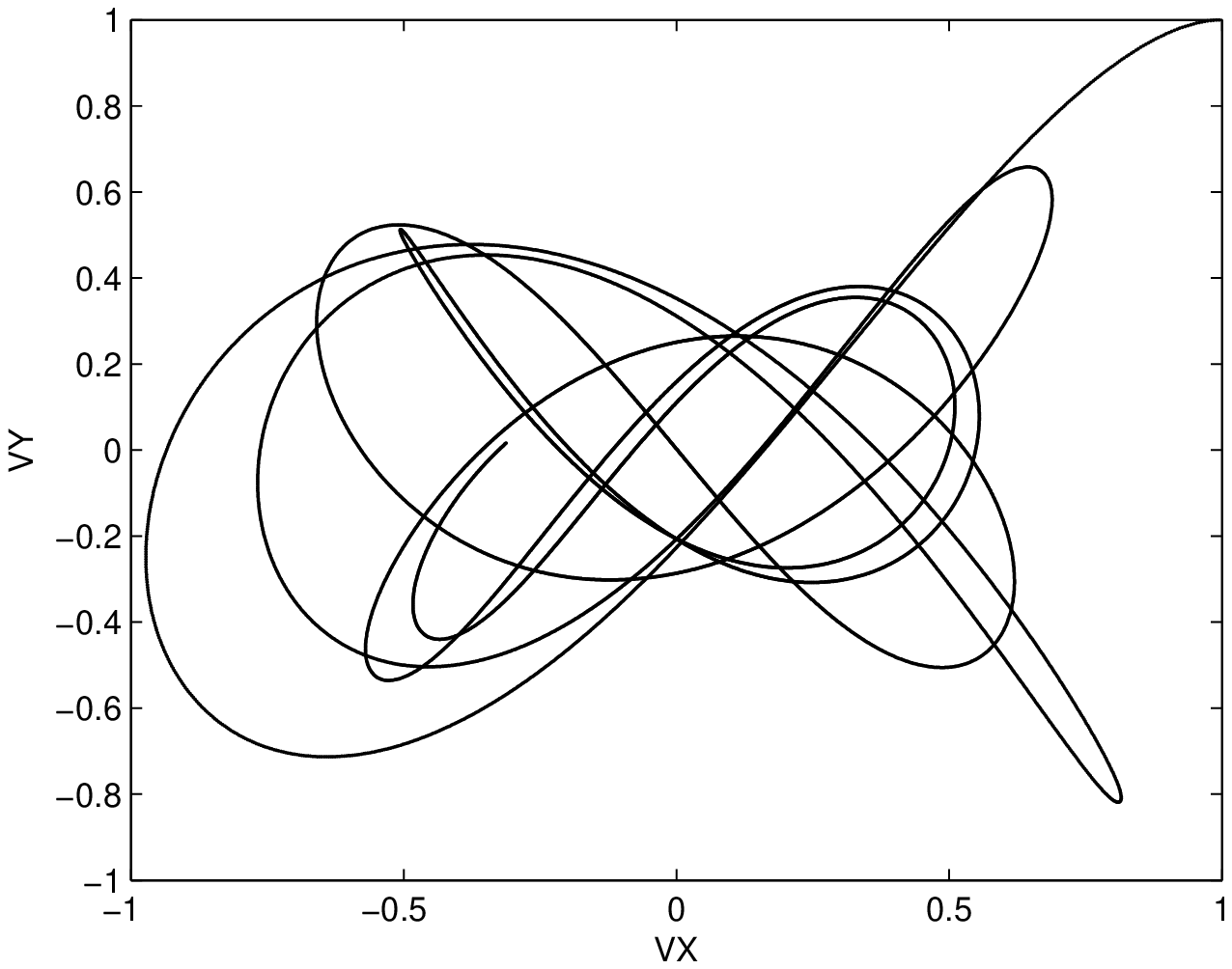}
\includegraphics[width=0.35\columnwidth]{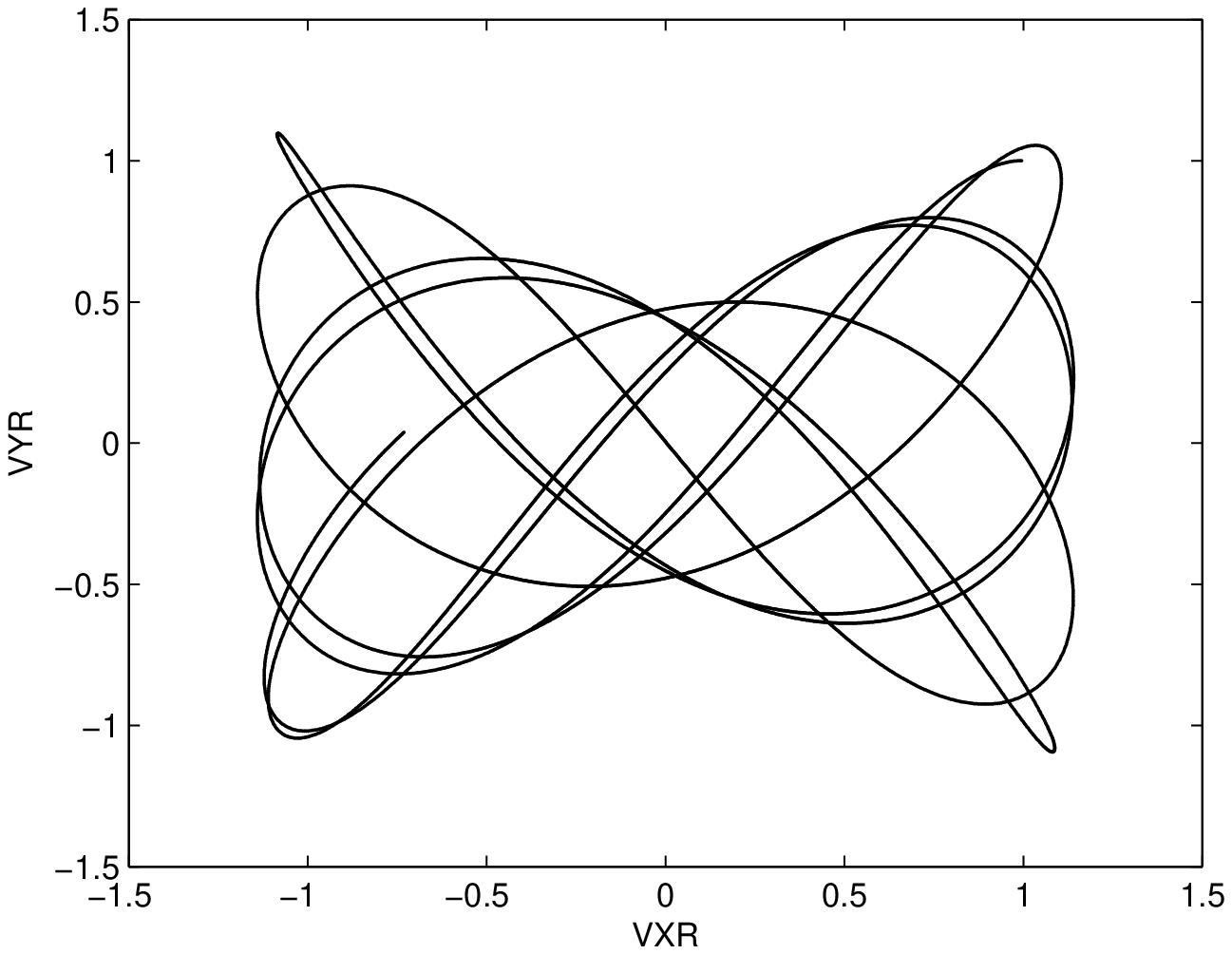}
\includegraphics[width=0.35\columnwidth]{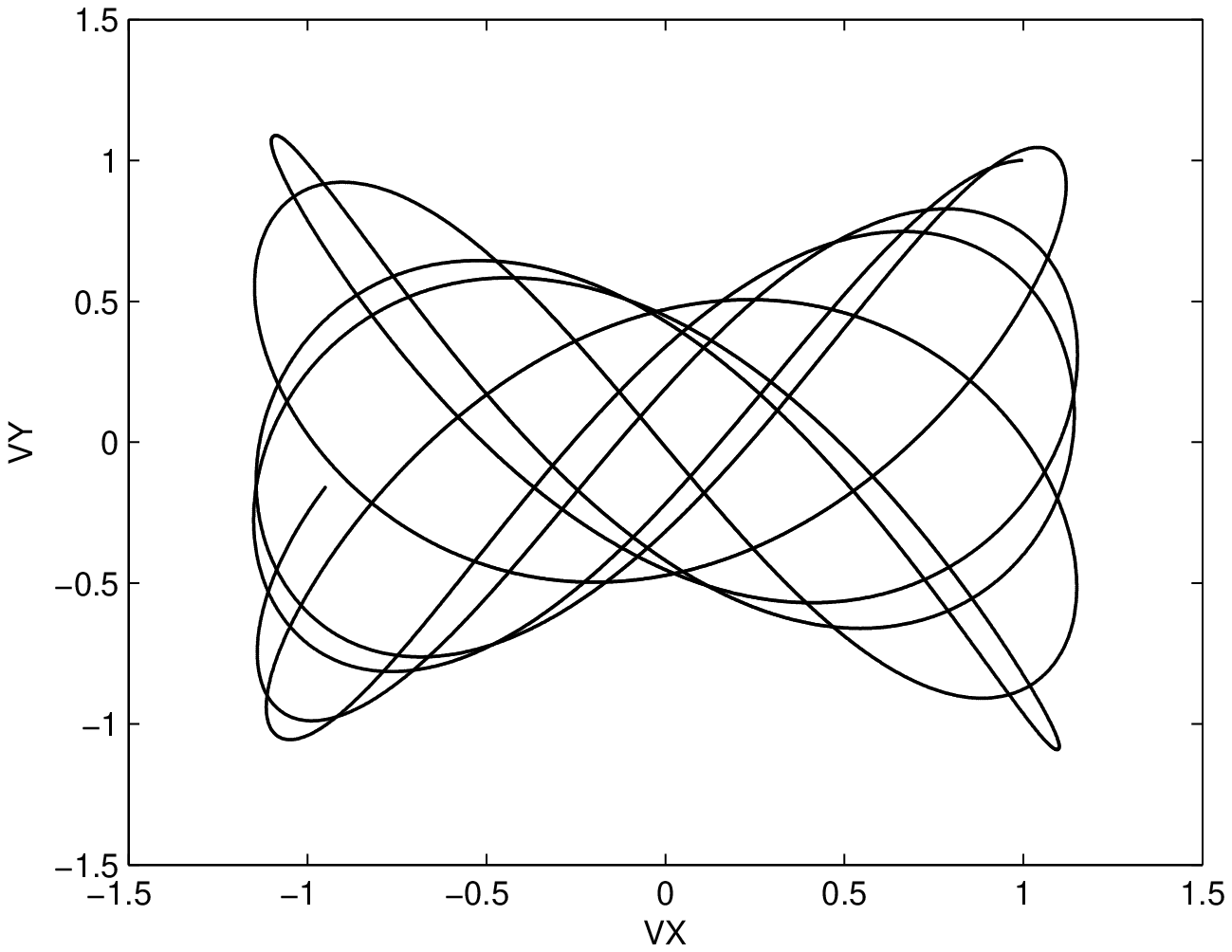}
\end{tabular}
\newpage
\begin{tabular}{lll}
\includegraphics[width=0.35\columnwidth]{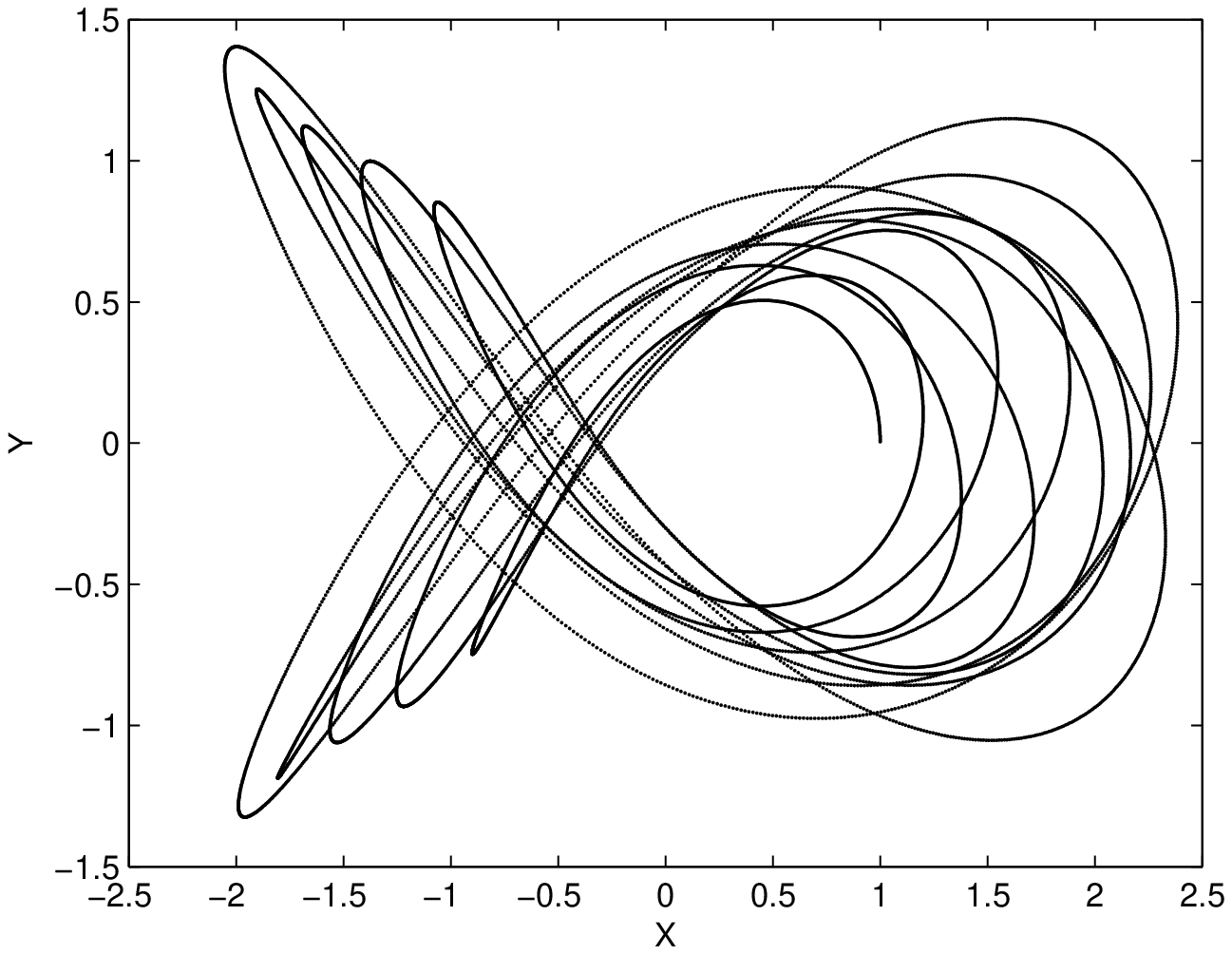}
\includegraphics[width=0.35\columnwidth]{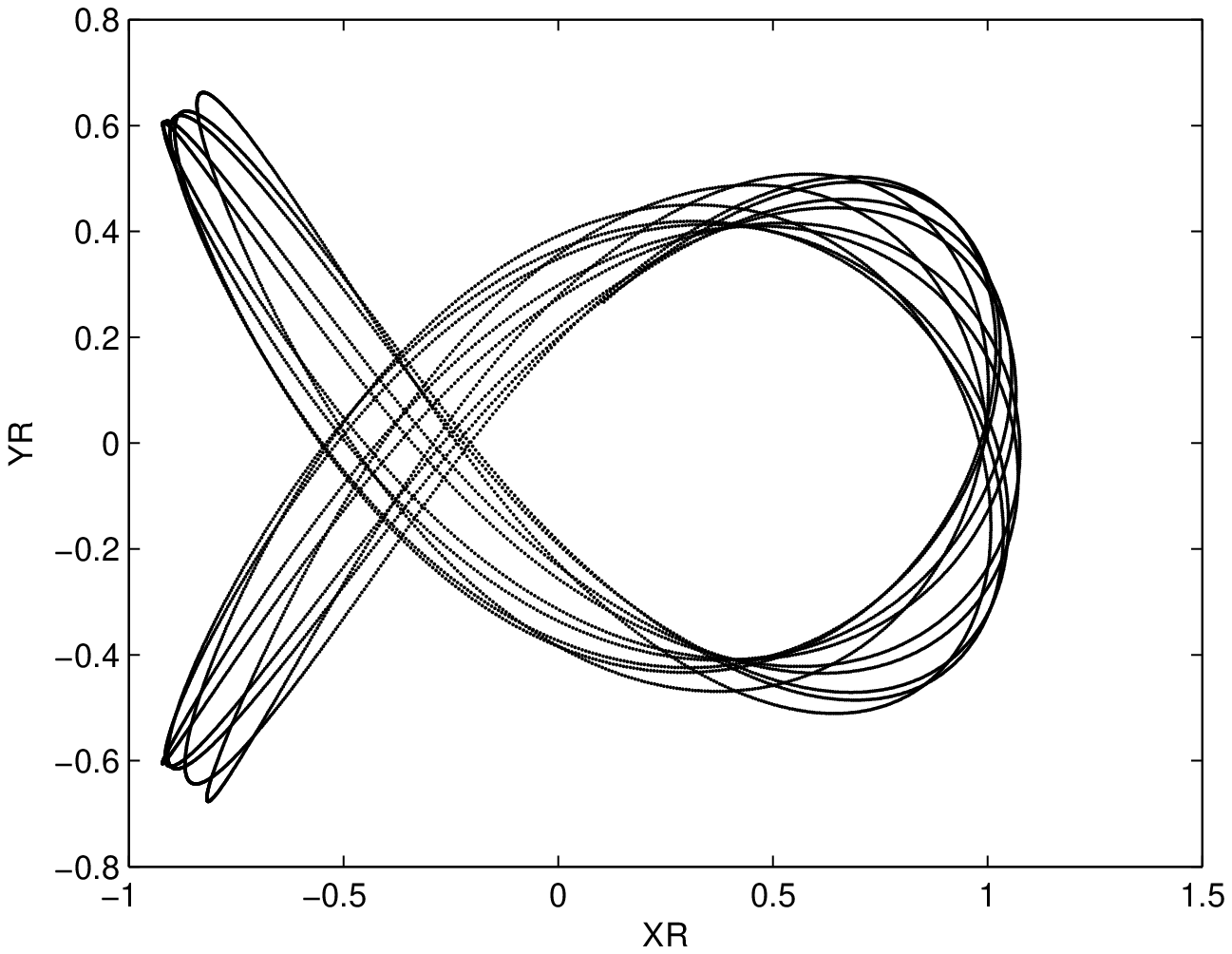}
\includegraphics[width=0.35\columnwidth]{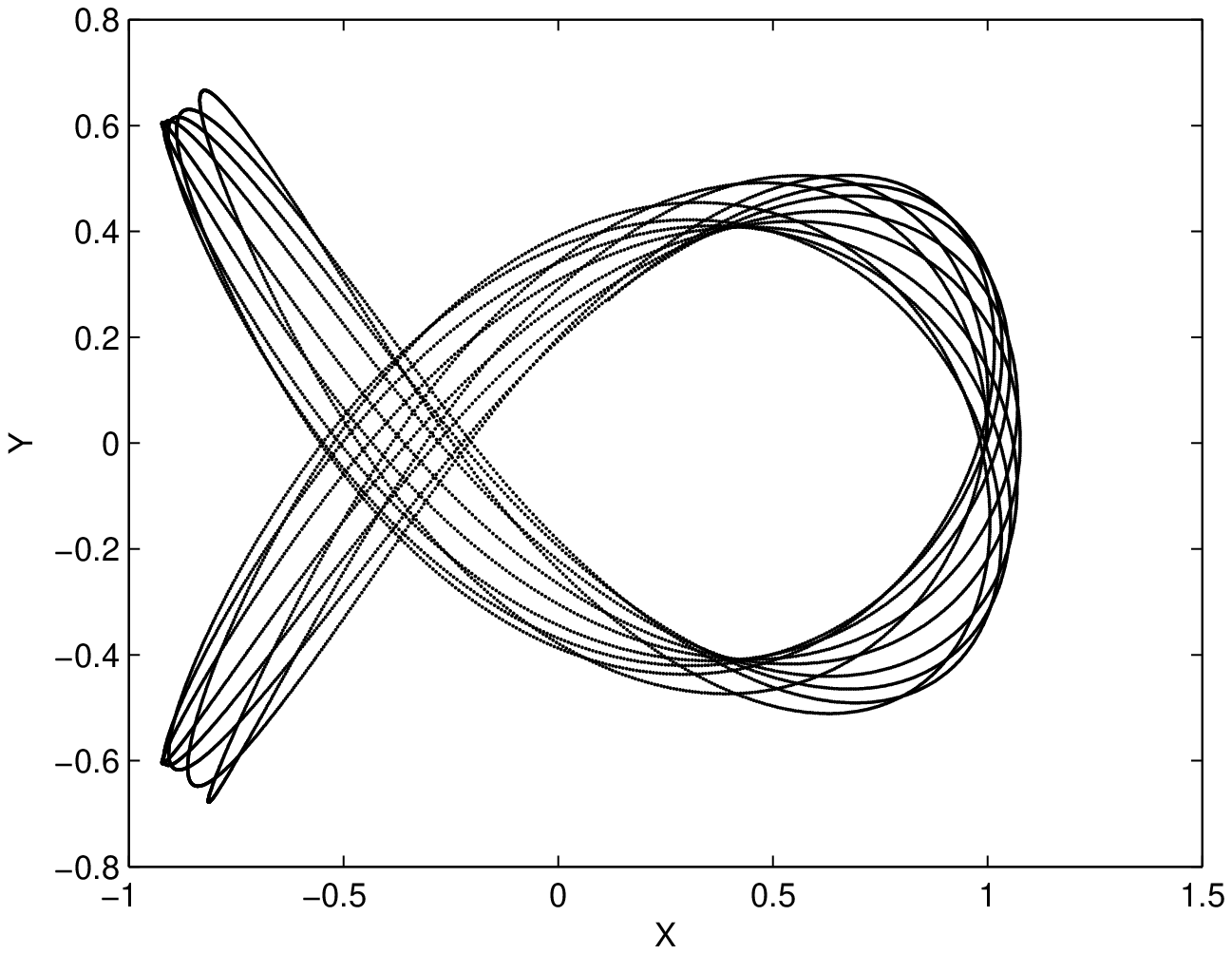}
\end{tabular}
\begin{tabular}{lll}
\includegraphics[width=0.35\columnwidth]{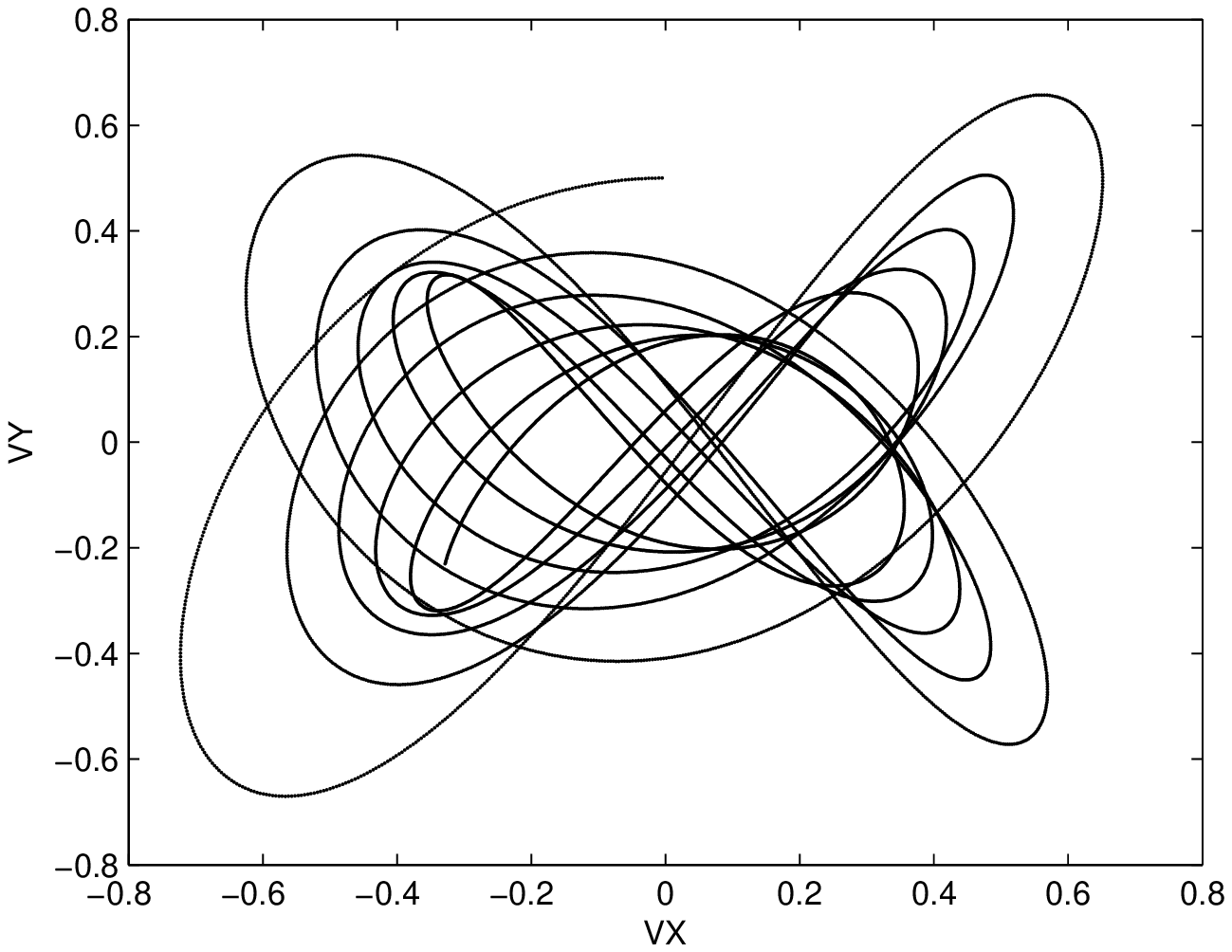}
\includegraphics[width=0.35\columnwidth]{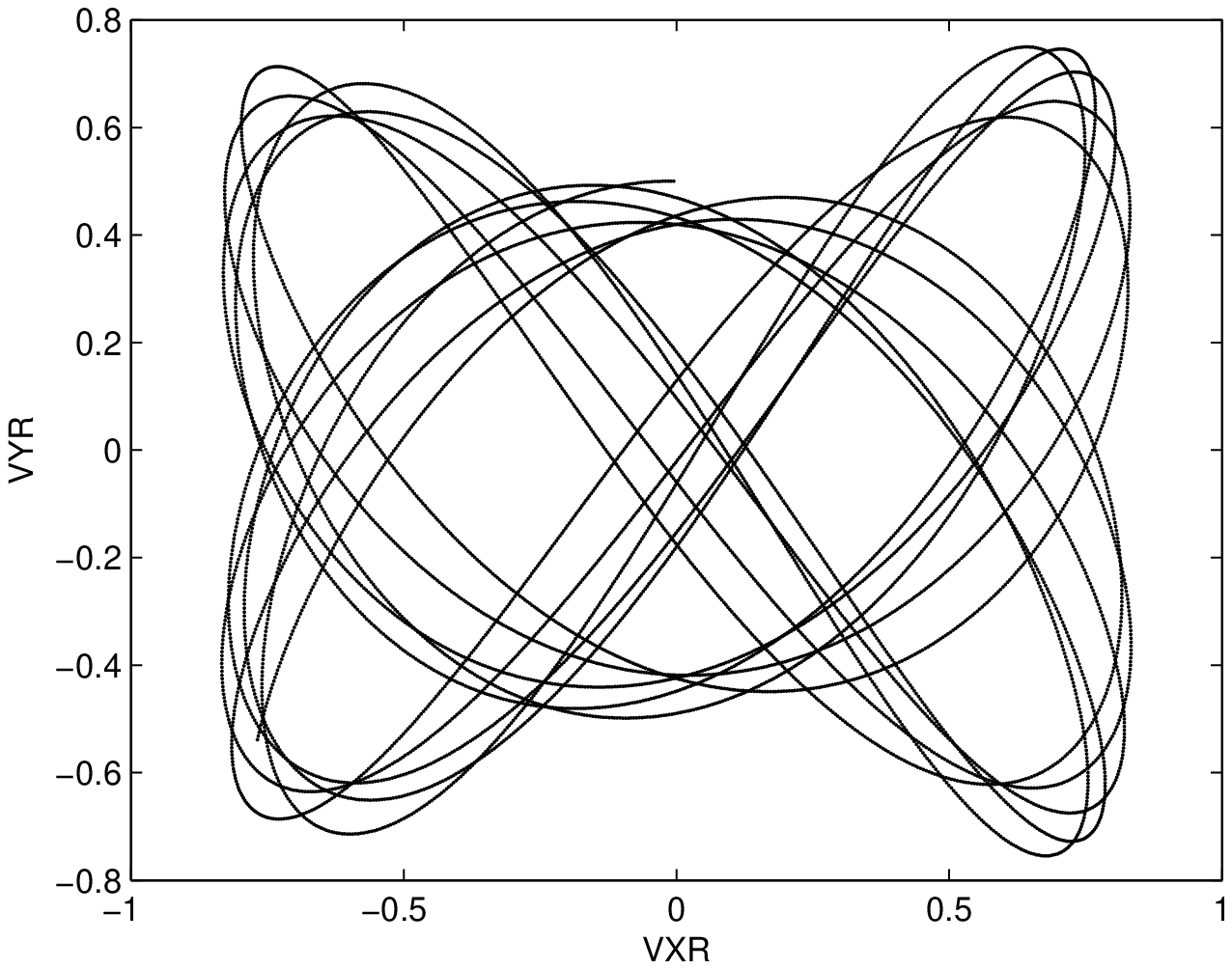}
\includegraphics[width=0.35\columnwidth]{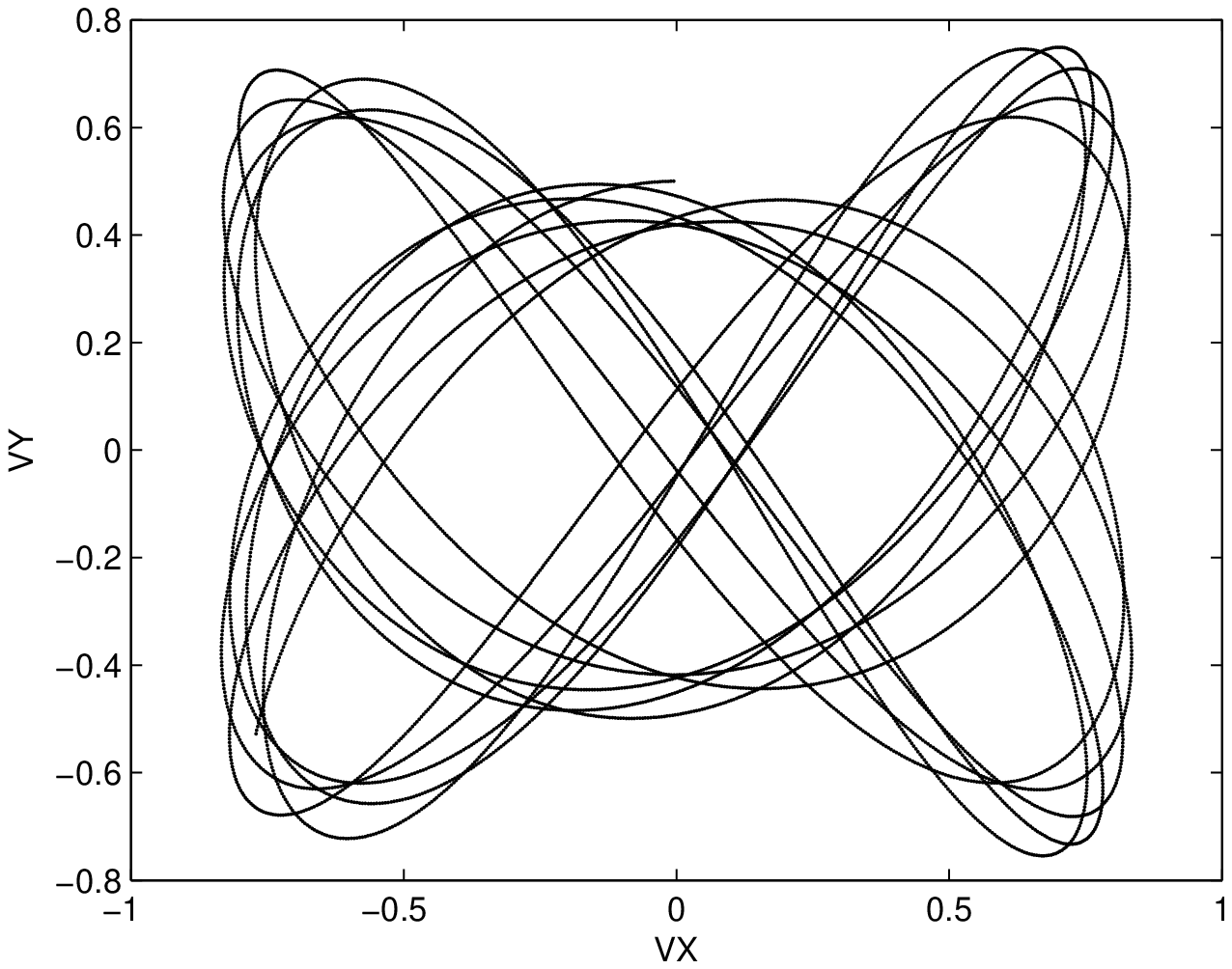}
\end{tabular}
\caption{Same as Fig \ref{fig1} for two orbits in the field of a DML isothermal, biaxial, isotropic sphere, with axes ratio $q_1^{1/2}=\sqrt{2}$. Rows 1-2: initial conditions are $x=0.5,~y=0,~z=0,~V_x=1,~V_y=1,~V_z=0$.
Rows 3-4: initial conditions are $x=1,~y=0,~z=0,~V_x=0,~V_y=0.5,~V_z=0$.}\label{fig2}
\end{figure}
\begin{figure}

\begin{tabular}{lll}
\includegraphics[width=3.8cm,height=3.6cm]{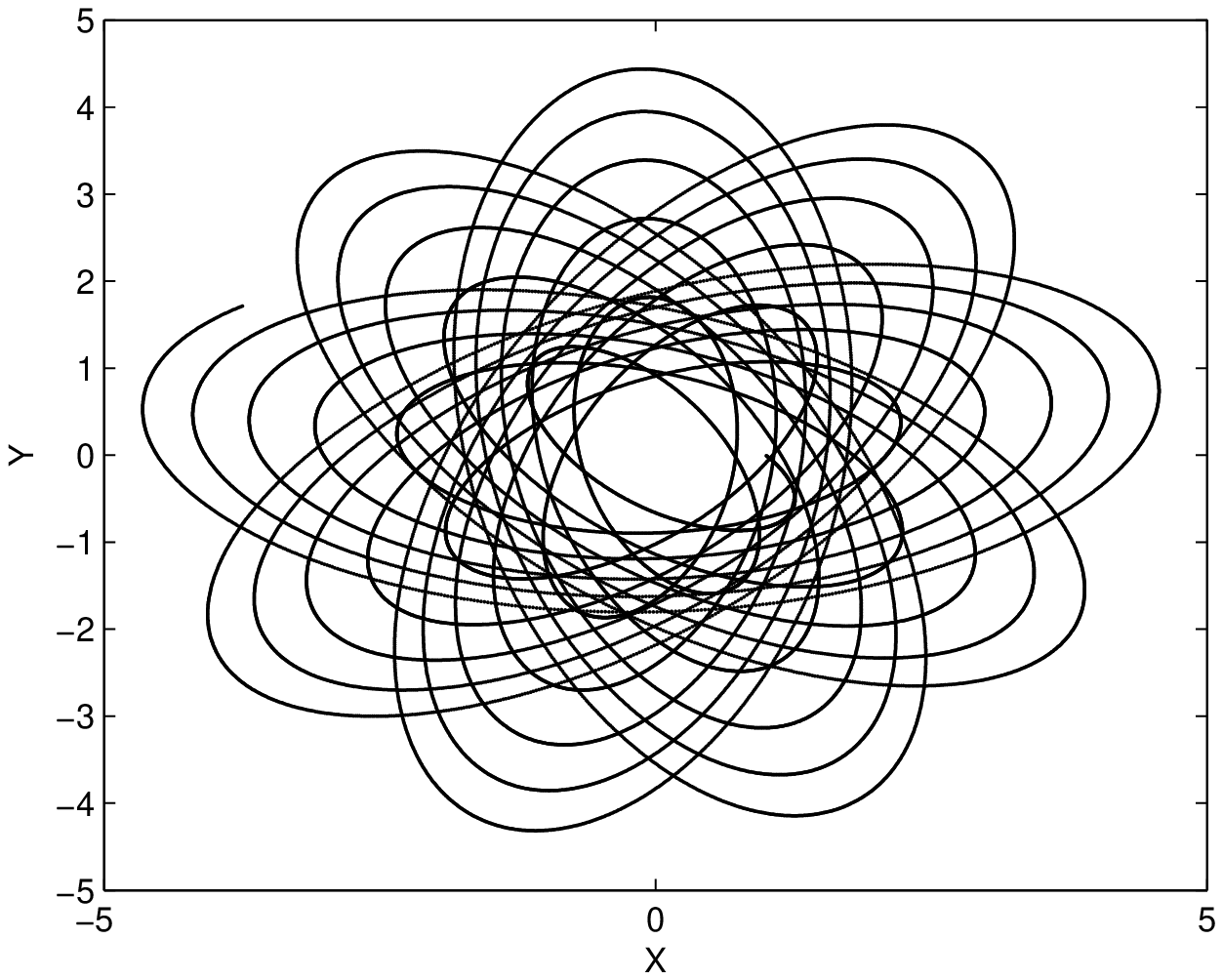}
\includegraphics[width=3.8cm,height=3.6cm]{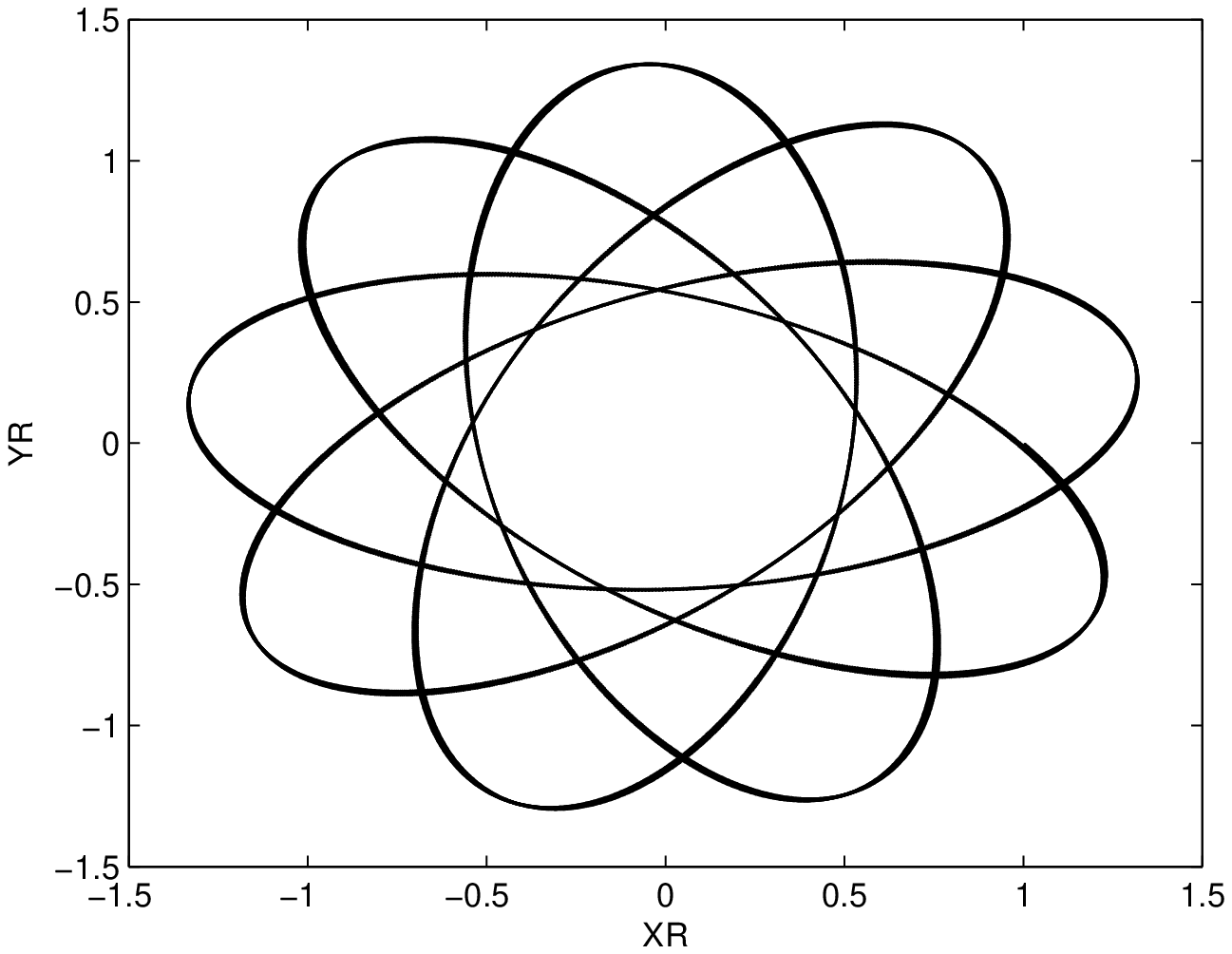}
\includegraphics[width=3.8cm,height=3.6cm]{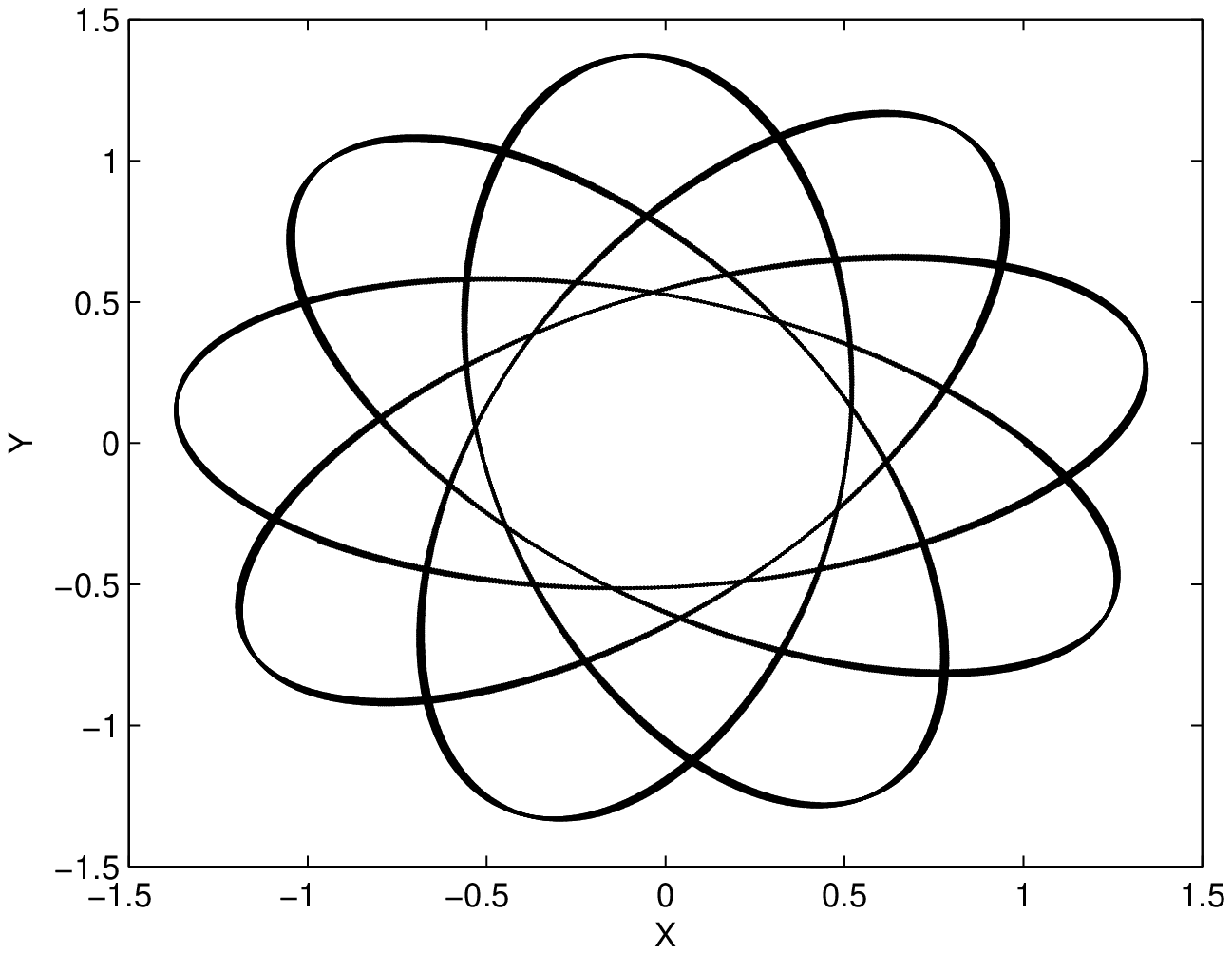}
\end{tabular}
\begin{tabular}{lll}
\includegraphics[width=3.8cm,height=3.6cm]{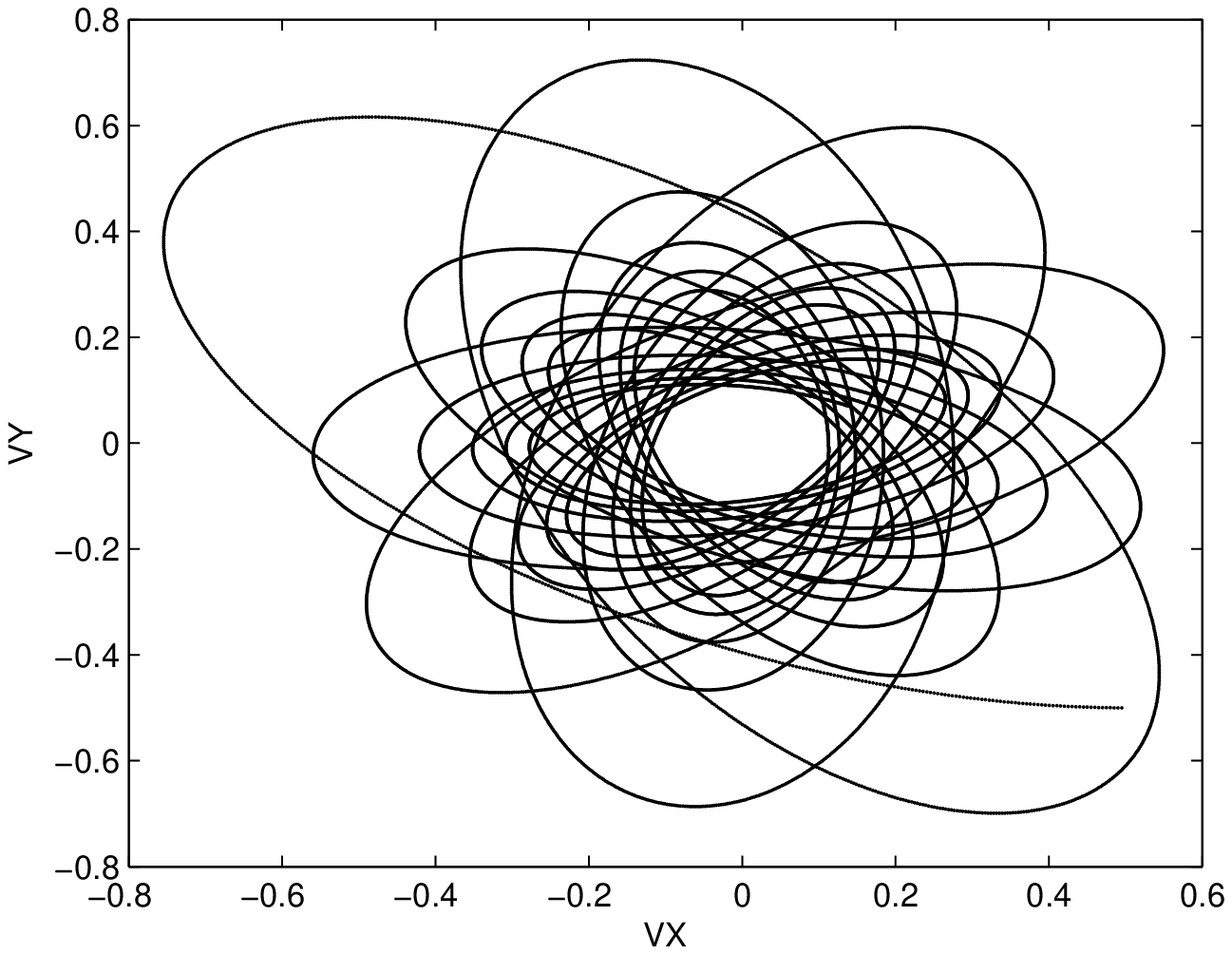}
\includegraphics[width=3.8cm,height=3.6cm]{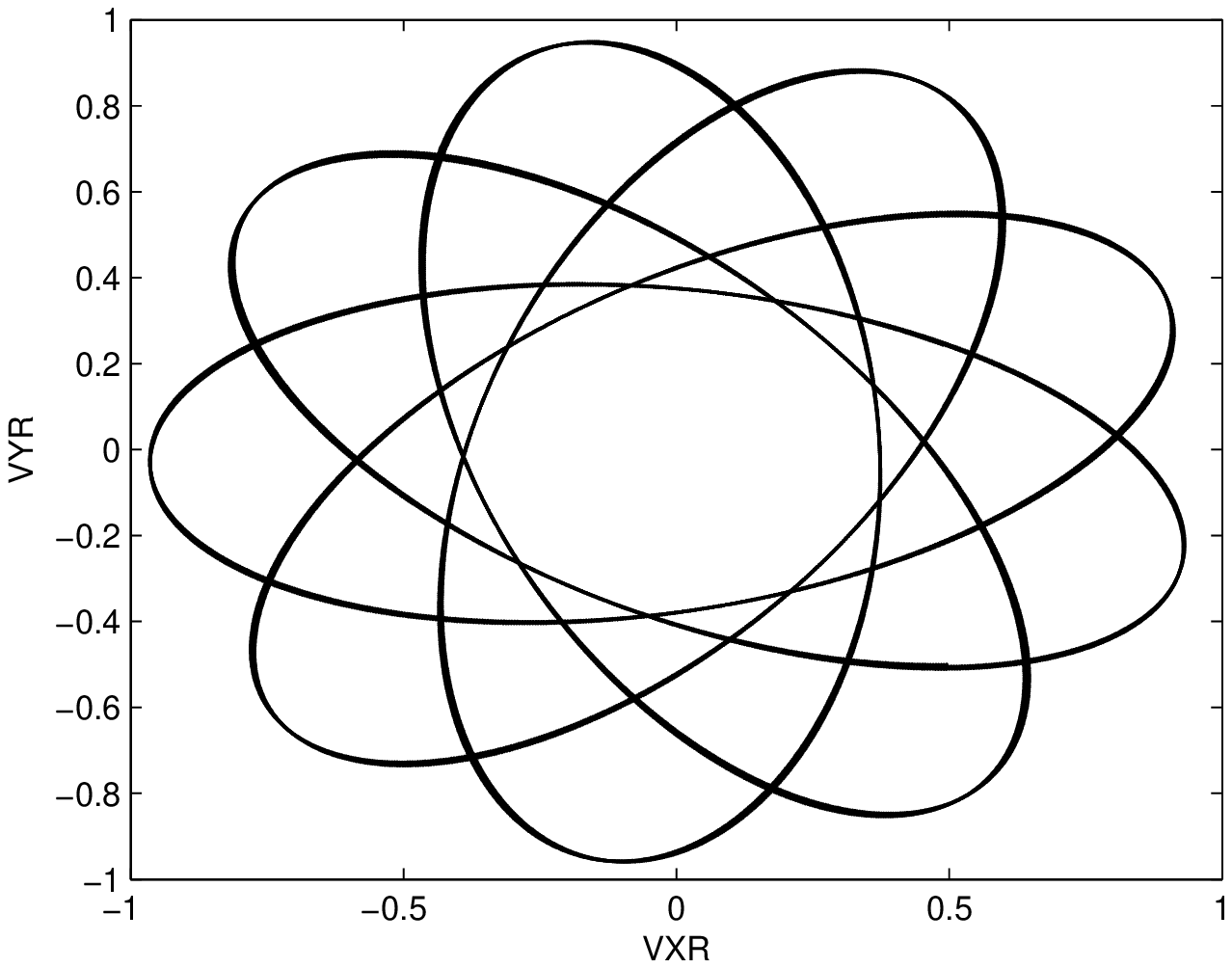}
\includegraphics[width=3.8cm,height=3.6cm]{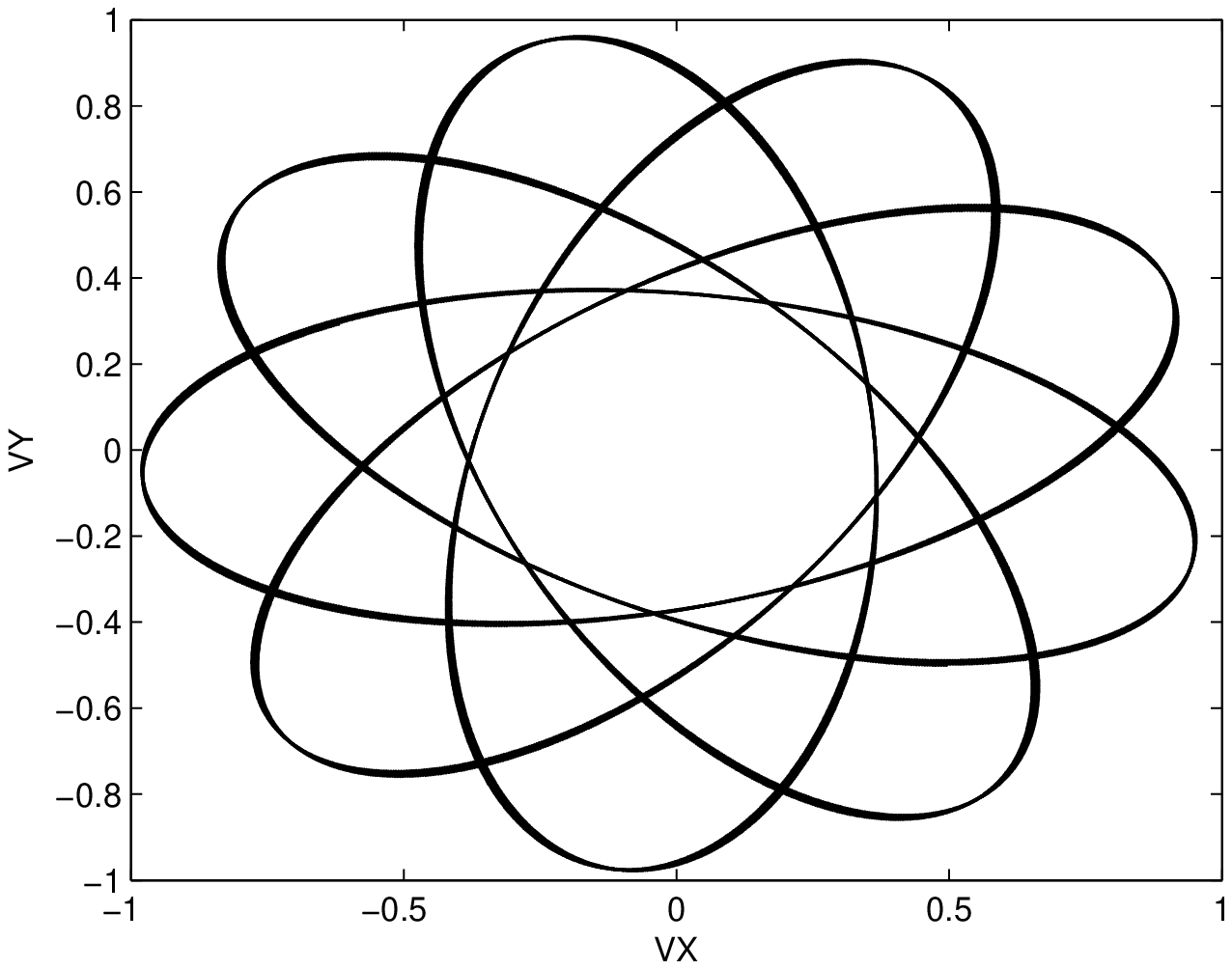}
\end{tabular}
\begin{tabular}{lll}
\includegraphics[width=3.8cm,height=3.6cm]{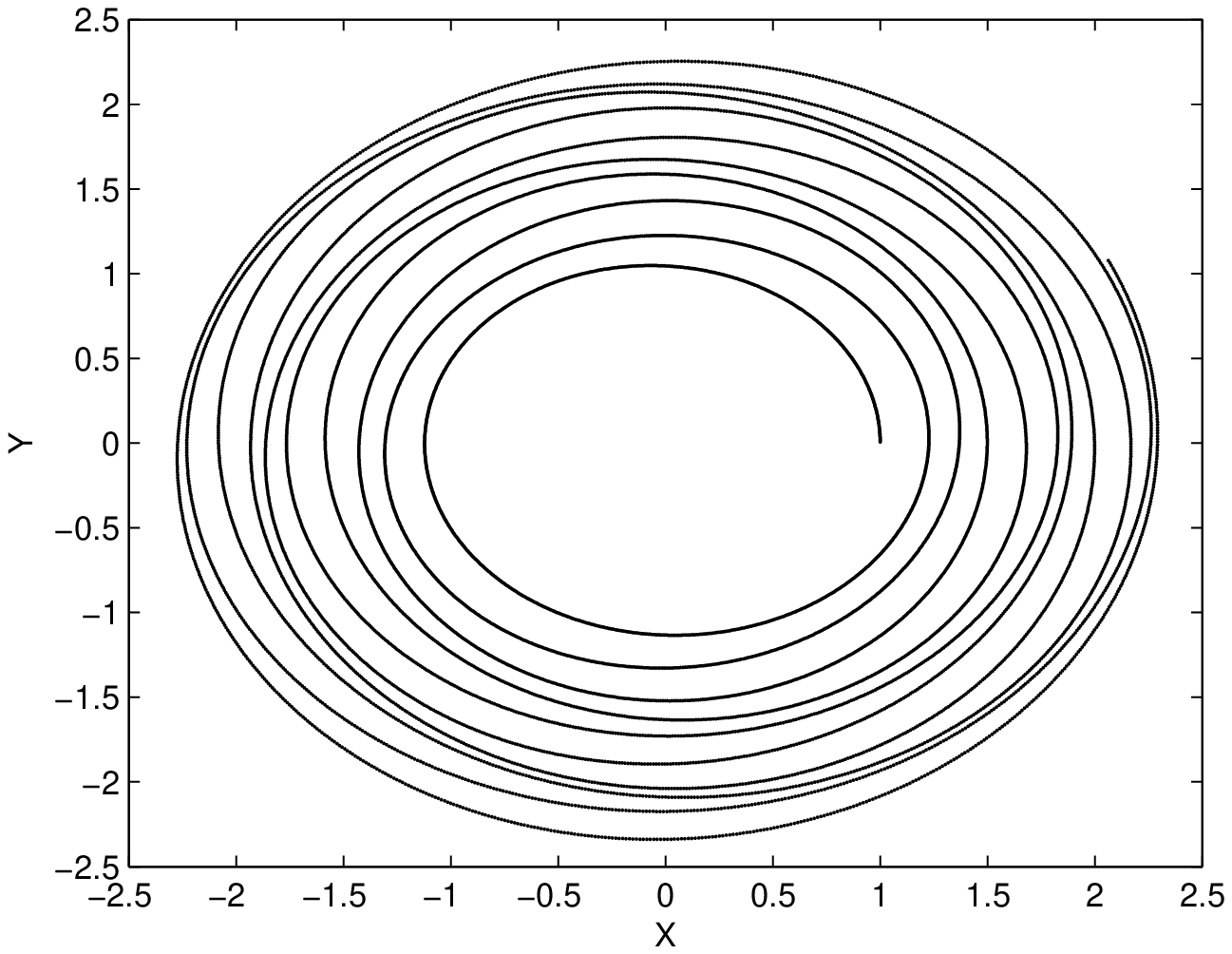}
\includegraphics[width=3.8cm,height=3.6cm]{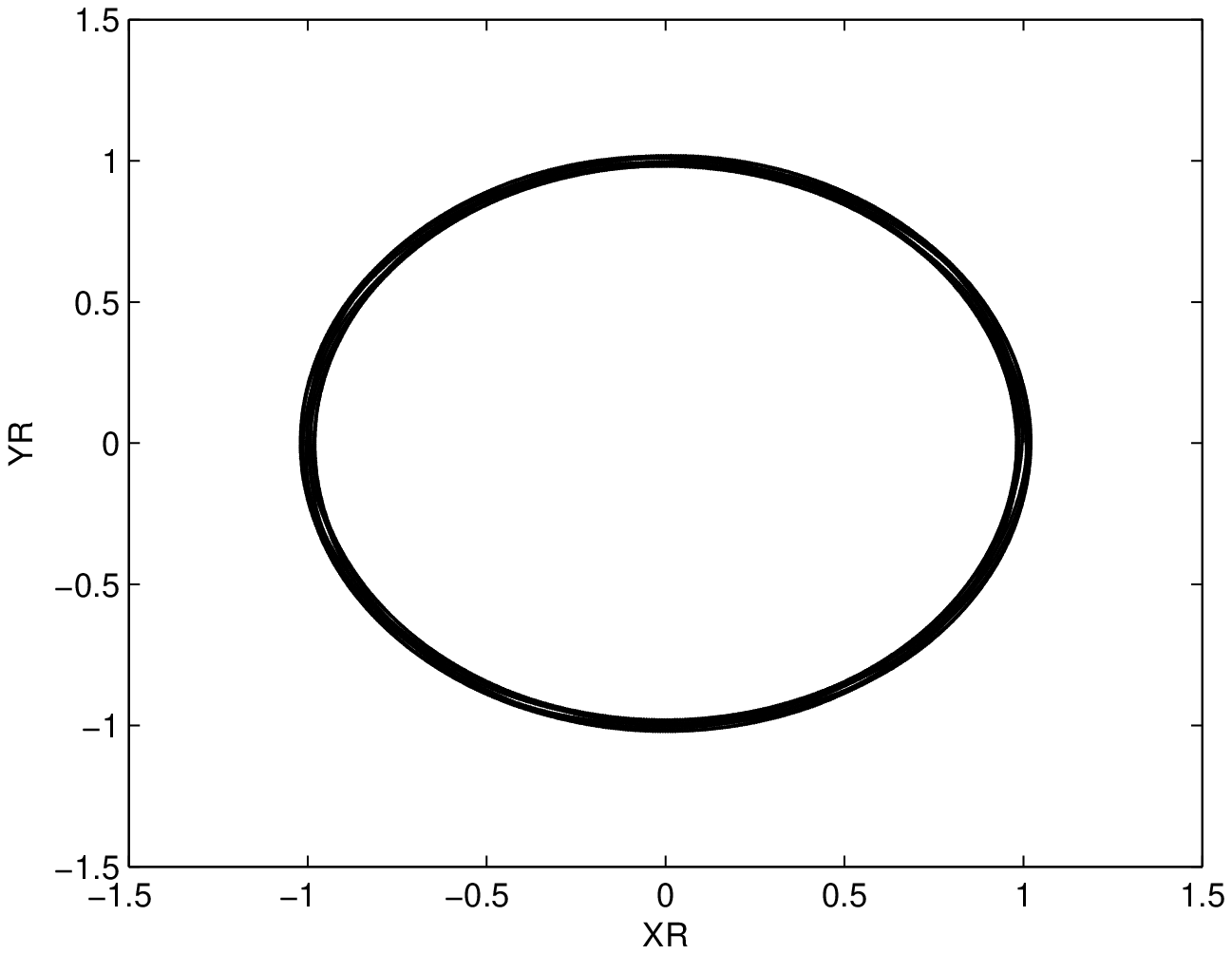}
\includegraphics[width=3.8cm,height=3.6cm]{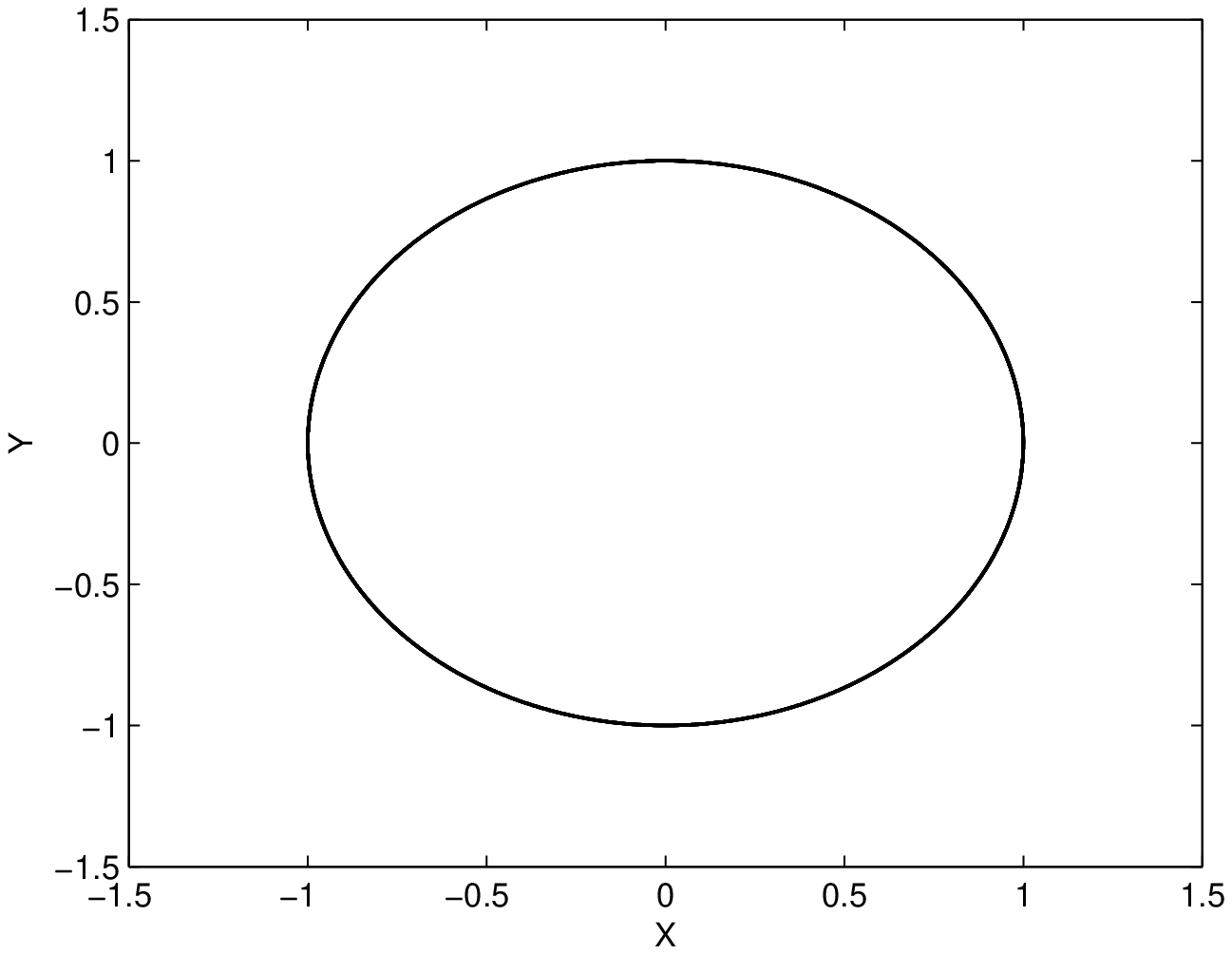}
\end{tabular}
\begin{tabular}{lll}
\includegraphics[width=3.8cm,height=3.6cm]{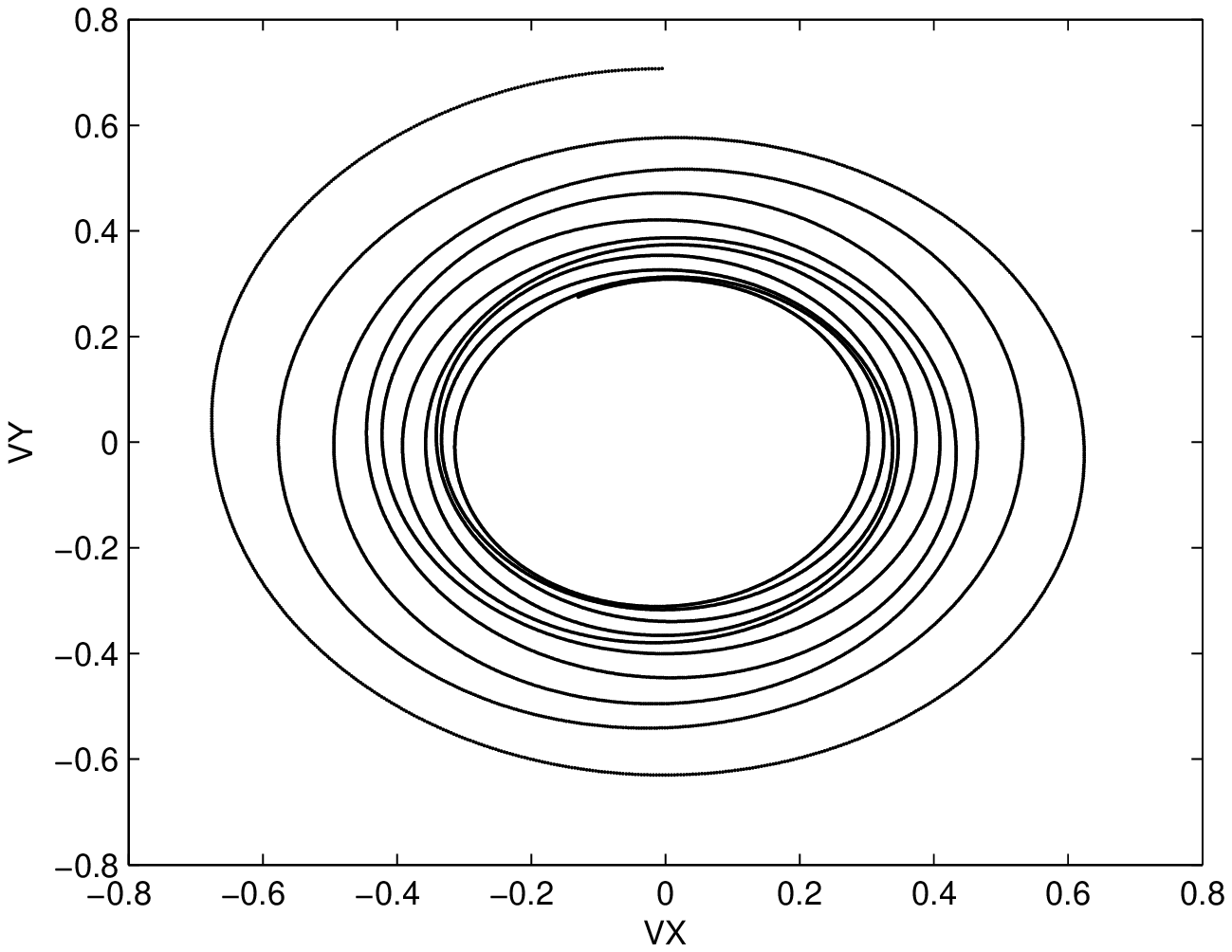}
\includegraphics[width=3.8cm,height=3.6cm]{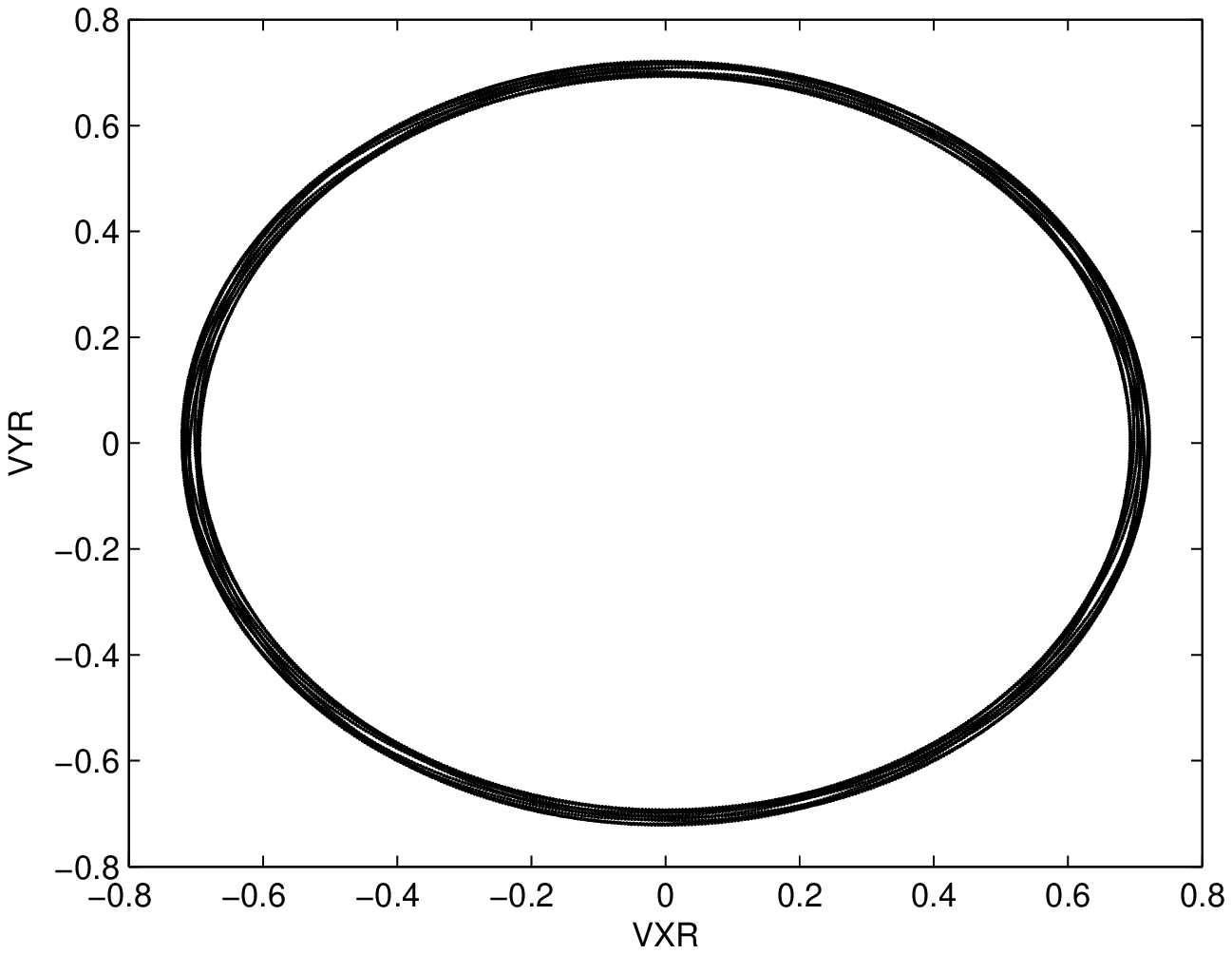}
\includegraphics[width=3.8cm,height=3.6cm]{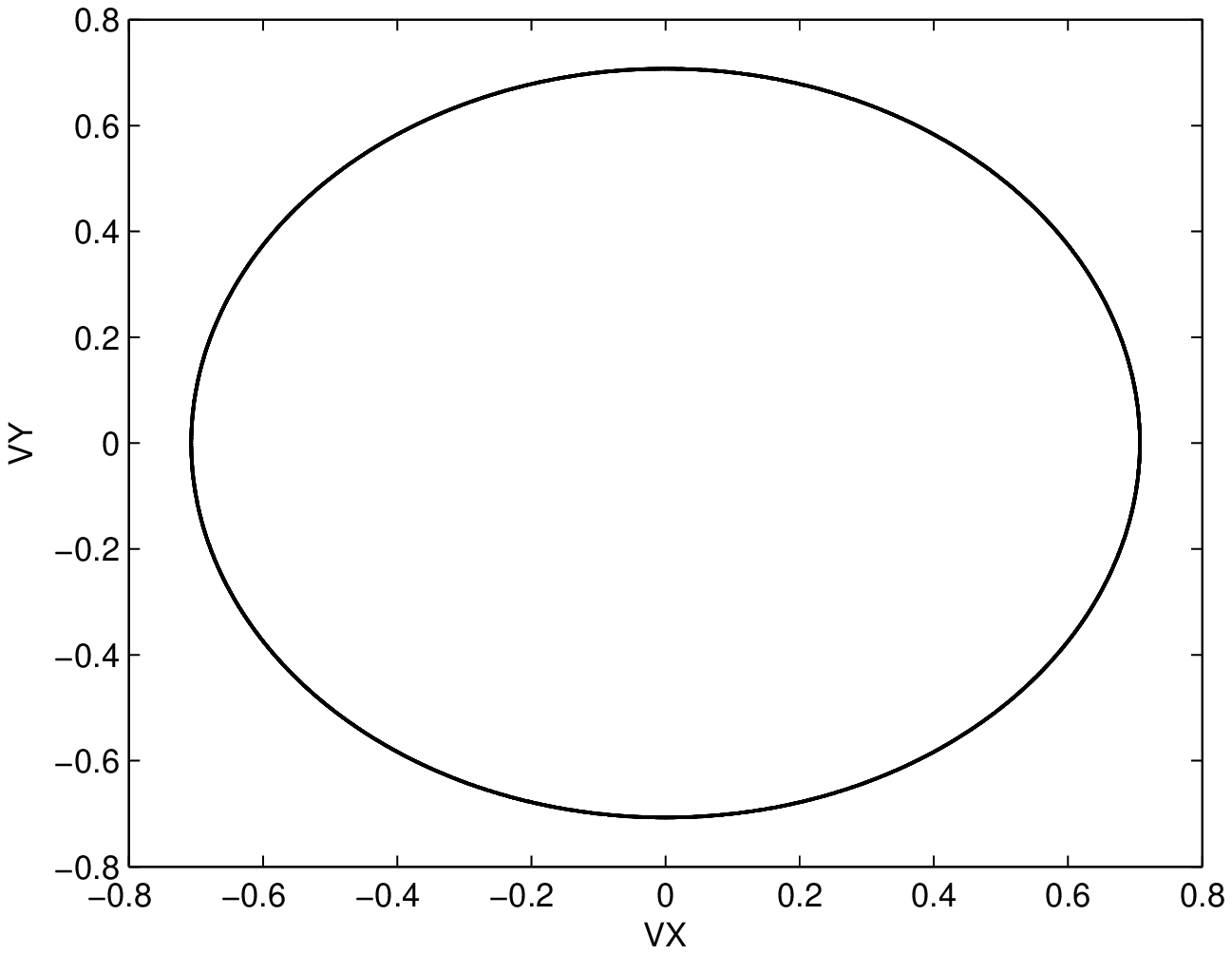}
\end{tabular}
\begin{tabular}{lll}
\includegraphics[width=3.8cm,height=3.6cm]{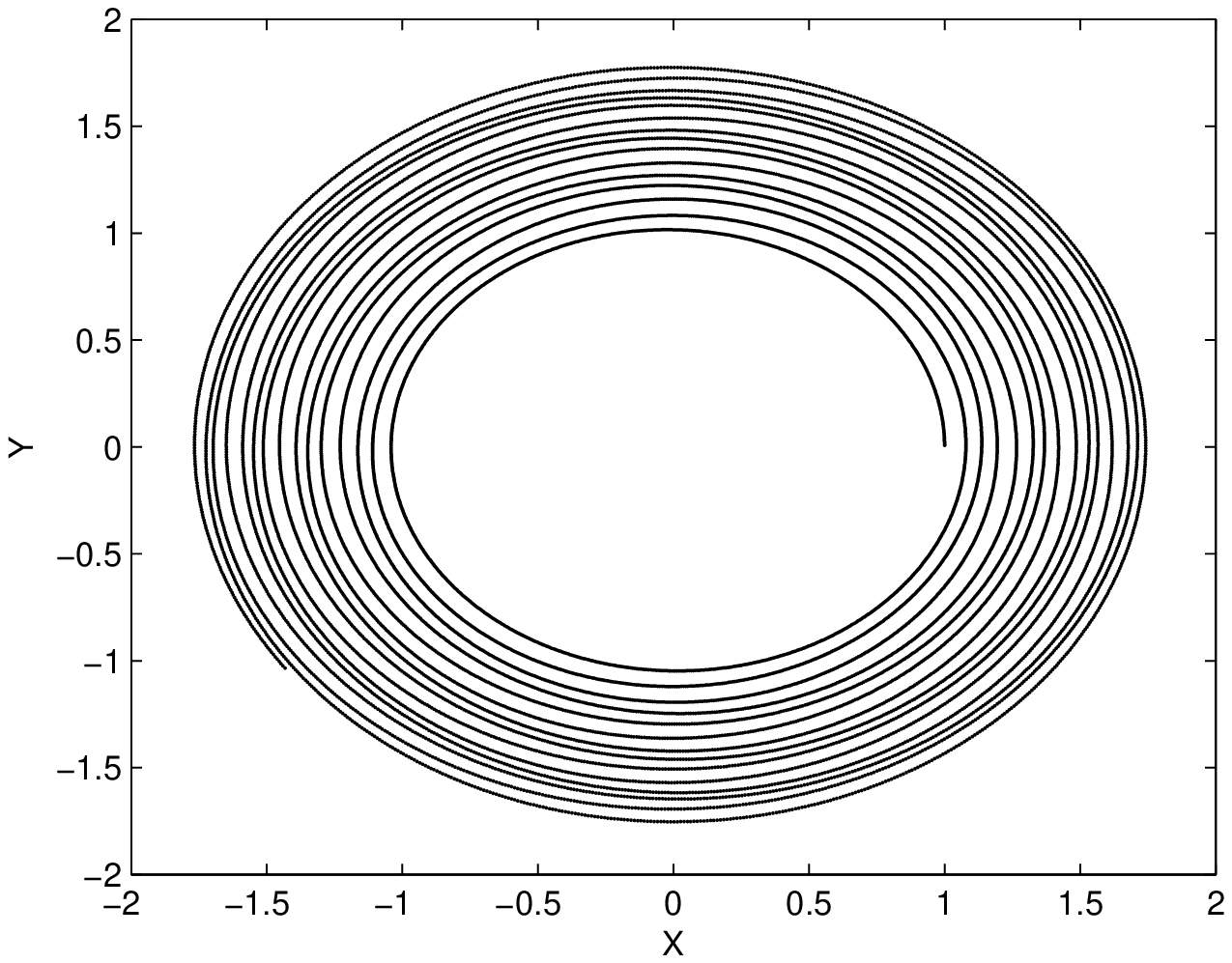}
\includegraphics[width=3.8cm,height=3.6cm]{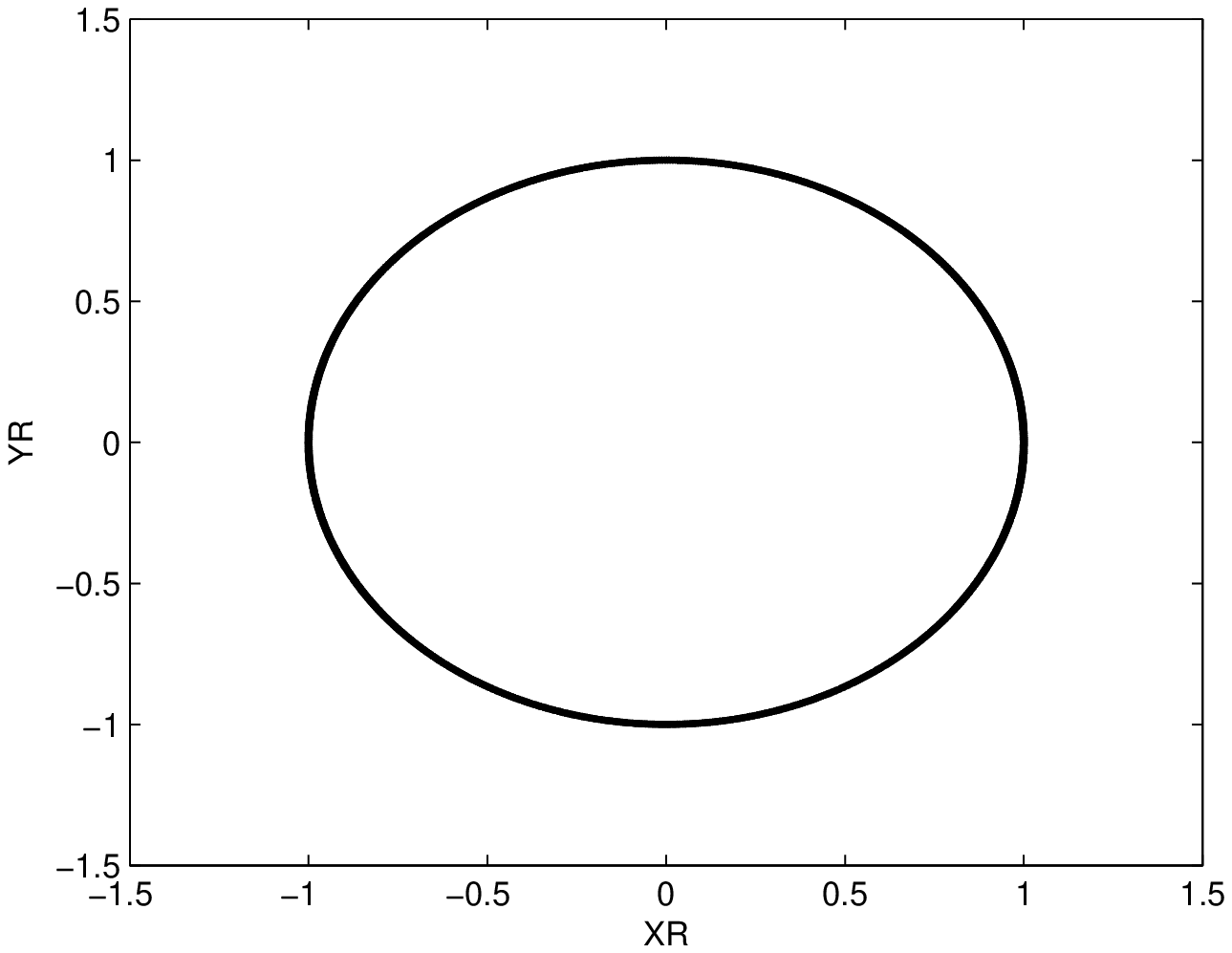}
\includegraphics[width=3.8cm,height=3.6cm]{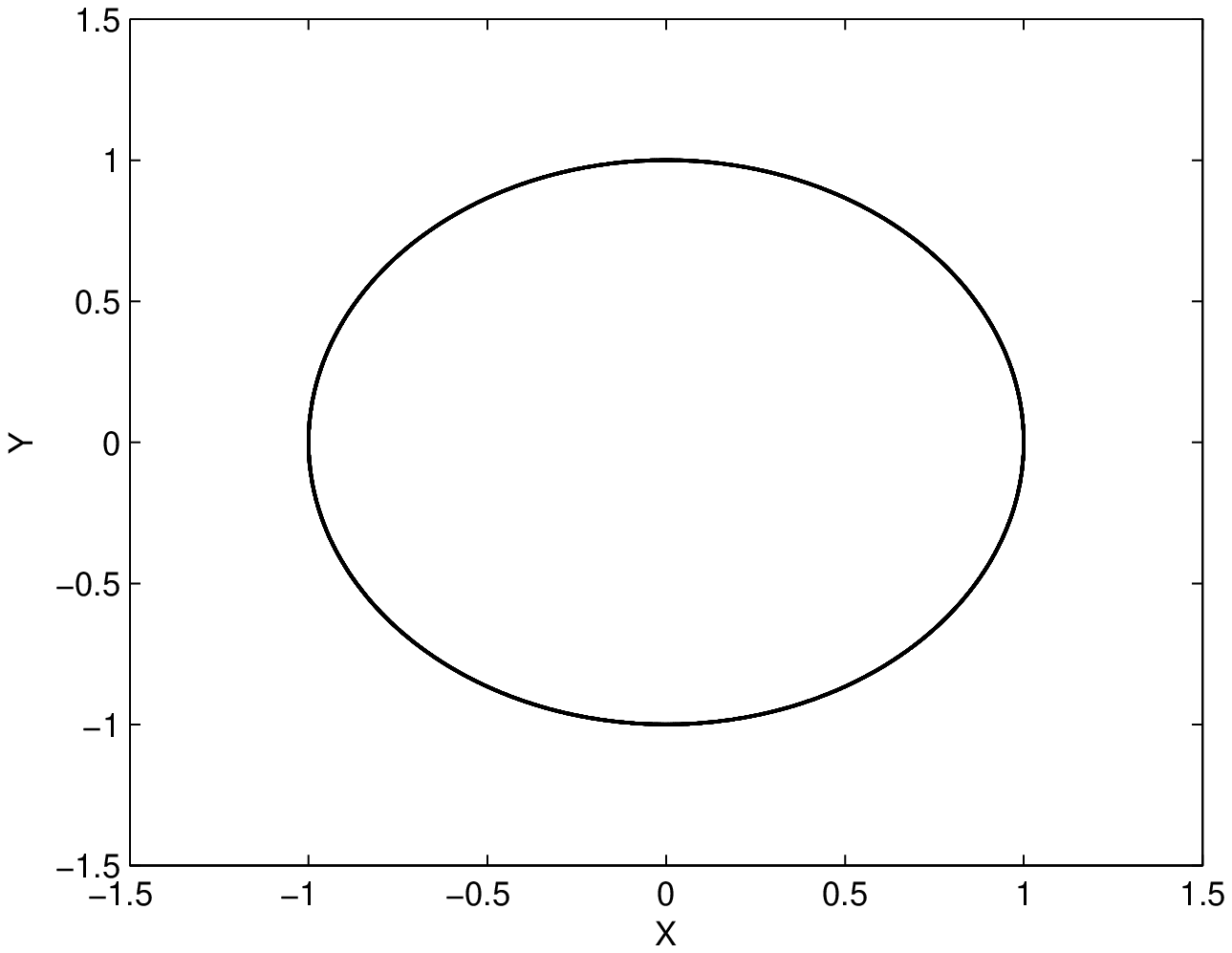}
\end{tabular}
\begin{tabular}{lll}
\includegraphics[width=3.8cm,height=3.6cm]{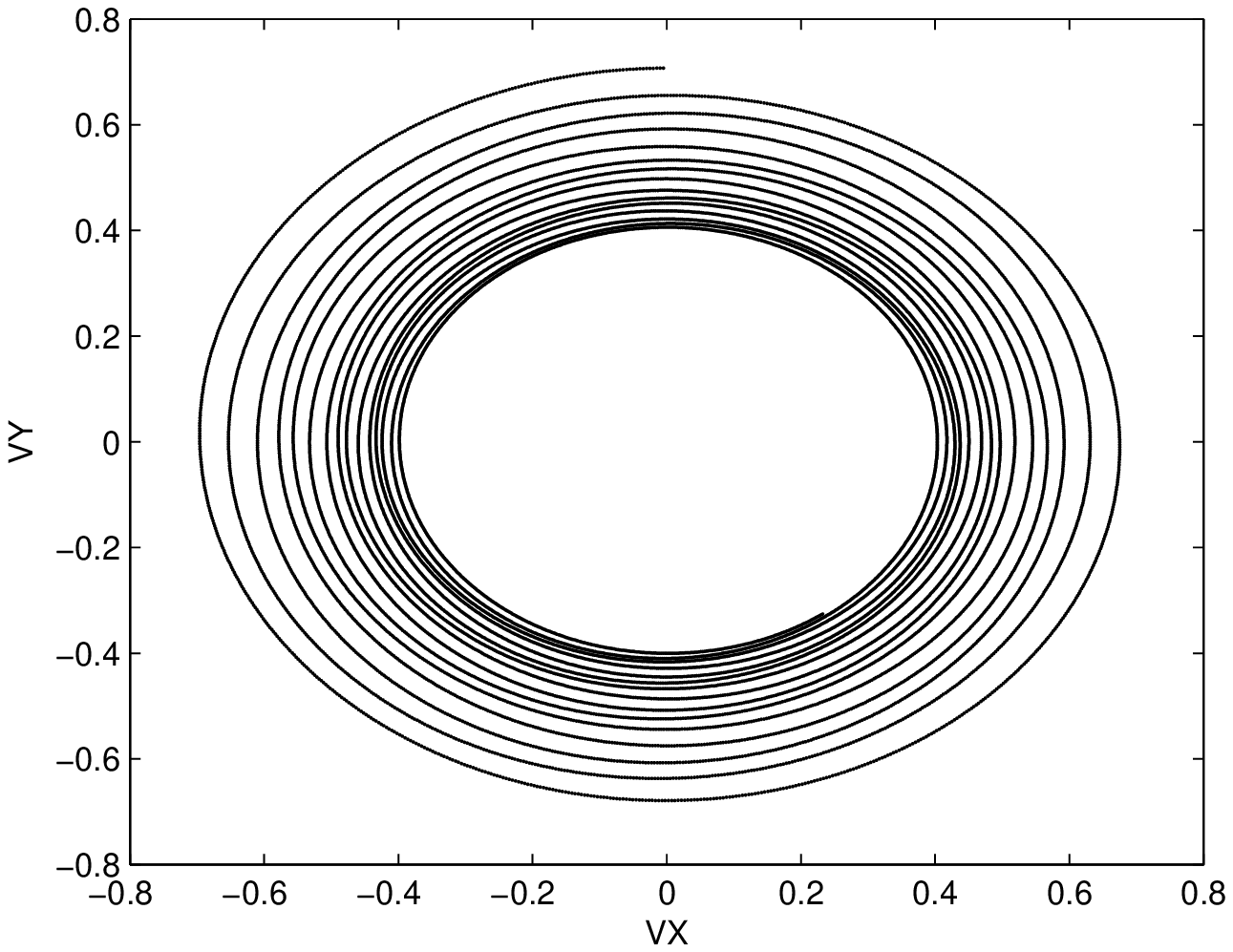}
\includegraphics[width=3.8cm,height=3.6cm]{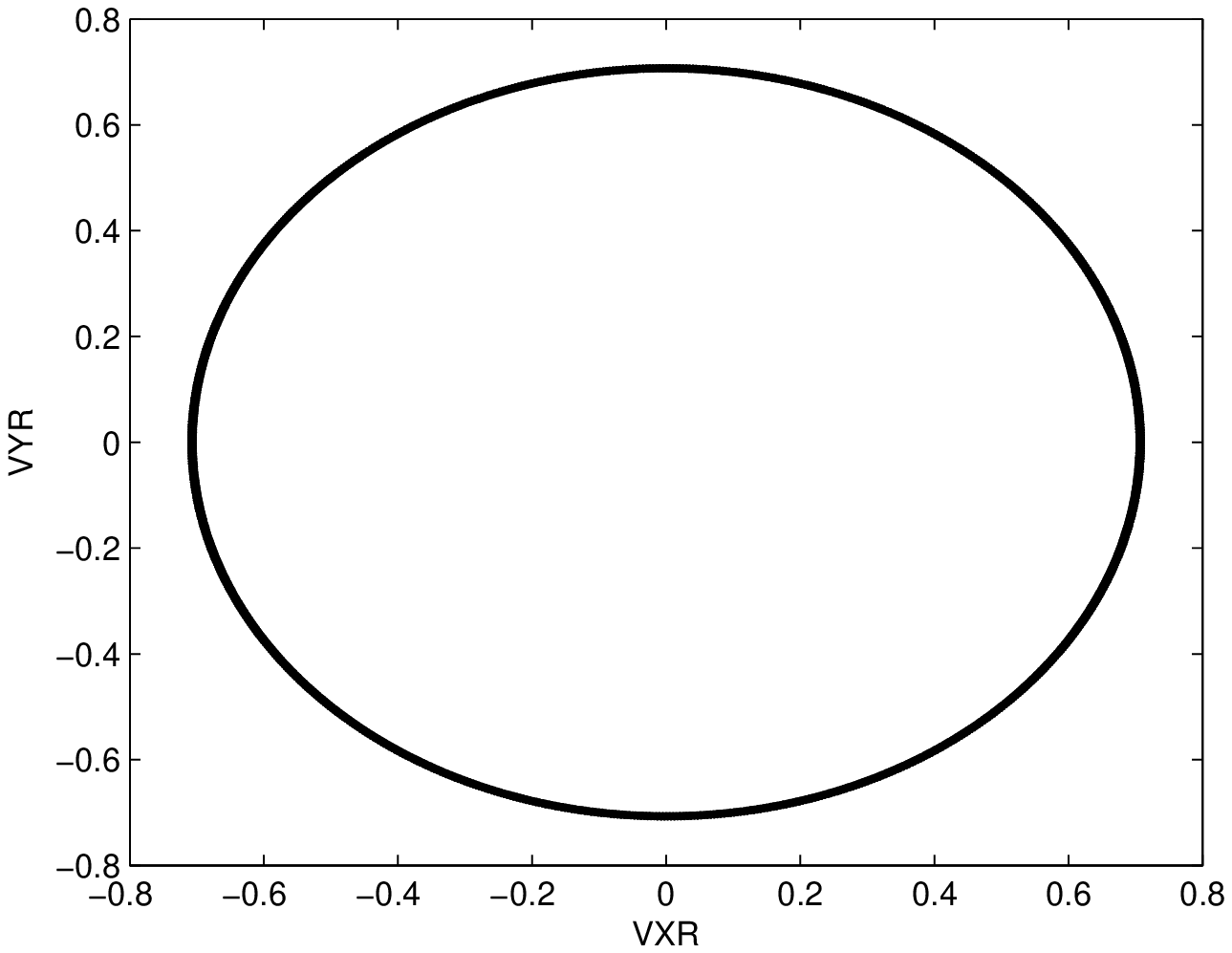}
\includegraphics[width=3.8cm,height=3.6cm]{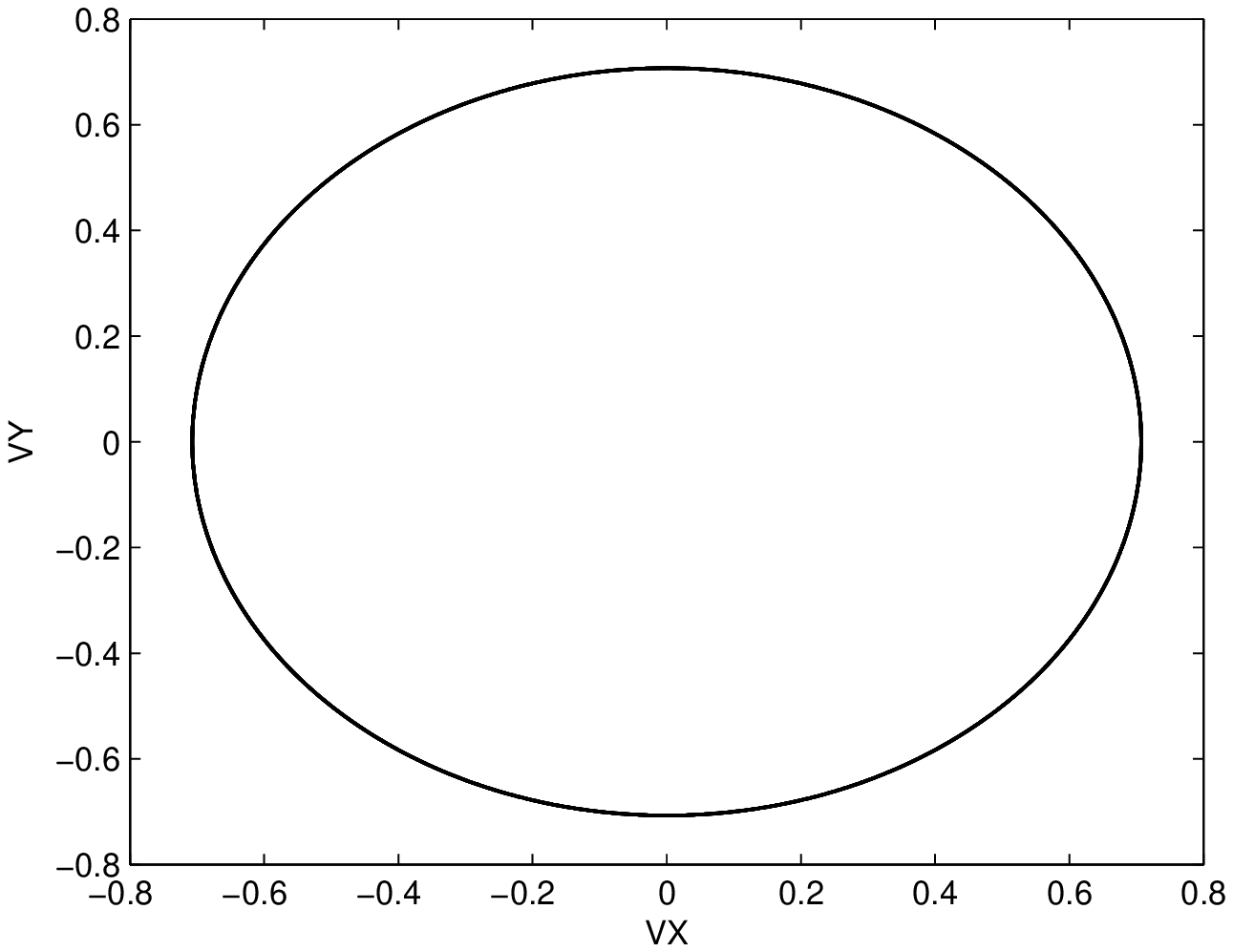}
\end{tabular}
\caption{Same as Fig \ref{fig1} for two orbits in the midplane of a DML, axisymmetric, Kuzmin disc. Rows 1-2: initial conditions are $x=1,~y=0,~z=0,~V_x=0.5,~V_y=-0.5,~V_z=0$. But here $t\_f=1500$ to show long-time behavior.
Rows 3-4: initial conditions, corresponding to a circular orbit for constant $\azg$, are $x=1,~y=0,~z=0,~V_x=0,~V_y=2^{-1/2},~V_z=0$ ($t\_f=300$).
Rows 5-6: the same initial conditions as rows 3-4, but with $t\_i=30$ and $t\_f=300$ to show the effect of increased adiabaticity parameter.}\label{fig3}
\end{figure}
\begin{figure}

\begin{tabular}{lll}
\includegraphics[width=0.35\columnwidth]{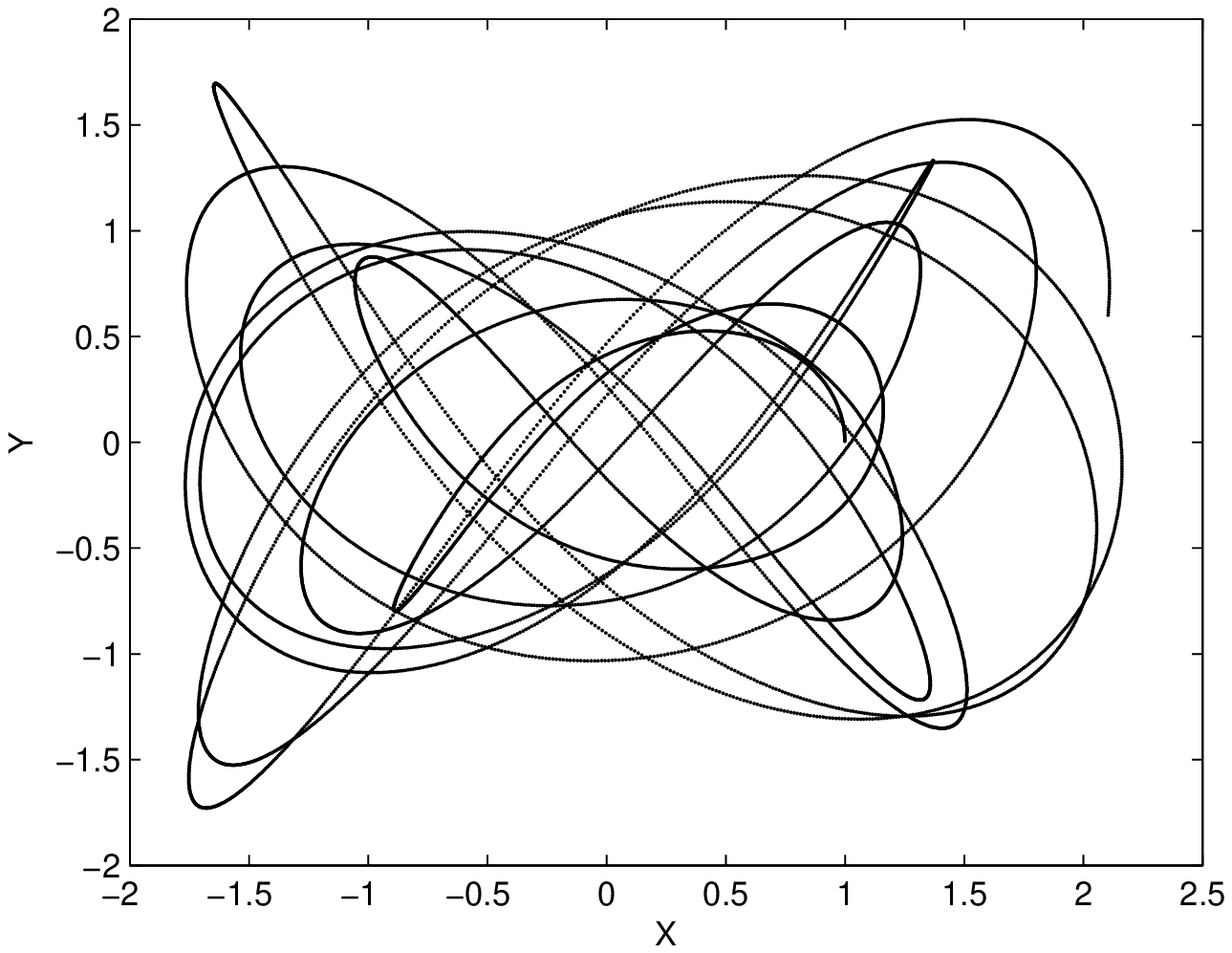}
\includegraphics[width=0.35\columnwidth]{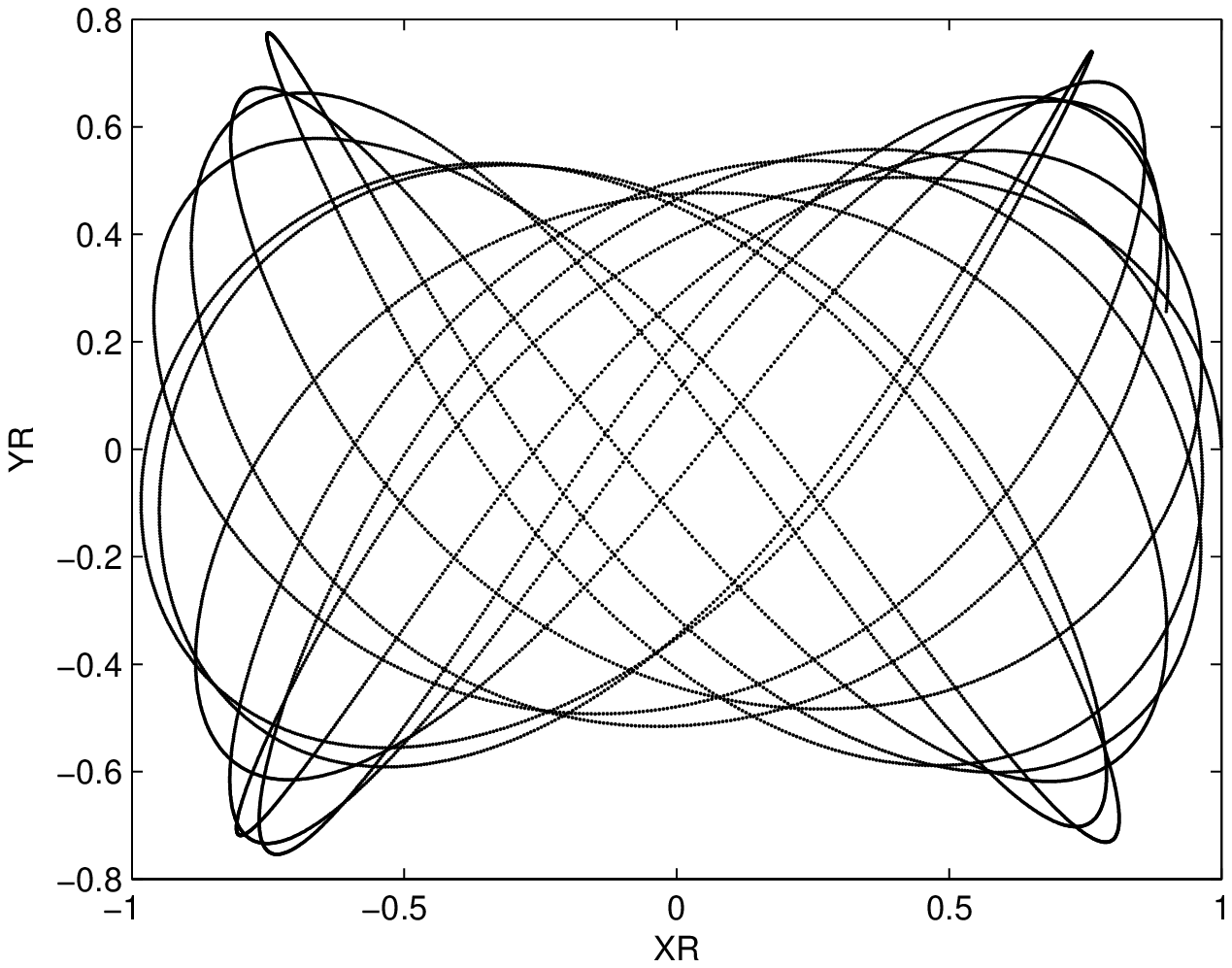}
\includegraphics[width=0.35\columnwidth]{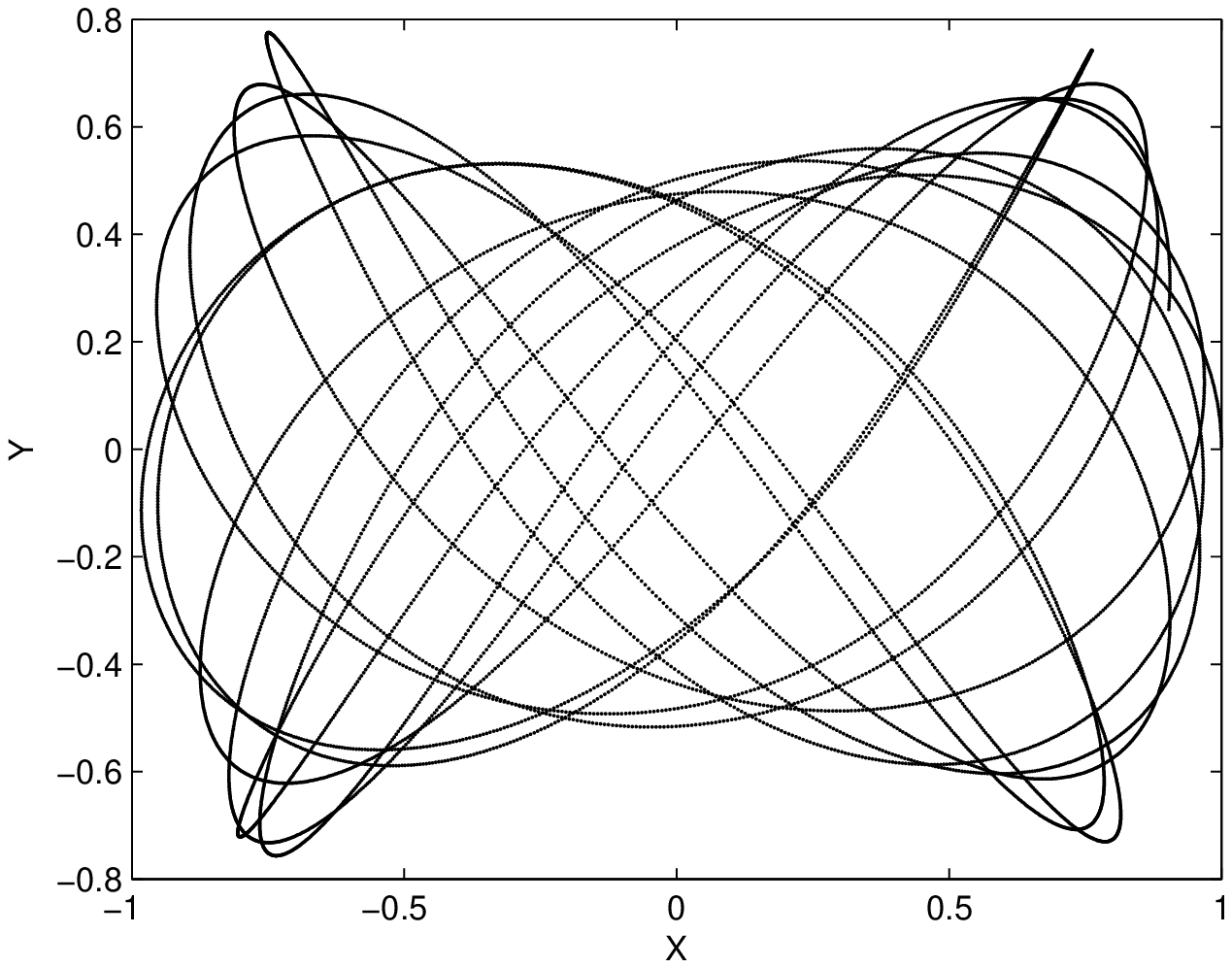}
\end{tabular}
\begin{tabular}{lll}
\includegraphics[width=0.35\columnwidth]{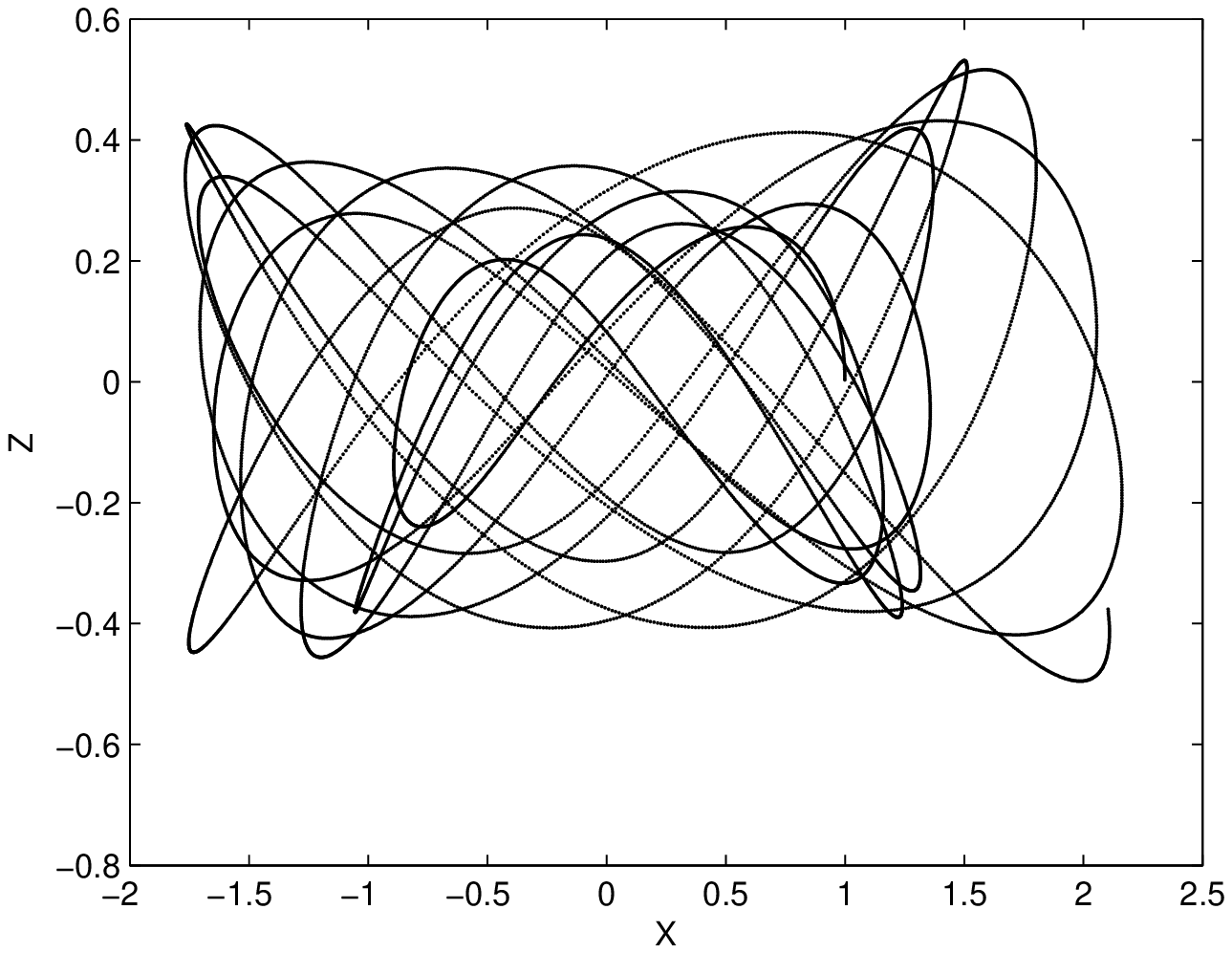}
\includegraphics[width=0.35\columnwidth]{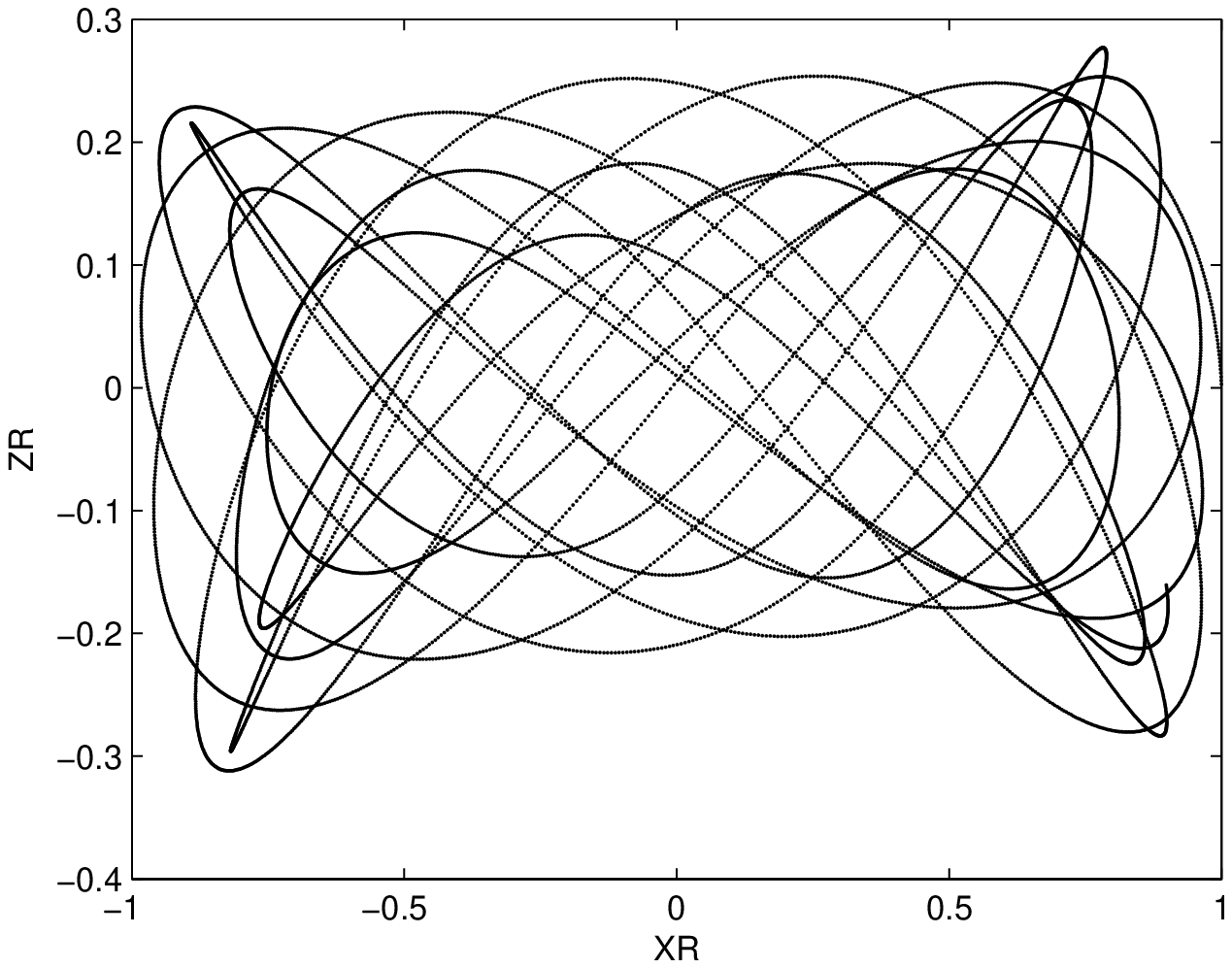}
\includegraphics[width=0.35\columnwidth]{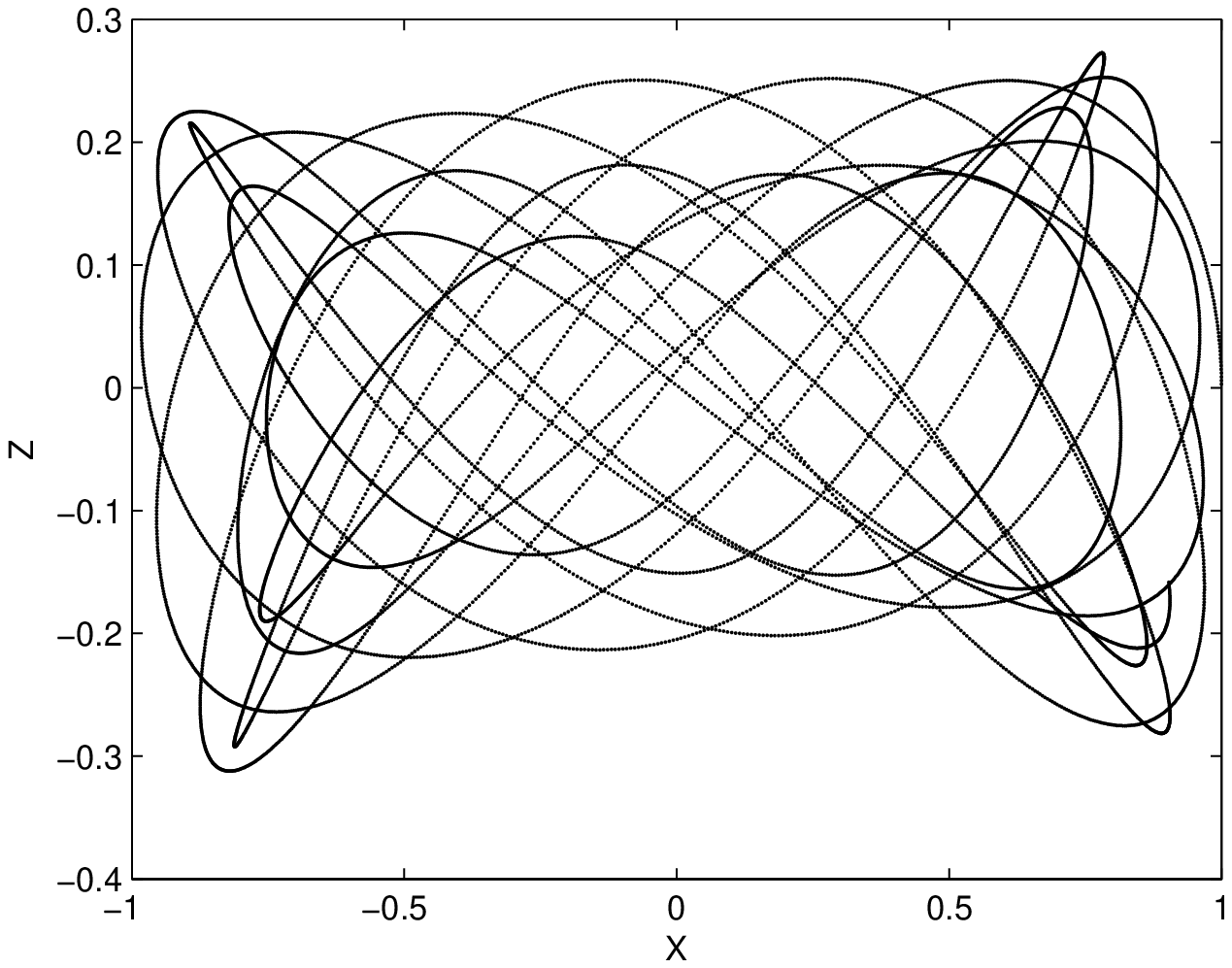}
\end{tabular}
\begin{tabular}{lll}
\includegraphics[width=0.35\columnwidth]{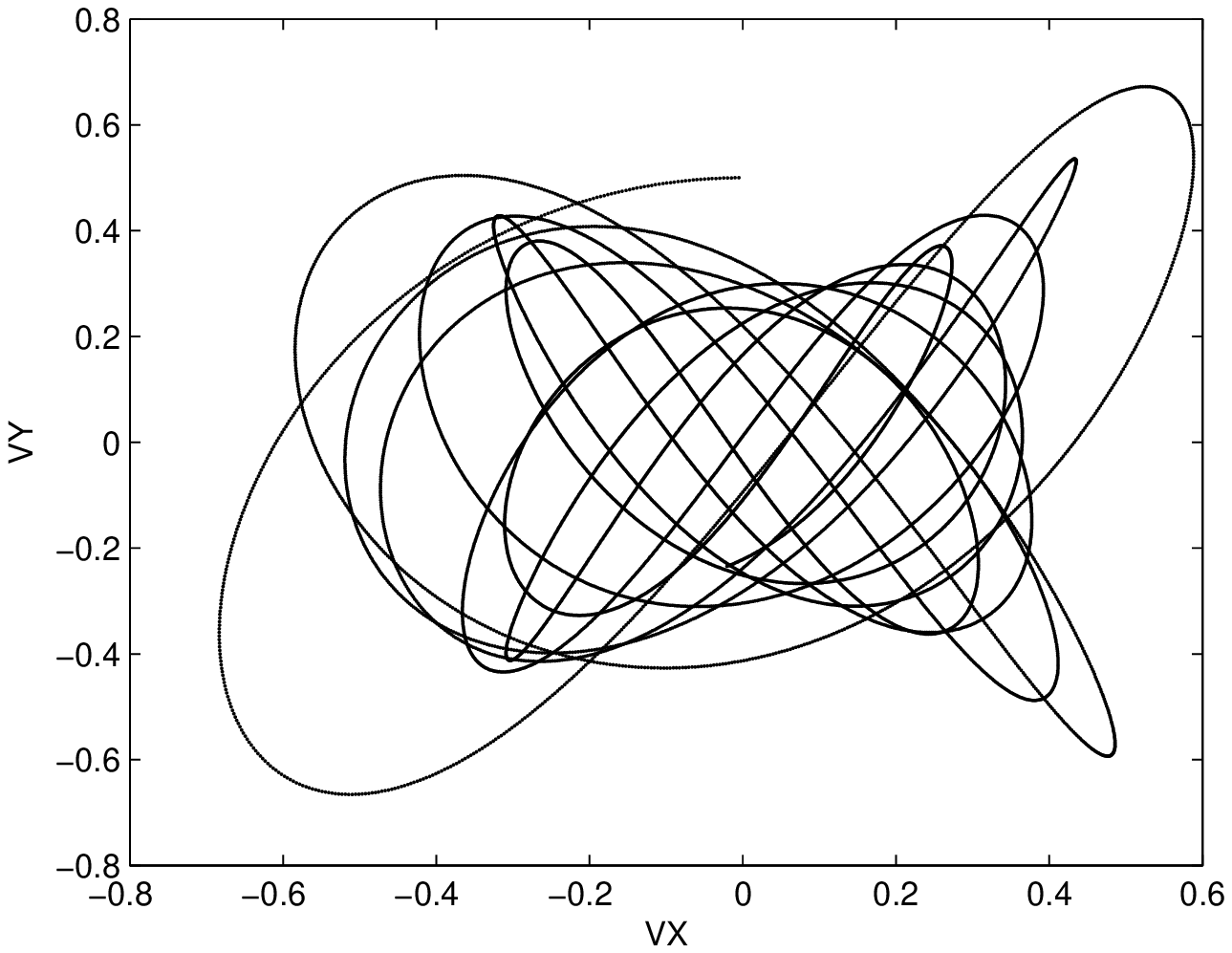}
\includegraphics[width=0.35\columnwidth]{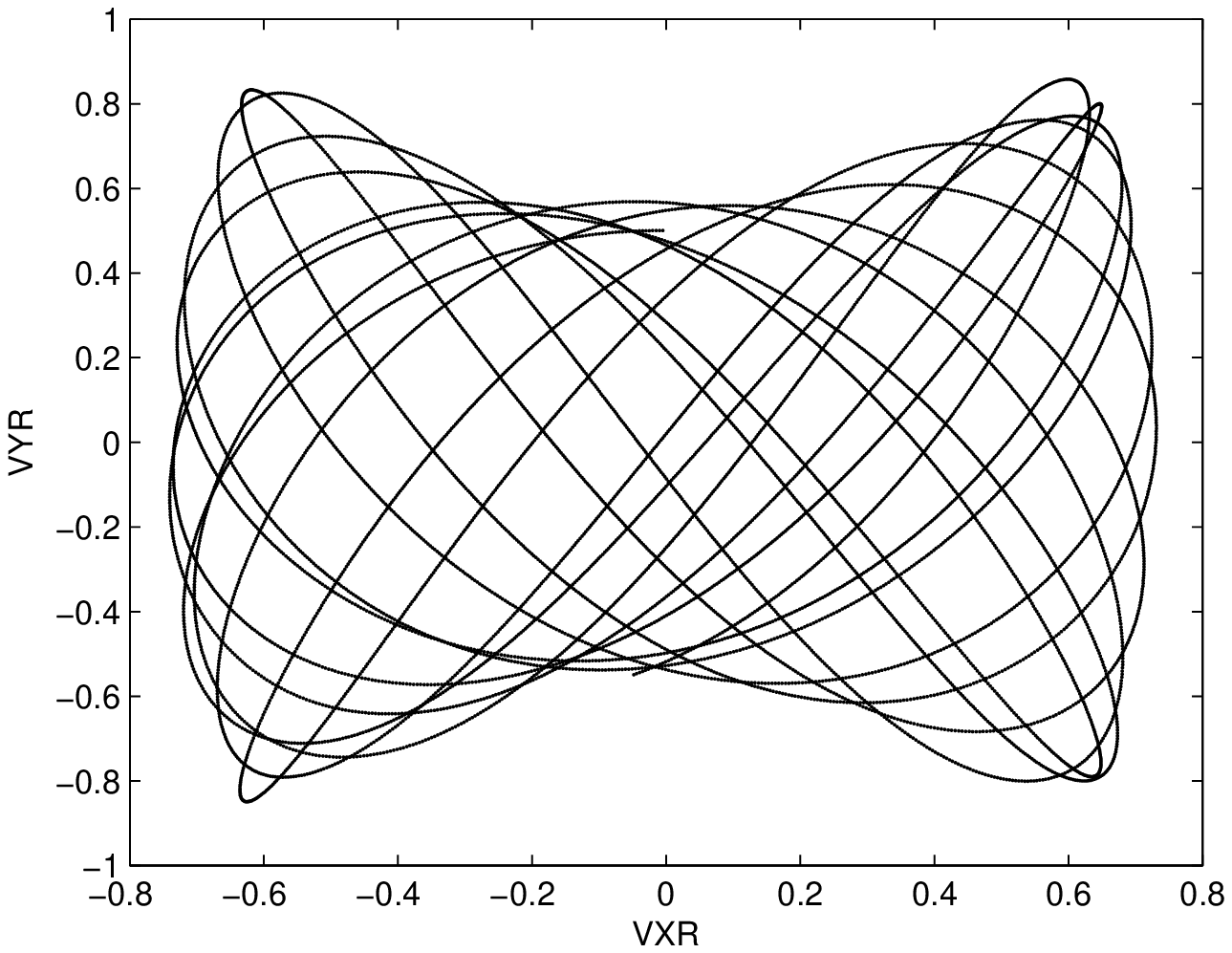}
\includegraphics[width=0.35\columnwidth]{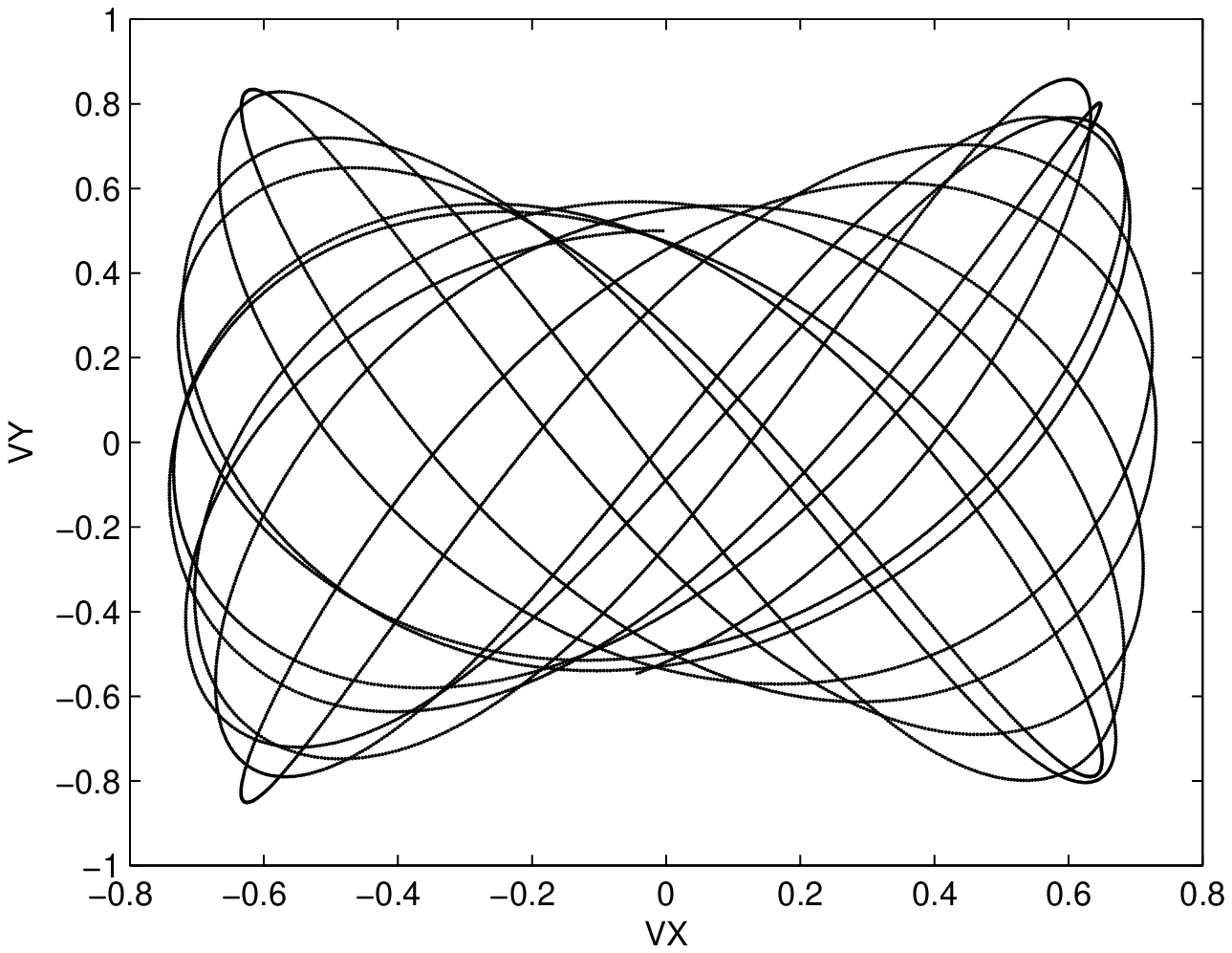}
\end{tabular}

\caption{Three-dimensional orbits. The $x-y$, $x-z$ and $V_x-V_y$ projections of a phase-space orbit in a triaxial isothermal sphere potential with axes parameters $q_1=2$, $q_3=3$. Initial conditions are:
$x=1,~y=0,~z=0,~V_x=0,~V_y=0.5,~V_z=0.3$. The initial and final times and the meaning of the columns are as in Fig. \ref{fig1}.}\label{fig4}
\end{figure}

I show in Figs. \ref{fig1}-\ref{fig4} examples of some of the calculations I ran (all for nonrotating systems).
I show results for three mass distributions for which the DML potential is known analytically:
(1) A point mass for which
 \beq\f\^*(R)=(M\azg)^{1/2}{\rm ln}(R).\eeqno{point}
(I drop the stars on $\vr$ and $R$.) This potential pertains also to the DML, two-body problem where the force is known to be inversely proportional to the separation.
It also pertains to the asymptotic field for any isolated, bounded mass distribution.
(2) A  DML isothermal sphere of mass $M$, with isotropic velocity distribution, for which \cite{milgrom84} \beq\f\^*(R)=(M\azg)^{1/2}{\rm ln}[(R^{3/2}+h^{3/2})^{2/3}]. \eeqno{isoth}
(3) A DML Kuzmin disc of mass $M$, for which the potential in known modified-gravity MOND theories--such as the nonlinear Poisson extension and QUMOND--is \cite{brada95}
\beq\f\^*(\vr)=(M\azg)^{1/2}{\rm ln}\{[r^{2}+(|z|+h)^{2}]^{1/2}\}. \eeqno{kuzma}
[Here, $r=(x^2+y^2)^{1/2}$,  $R=(r^2+z^2)^{1/2}$.]
In some cases biaxiality or triaxiality are introduced for the first two potentials by replacing $R$ by $(x^2+q_1y^2+q_2z^2)^{1/2}$.
\par
I work in units where $M=1$ and $\azgs=1$, and velocities are in units of $(M\azgs)^{1/4}=1$; length units are such that the initial scale length $h(0)=1$. Lengths are in arbitrary units for the point mass potential, and in units of $h(0)=1$ for the other two potentials. So all quantities appearing are dimensionless.
Initial conditions were not chosen with a particular aim, except to produce some variety of orbits.
\par
In the examples shown, I took $\azg(t)=t\_i/t$, where $t\_i$ is the initial time. Thus the initial variation time scale of $\azg$ is $t\_i$. The initial orbital frequency is $\o_o=V/R\sim 1$. So, formally the initial adiabaticity parameter is $\theta_a(0)\sim t\_i$, and then increases as $\theta_a\sim (tt\_i)^{1/2}$.
In most of the results I show, I took $t\_i=10$ and ran the orbits to $t\_f=300$ (with two exceptions to show certain variations). Thus $\azg$ decreased by a factor 30. Length are then expected to increase, and velocities to decrease, 2.3 fold.
These parameters would correspond roughly to a final-state galaxy with ratio of the Hubble-to-dynamical time  of order 60, started from $z\sim 9$, and $\azg$ varying as $H$.
\par
Note, however, that the dynamical time is not the only time scale characterizing the orbit. This
is not, in general, periodic with a frequency of the order of the inverse dynamical time; so, longer time scales, such as precession periods, etc. can appear in its description.
So the $\azg$ variations I study might not be strongly adiabatic in all regards, another reason to check numerically if, and under what conditions, our conjecture holds.
\par
The figures show various phase-space projections of the orbits $\vr$,
$\bar\vr\^*$, and $\vr\^*$ each for the appropriate time interval (note the different scales in the figures for the first one and the last two). If the last two orbits agree with each other our conjecture is supported.
\par
We see from the figures that this is indeed the case for the examples shown. It is also the case in other cases I studied, as long as the adiabaticity parameter is not too small. The agreement, while very good, is not quite perfect. However, the small differences between $\bar\vr\^*$ and $\vr\^*$ are negligible compared with those between them and $\vr$, which is the important fact. These small differences are attributable to departure from ideal adiabaticity for the parameters and orbits I took.
Indeed, I found that increasing the adiabaticity parameter (here, by taking a larger value to $t\_i$) greatly improves the agreement. This is demonstrated in the last two rows of  Fig. \ref{fig3}.

\par
Some satellite systems can be treated as detached from their mother system, whose effect on the satellite's internal dynamics is accounted for, to a good approximation, by embedding the satellite in a constant acceleration field due to the mother system. Because of the nonlinearity of MOND, the latter can greatly affect
the internal dynamics. This is known as the MOND external-field effect, and was discussed many times, starting from Ref. \cite{milgrom83}. In particular, when the external acceleration, $g\_{ex}$, dominates over the internal ones, the internal dynamics are quasi-Newtonian, with only one dimensioned constant, $G\_{eff}$, appearing in the effective theory. When the external field is also in the DML, $g\_{ex}\ll\az$, we have $G\_{eff}\approx G\az/g\_{ex}$. Examples of such systems are some of the dwarf spheroidal satellites-- such as the Milky Way's or Andromeda's. Our adiabatic scaling behavior applies in such a description as well since the external field itself, being in the DML, scales as $g\_{ex}\propto \eta^3$. So the general MOND adiabatic scaling behavior holds.
Seen differently, since we are in a world where $G$ does not vary, and $\az\propto \eta^4$, we see that $G\_{eff}\propto\eta$. This corresponds to the situation described at the end of footnote \ref{gvar}, and, again, implies the validity of the adiabatic scaling behavior.

\section{\label{newt}The partly Newtonian phase}
Many systems may have started the period of their existence relevant to our discussion with some of their parts in the Newtonian, not the deep-MOND, regime. Even systems that started wholly in the DML, are pushed by the above evolution towards the Newtonian regime. This is because accelerations everywhere in the system scale as $g\propto\azg^{3/4}$, which, with $G$ fixed, corresponds to $g\propto\az^{3/4}$; so $g/\az$ increase as $\az^{-1/4}$ (see footnote \ref{jama}). This increases by a factor of $\sim 2.5$ in the examples shown in Sec. \ref{DML}.
\par
For example, in a DML Kuzmin disc of mass $M$, the acceleration in the plane, in modified-gravity MOND theories, is \cite{brada95}
\beq\frac{g(r)}{\az(t)}=\frac{r\^*\_M}{\eta}\frac{\eta r}{1+(\eta r)^2}.\eeqno{gamulas}
Here, $\eta=\eta(t)$, $r\_M\equiv (MG/\az)^{1/2}$ is the MOND radius of the whole mass, and $r\^*\_M$ its initial value, when $\eta=1$. All radii (including $r\_M$) are in units of the initial scale length of the mass distribution, $h\^*$ [see eqs. (\ref{isoth}-\ref{kuzma})].\footnote{In so-called ``modified inertia'' formulations of the type discussed in Ref. \cite{milgrom94a}, the acceleration in the plane is $g(r)/\az=(r\^*\_M/\eta)(\eta r)^{1/2}/[1+(\eta r)^2]^{3/4}$.}
The maximum acceleration is at $\eta r=1$, where $g(r)/\az(t)=r\^*\_M/2\eta(t)$.
So, even if we start with a DML disc, with $r\^*\_M/2\ll 1$, the region around this radius becomes Newtonian around the time when $\eta\sim r\^*\_M/2$.
\par
Once parts of the system are not in the DML, it is difficult to divine the exact response to adiabatic $\az$ variations. To answer these question requires numerical studies for various systems from discs to spherical and with various initial conditions.
Here I can only make some qualitative guiding observations of what can be expected. The transition from the DML to ND is expected to be gradual. But to simplify matters somewhat I write below, heuristically, as if the transition is sharp.

Since $\az$ does not appear in ND, its variation does not affect, of course, systems that are fully Newtonian.\footnote{There are no isolated, fully Newtonian systems since, when probed far enough from their center--e.g., by gravitational lensing--DML aspects must show up. But if we consider a system embedded in a Newtonian external field, or we are interested only in the in-system dynamics (of, e.g., a star), we can discuss strict ND.} However, in a mixed system all regions are affected, not only the MONDian regions. Generally, none is described by the simple adiabatic scaling laws discussed in Sec. \ref{DML}. The reason for this is generally twofold: (1) In general, the mass distribution anywhere can affect the potential field everywhere. Thus we cannot generally have the scaling of the potential field even in the DML regions. (2) Constituents on their orbits may traverse both Newtonian and DML regions. So we cannot generally retain the scaling of the orbits, and this in general breaks the scaling behavior in the DML regions as well. These two effects are coupled, each affecting the way the other acts.
\par
An acute example of the first point concerns systems that are subject to a dominant external acceleration field. In this case it will be the external field that becomes Newtonian first. According to MOND this will render the whole system Newtonian at once with the external field.
\par
These obstacles to scaling behavior are not present in some interesting special cases. For example, in a spherical system--where in known modified gravity MOND theories the gravitational field depends only on the total enclosed mass--the first obstacle does not exist, since adiabatic variations of $\az$ cannot, in themselves, induce shall crossing. The second effect does not exist if there are no orbits that traverse both DML and Newtonian regions, for example, if the orbits are circular as in a disc galaxy.
For the same reasons, aspects related to the asymptotic dynamics of an isolated system are not affected by some parts of the system being Newtonian, as only the total mass enters them. For example, the asymptotic rotation speed always continues to be $\propto \az^{1/4}$, and large-impact-parameter-light-bending angles always scale as $\az^{1/2}$.
\par
Beyond equilibrium dynamics, stability properties are also expected to change as we cross from the DML to the Newtonian regime. It has been well discussed (e.g.,
\cite{milgrom89a,christ91,brada99}) that MOND endows self-gravitating systems with added stability relative to ND. The reason for this is that in the DML, accelerations scale as the square root of densities (for a given size), while in ND they scale as densities. Thus, density perturbations produce only half the acceleration perturbation in the DML than in ND. This has been suggested as the reason for the observed upper limit of $\sim\az/G$ on the mean surface density of galactic discs (underlying the so-called ``Freeman law''). The idea is that discs with higher surface densities are Newtonian and may be unstable to bar formation, with a subsequent churning of disc material to a (pseudo) bulge (see, e.g., Ref. \cite{kc04}) always leaving an almost MONDian disc.
\par
While equilibrium properties change smoothly as we go from the DML to the Newtonian regime, stability properties change more abruptly, as they hinge on the derivative of the transition, since they involve the response to small changes. For example, in MOND theories that involve an interpolating function, the equilibrium properties involve this function while stability properties involve its derivative. This can be demonstrated clearly if we take the transition to be sharp. Then we can have, for example, two disc galaxies
with a mass distribution that produces a constant acceleration in the plane (within the material disc). For one, this acceleration is just above $\az$ and for the other just below. The two models have the same equilibrium structure, but very different stability properties, the MONDian one being ``twice'' more stable.\footnote{We are comparing here the stability in MOND of two baryonic galaxies, one in the DML, the other Newtonian. Galaxies with a dark matter halo that is required to explain the equilibrium properties in ND may be much as stable (too stable??) than DML discs, since, unlike halos, MOND enhances stability only by a limited factor.}
Thus, even if equilibrium considerations tell us that $g/\az$ should continue to grow after crossing to the Newtonian regime, instabilities, which may show up, may limit this growth. This would be a good issue to check numerically.
\par
Spheroidal systems also exhibit a similar upper limit on their mean surface densities. This has no known explanation in standard dynamics. In MOND it may be related to the fact that self-gravitating, finite-mass, isothermal spheres (which do not exist in Newtonian dynamics) can exist only if they obey this constraint.
How this argument survives adiabatic transit of such systems into the Newtonian regime would also necessitate numerical studies, again allowing for the role of the onset of instabilities (Ref. \cite{wu09} discussed stability of MOND models of triaxial galaxies with a fixed $\az$).

\clearpage
\end{document}